\documentclass[fleqn,usenatbib]{mnras}
\usepackage{newtxtext,newtxmath}
\usepackage{comment}
\usepackage{multirow}
\usepackage{mathtools}
\usepackage[T1]{fontenc}
\usepackage{ae,aecompl}

\usepackage{graphicx}	% Including figure files
\usepackage{amsmath}	% Advanced maths commands
\usepackage[ruled,vlined]{algorithm2e}
\usepackage{algpseudocode}
\usepackage{enumitem} % items tools
\usepackage{xfrac} % fractions

\setlist[itemize]{leftmargin=*}
\newcommand{\bigO}[1]{\mathcal{O}(#1)}

\newcommand{\ud}{\mathrm{d}}

\newcommand{\derfrac}[2]{\frac{\ud #1}{\ud #2}}

\newcommand{\Rsol}[1]{{#1}\:\mathrm{R_\odot}}
\newcommand{\msol}[1]{{#1}\:\mathrm{M_\odot}}
\newcommand{\zsol}[1]{{#1}\:\mathrm{Z_\odot}}
\newcommand{\Sigmasol}[1]{{#1}\:\mathrm{M_\odot\:pc^{-2}}}
\newcommand{\rhosol}[1]{{#1}\:\mathrm{M_\odot\:pc^{-3}}}

\newcommand{\ratesol}[1]{{#1}\:\mathrm{M_\odot yr^{-1}}}
\newcommand{\collratesol}[1]{{#1}\:\mathrm{yr^{-1}}}
\newcommand{\msolrange}[3]{{#1}\:\mathrm{M_\odot\lesssim}\:{#2}\lesssim{#3}\:\mathrm{M_\odot}}
\newcommand{\zsolrange}[3]{{#1}\:\mathrm{Z_\odot\lesssim}\:{#2}\lesssim{#3}\:\mathrm{Z_\odot}}
\newcommand{\Sigmasolrange}[3]{{#1}\:\mathrm{M_\odot} \mathrm{pc}^{-2} \lesssim \:{#2}\lesssim{#3}\:\mathrm{M_\odot} \mathrm{pc}^{-2}}

\newcommand{\sevn}{\texttt{SEVN}}

\newcommand{\bifrost}{\texttt{BIFROST}}
\newcommand{\ketju}{\texttt{KETJU}}

\newcommand{\nbody}{\textit{N}-body}

\title[Stellar collisions, metallicity \& IMBHs]{{FROST-CLUSTERS -- III. Metallicity-dependent intermediate mass black hole formation by runaway collisions in dense star clusters}}
\author[A. Rantala et al.]{Antti Rantala$^{1,2,3}$\thanks{E-mail: anttiran@mpa-garching.mpg.de}, Thorsten Naab$^{1}$, Natalia Lahén$^{1,4}$, Klaus Reuter$^{5}$, Markus Rampp$^{5}$, \newauthor Martyna Chru{\'s}li{\'n}ska$^{6}$, Bastián Reinoso$^{7}$\\
$^{1}$Max-Planck-Institut f\"ur Astrophysik, Karl-Schwarzschild-Str. 1, 
D-85748, Garching, Germany\\
$^{2}$Institute of Astronomy, University of Cambridge, Madingley Road, Cambridge CB3 0HA, UK\\
$^{3}$Kavli Institute for Cosmology, Cambridge (KICC), University of Cambridge, Madingley Road, Cambridge CB3 0HA, UK\\
$^{4}$ Zentrum f\"ur Astronomie der Universit\"at Heidelberg, Astronomisches Rechen-Institut, M\"onchhofstr. 12-14, D-69120 Heidelberg\\
$^{5}$Max Planck Computing and Data Facility (MPCDF), D-85748, Garching, Germany\\
$^{6}$European Southern Observatory, Karl-Schwarzschild-Str. 2, D-85748 Garching, Germany\\
$^{7}$Department of Physics, University of Helsinki, P.O. Box 64, Gustaf Hällströmin katu 2, FI-00014, University of Helsinki, Finland\\
}
\date{Accepted XXX. Received YYY; in original form ZZZ}
\pubyear{2024}

\begin{document}
\label{firstpage}
\pagerange{\pageref{firstpage}--\pageref{lastpage}}
\maketitle

\begin{abstract}

We explore the formation of intermediate mass black holes (IMBHs), potential seeds for supermassive black holes (SMBHs), via runaway stellar collisions for a wide range of star cluster (surface) densities ($\Sigmasol{4\times10^3} \lesssim \Sigma_\mathrm{h} \lesssim \Sigmasol{4\times10^6}$) and metallicities $(\zsolrange{0.01}{Z}{1.0})$. Our sample of isolated (>1400) and hierarchical (30) simulations of young, massive star clusters with up to $N=1.8\times10^6$ stars includes collisional stellar dynamics, stellar evolution, and post-Newtonian equations of motion for black holes using the \bifrost{} code. High stellar wind rates suppress IMBH formation at high metallicities ($Z\gtrsim\zsol{0.2}$) and low collision rates prevent their formation at low densities ($\Sigma_\mathrm{h}\lesssim\Sigmasol{3\times10^4}$). The assumptions about stellar wind loss rates strongly affect the maximum final IMBH masses ($M_\bullet\sim\msol{6000}$ vs. $\msol{25000}$). The total stellar mass loss from collisions and collisionally boosted winds before $t=3$ Myr can together reach up to $5$--$10\%$ of the final cluster mass. We present fitting formulae for IMBH masses as a function of host star cluster $\Sigma_\mathrm{h}$ and $Z$ which can be used to seed SMBHs in high resolution cosmological hydrodynamical simulations and in semi-analytic models for galaxy formation. Our results favour IMBH formation in dense low metallicity environments similar to $z\sim10$ James Webb Space Telescope (\textit{JWST}) proto globular clusters. IMBH formation is suppressed in the high metallicity and low density conditions of the local Universe.
\end{abstract}

% At high cluster densities and low metallicities IMBHs up to $\msol{6000}$ can form.


%%%%%%%%%%%%%%%%%%%%%%%%%%%%%%%%%%%%%%%%%%%%%%%%%%
%%%%%%%%%%%%%%%%% BODY OF PAPER %%%%%%%%%%%%%%%%%%
%%%%%%%%%%%%%%%%%%%%%%%%%%%%%%%%%%%%%%%%%%%%%%%%%%

\section{Introduction}\label{section: 1}

The origin of supermassive black holes (SMBHs; $M_\bullet \geq \msol{10^6}$) remains a major unsolved puzzle in modern astrophysics \citep{Rees1984,Inayoshi2020,Volonteri2021}. Recently, the \textit{James Webb Space Telescope} (\textit{JWST}) observations of active galactic nuclei (AGN) powered by accreting SMBHs at redshifts above $z \gtrsim 6$--$10$ \citep{Maiolino2024a,Maiolino2024b,Juodzbalis2024,Scholtz2024,Übler2024} have only deepened the conundrum of the SMBH origins. The majority of theoretical SMBH seed formation scenarios (Pop-III stars, direct collapse clouds, runaway Pop-II collisions) rely on the low metallicity environments of the high redshift ($z>10$) galaxies. Primordial Pop-III stars can only occur at metallicities below $Z\lesssim\zsol{10^{-4}}$ \citep{Bromm2001}, direct collapse gas clouds require low metallicities to avoid fragmentation (e.g. \citealt{Regan2017}) while massive stars built up by runaway Pop-II stellar collisions \citep{PortegiesZwart2004} must avoid high wind loss rates (e.g. \citealt{Mapelli2016}) to produce a massive black hole remnant. In most SMBH and galaxy stellar mass build up scenarios, possibly excluding primordial black holes \citep{Dayal2025}, the galaxies themselves are expected to become metal enriched (e.g. \citealt{Tassis2012}) during their assembly. From this point of view, the recently observed near pristine \textit{JWST} SMBH host galaxies \citep{Maiolino2025,Juodzbalis2025} appear especially puzzling. Nevertheless, it is not uncommon for the early ($6\lesssim z \lesssim 11$) \textit{JWST} AGN and galaxies to show metallicities of $Z\sim\zsol{0.10}$ or even higher \citep{Bunker2023,Isobe2023}. The observed nitrogen enrichment at low metallicities and high star formation densities of the early \textit{JWST} galaxies and AGN (e.g. \citealt{Bunker2023,Cameron2023,Ji2024,Schaerer2024,Isobe2025,Ji2025,Naidu2025}) may point to dense low-metallicity environments in
which stellar collisions can frequently occur \citep{Gieles2018,Charbonnel2023,MarquesChaves2024,Schaerer2025,Ebihara2026}, indicating the stellar collisional picture for the SMBH seed formation may be indeed feasible for at least a subset of early galaxies. Moreover, archaeological studies of the initial build-up phase of the Milky Way suggest highly clumpy and clustered formation environments for the early low metallicity ([Fe/H]$<-1.3$) stars of our Galaxy \citep{Belokurov2022}.

While especially common during the first gigayear, clumpy low metallicity star formation (\citealt{Adamo2024, Fujimoto2024, Mowla2024, Bradac2025, Abdurrouf2025,Yanagisawa2026, Claeyssens2026}, see also e.g. \citealt{vanDonkelaar2026}) is not restricted to the redshift $z>6$ Universe \citep{Claeyssens2025, Zhu2026}. Giant star forming clumps have been observed to be very common in more evolved massive gas-rich disk galaxies around cosmic noon at redshifts 1$\lesssim$ z $\lesssim$ 3 \citep[e.g.][]{2005ApJ...627..632E,2008ApJ...687...59G,2009ApJ...692...12E,Genzel2011}. Cosmological simulations suggest that $20\%$ of the stars with metallicities $Z<\zsol{0.10}$ have formed since $z=2$ \citep{Pakmor2022}. However, for $Z<\zsol{0.01}$ the fraction is considerably lower, only $3\%$. In general, the simulation models are in broad agreement with observation-based frameworks (e.g. \citealt{Chruslinska2019}). Recent observation-based models \citep{Chruslinska2024b,Chruslinska2025} indicate that up to $\sim6$--$60\%$ of the total stellar mass formed at low metallicities below $Z<\zsol{0.1}$ since $z=10$. The relatively large model uncertainties originate from the ambiguities in the absolute metallicity scale, the evolution in the star formation--mass relation and the low-mass end of the galaxy mass function. In any case the overall fraction of extremely low metallicity stars ($Z\lesssim\zsol{0.01}$) is less than $<5\%$.

Today, the star cluster population of the Milky Way and the Magellanic clouds consists of two broad sub-populations. Ancient ($>10$ Gyr) globular clusters (GCs) show very low metallicities below $Z\lesssim\zsol{0.04}$ while young massive star clusters (YMCs) with ages less than $1$ Gyr show a higher and somewhat broader metallicity distribution in the range of $\zsol{0.10}\lesssim Z \lesssim \zsol{1.0}$ (e.g. \citealt{Mackey2003}). The local star clusters follow a relatively shallow mass-radius relation \citep{Brown2021} while older clusters typically show extended sizes compared to their recently formed counterparts \citep{PortegiesZwart2010} attributed to internal and tidally driven star cluster evolution. However, the typical present-day stellar (surface) densities of the local YMCs are several orders of magnitude lower than in the $z\sim10$ \textit{JWST} proto GC star clusters \citep{Adamo2024,Abdurrouf2025} that can reach surface densities of $\Sigma_\mathrm{h} \gtrsim \Sigmasol{10^5}$. Despite their relatively low densities, the local YMCs contain some of the most massive known individual stars: the R136 star cluster in the Large Magellanic Cloud (LMC) with a half mass surface density of $\Sigma_\mathrm{h} \sim \Sigmasol{3.6\times10^2}$ \citep{McLaughlin2005} at $Z\sim\zsol{0.5}$ hosts several very massive stars with masses likely above $\gtrsim\msol{150}$ \citep{Crowther2010}.

Metallicity plays a defining role in stellar physics due to its interconnected effects on stellar nuclear reaction pathways, gas opacities and wind loss rates (e.g. \citealt{Kippenhahn2013}). Consequently, metallicity has a major impact on the astrophysics of star clusters and black hole remnant formation as well. The earliest metal-free Pop-III star formation might have produced light SMBH seeds \citep{Madau2001,Schneider2002} from early stellar populations with a top-heavy or even logarithmically flat initial mass function (IMF; \citealt{Marks2012b,Chon2022}) and very weak winds \citep{Muijres2012}. For the metal-enriched Pop-II stellar population, the properties of the compact remnant mass function depend on the metallicity of the stars \citep{Fryer2001,Fryer2012} with more massive stellar black holes (BHs) forming at low metallicities \citep{Belczynski2010,Spera2017,Belczynski2020b,Dorozsmai2024}. This directly results in the formation of more massive BH-BH binaries in low metallicity stellar populations \citep{Mapelli2013,Ziosi2014,Chatterjee2017,Cao2018,Giacobbo2018,Giacobbo2018b,DiCarlo2020}. Thus, gravitational wave (GW) observations of merging BHs with component masses $>\msol{30}$ suggest that their stellar progenitors may have formed in low metallicity environments. Together with low measured redshifts of a number of massive GW events, such as GW150914 and GW170104 ($z\sim0.1$--$0.2$), the GW events imply either ongoing low metallicity star formation at low redshifts, long delay times for BH-BH mergers or their hierarchical dynamical assembly \citep{Abbott2017b}. The GW merger population properties may also evolve with redshift (e.g. \citealt{Rinaldi2024}), which, if confirmed with future data, would suggest a metallicity trend or contribution from multiple channels for stellar BH-BH mergers.

Besides individual stars and stellar binaries, metallicity can affect the global evolution of entire star clusters as well \citep{Chattopadhyay2022}. If the cluster core collapse timescale $t_\mathrm{cc}$ exceeds the lifetimes of massive stars ($t_\mathrm{life}\lesssim3$ Myr), the post core collapse bounce is weaker at low metallicities \citep{Mapelli2013b}. This is due to the lower wind mass losses, and results in somewhat higher central post collapse densities of the low metallicity clusters. In initially denser clusters, $t_\mathrm{cc}$ may be shorter than $t_\mathrm{life}$, promoting stronger dynamical interactions, and heating of the cluster core by more massive metal-poor stars that have experienced weaker wind losses \citep{Trani2014}.

Analytic arguments, cluster Monte Carlo models \citep{Henon1971} and direct \nbody{} simulations \citep{Aarseth2003,Spurzem2023} have demonstrated that massive star clusters ($M_\mathrm{cl}\gtrsim\msol{10^4}$) that have half mass surface densities above $\Sigma_\mathrm{h} \gtrsim \Sigmasol{10^4}$--$\Sigmasol{10^5}$ and low metallicities ($Z \lesssim \zsol{0.1}$) can build up increasingly massive stars by Pop-II 
\citep{PortegiesZwart1999,Miller2002,PortegiesZwart2004,Fregeau2004,Gurkan2004,Freitag2006a,Goswami2012,Fujii2013,Mapelli2016,Reinoso2018,Rizzuto2021,GonzalezPrieto2022,ArcaSedda-DRAGON2a,Rantala2024b,Barber2025,Ugolini2025,Vergara2025} and Pop-III \citep{Sakurai2017, Wang2022, Reinoso2023} stellar collisions. Following the massive star taxonomy of \cite{Gieles2025}, we use the following classification of collisionally grown stars:  very massive stars: (VMSs; $\msolrange{100}{m_\star}{1000}$), extremely massive stars (EMSs; $\msolrange{1000}{m_\star}{10000}$ and finally supermassive stars (SMSs; $m_\star \geq \msol{10000}$). Numerical simulations indicate that the final mass of the collisionally grown stars sensitively depends on the mass, density, rotation, virial state and degree of fractality of its host cluster as well as its stellar population including the IMF, metallicity and binary fraction. In addition, gas likely plays a role in the collisional growth process as well (e.g. \citealt{Reinoso2023,Fujii2024,Lahen2025b}).

At the end of their lives, massive stars, formed either through single-star evolution or via collision cascades, may collapse into massive stellar black holes \citep{DiCarlo2020b,Kremer2020b,Costa2022} in the pulsational pair-instability ((P)PISN) mass gap ($50$--$\msol{70} \lesssim M_\bullet \lesssim 130$--$\msol{160}$; \citealt{Woosley2021}) or above it as intermediate-mass black holes (IMBHs; \citealt{Greene2020}; $\msolrange{10^2}{M_\bullet}{10^5}$). However, considerable uncertainties surround modelling the collisions and winds of VMSs and especially EMSs and SMSs. For a number of wind loss prescriptions the stellar mass loss rates can be high even at low metallicities, which may have an
impact on the later evolution of the massive stars (e.g. \citealt{Sabhahit2023,Shepherd2025,Simonato2025,Torniamenti2025}). In the extreme case very high wind mass loss rates might inhibit the runaway stellar collisional IMBH formation channel altogether \citep{Belkus2007,Yungelson2008,Glebbeek2009,Pauldrach2012}. In addition, especially extended and weakly bound EMSs and SMSs might suffer from catastrophic mass loss in a subset of stellar collisions \citep{RamirezGaleano2025} that result in net mass loss, not growth, in the collisions.

Even though theoretical arguments suggest that SMS formation could be possible up to $Z\sim\zsol{0.01}$ \citep{Nandal2025}, thus far no supermassive star candidates have been observed in the local Universe \citep{Kuruvanthodi2023}. Similarly, despite a growing number of (P)PISN candidates (e.g. \citealt{Schulze2024}), none of them are thus far unambiguous (P)PISN events \citep{Angus2024}. Notwithstanding substantial observational effort \citep{Gebhardt2002, Ibata2009, Noyola2010, vanderMarel2010, Lutzgendorf2011, Jalali2012,Lanzoni2013, Kamann2016,Baumgardt2017,Pechetti2022,DellaCroce2024}, IMBHs in bona fide GCs of the Local Group also remain elusive. The strongest IMBH candidate thus far \citep{Häberle2024} is located at the centre of $\omega$ Centauri, a likely stripped nucleus of a dwarf galaxy (e.g. \citealt{Lee1999, Bekki2003}). This dynamically detected IMBH candidate with $M_\bullet>\msol{8200}$ lies in the mass range of the inferred IMBH population in X-ray selected AGN in dwarf galaxies \citep{Greene2007, Dong2012, Chilingarian2018, Mezcua2018, Greene2020, Reines2022} and overlaps the proposed mass range of $\msolrange{200}{M_\bullet}{10^5}$ of IMBHs \citep{Colbert1999,Kaaret2001,Matsumoto2001, Strohmayer2003, Patruno2006, Farrell2009,Mezcua2013,Pasham2014, Mezcua2015, Kim2020} in ultra or hyper-luminous X-ray emitters (ULXs, HLXs).

Supporting the electromagnetic searches for IMBHs, ground-based gravitational wave observations with \textit{Laser Interferometer Gravitational-Wave Observatory} (\textit{LIGO}), Virgo and \textit{Kamioka Gravitational Wave Detector} (\textit{KAGRA}) continue to reveal increasingly massive BH-BH mergers events. Still, the thus far heaviest GW mergers with the total mass of $M_\mathrm{\bullet,tot} \sim \msol{190}$--$\msol{265}$ \citep{Abac2025b} are relatively close to the low mass end of the IMBH regime. At higher BH binary masses above $M_\mathrm{\bullet,tot} \gtrsim \msol{1000}$ the current detectors are not sufficiently sensitive due to noise levels below $\sim10$ Hz. However, IMBH mergers with a total mass of $\msolrange{500}{M_\mathrm{\bullet,tot}}{1000}$ could be observable with the current \textit{LIGO}/Virgo/\textit{KAGRA} detectors with a sufficient signal-to-noise ratio of $\mathrm{S/N}\sim8$ to cosmological distances up to $\sim$ a few Gpc, or to redshifts $z\lesssim 1$ (e.g. \citealt{Mazzolo2014, Haster2016a, Mehta2022,Fragione2022b}). Next generation ground-based GW detectors such as the Einstein Telescope and the Cosmic Explorer will be sensitive to GW mergers with total masses $M_\mathrm{\bullet,tot} \lesssim \msol{1000}$ up to $z\lesssim{10}$ \citep{Reali2024,Abac2025}. In addition, the future \textit{Laser Interferometer Space Antenna} (\textit{LISA}) will be able to observe merging high mass IMBHs with $M_\mathrm{\bullet,tot}>\msol{10^4}$ including high redshift mergers above $z>10$ \citep{Amaro-Seoane2012}.

A theoretical picture has emerged regarding IMBHs as leftover SMBH seeds that failed to grow at high redshifts $z\gtrsim6$ (e.g. \citealt{Mezcua2017}). Within this framework, finding and characterizing the local IMBH population and measuring their mass function could reveal invaluable information about the high redshift SMBH seeding mechanism. However, the picture relies on a number of crucial assumptions. First, the failed population of IMBHs has to be totally failed, i.e. not grown in mass since formation. While most of present day SMBH mass originates from gas accretion \citep{Soltan1982}, growing especially low mass IMBHs into the SMBH mass regime via accretion in low mass galaxies seems indeed challenging (e.g. \citealt{Partmann2025,Shin2025,Petersson2025}). Meanwhile, IMBH growth by tidal disruption events in dense stellar environments such as nuclear star clusters (NSCs) seems difficult to avoid \citep{Alexander2017,Rizzuto2023}. Second, the "failed seeds" picture assumes that all (or at least the vast majority of) SMBH seeds formed at high redshifts. This is certainly true at least for the progenitors of the $6 \lesssim z \lesssim 12$ AGN population (e.g. \citealt{Inayoshi2020}), however, it has been proposed that individual SMBHs might form in extreme conditions also at lower redshifts \citep{vanDokkum2025a,vanDokkum2025b}. Finally, a population of IMBHs that formed at redshifts $z\lesssim6$ might challenge, or at least complicate, the interpretation of the low redshift IMBH population as failed high redshift SMBH seeds. This is the key motivation of our study. We investigate whether IMBHs could form in young massive star clusters at redshifts below $z\lesssim6$--$10$ and in the local Universe besides the $z>10$ early proto GC like environments that we have explored in the previous studies of the FROST-CLUSTERS project (I: \citealt{Rantala2024b}; II: \citealt{Rantala2025b}).

In this work we explore if IMBHs can form in isolated and hierarchically assembling star clusters across a wide range of cluster densities up to $\rho_\mathrm{h} \sim \rhosol{6.5\times10^7}$ and metallicities ($\zsolrange{0.01}{Z}{1.0}$). The effect of metallicity on IMBH formation in individual cluster models has been examined in the literature \citep{Mapelli2016}. While the cluster mass and density parameter space has been thoroughly explored at least for isolated clusters (e.g. \citealt{ArcaSedda-DRAGON2a,GonzalezPrieto2024}), however, IMBH formation through runaway collisions has not been systematically studied in models that vary both the initial star cluster densities and metallicities in their initial conditions. 

We present an updated version of the \bifrost{} code including the support for AMD GPU hardware and new improved treatments for EMS and SMS models including their metallicity dependent radii and wind mass loss rates. Furthermore, we now include mass loss in stellar collisions from the colliding stars which can lead to substantial stellar mass loss in addition to the wind losses. Based on a novel sample of more than $1440$ \nbody{} simulations performed with the updated version of the \bifrost{} code, we provide practical fitting formulas for the IMBH masses $M_\bullet=M_\bullet(\Sigma_\mathrm{h},Z)$ as a function of their host cluster metallicity and surface density. Our results can be used to seed black holes into star clusters in high resolution cosmological zoom-in simulations, and in semi-analytic models of galaxy formation. 

The article is structured as follows. After the introduction we present the updated \bifrost{} code and our initial conditions in Section \ref{section: 2}. We detail the metallicity and density dependent results of our isolated and hierarchical models focusing on the masses of extremely massive stars and IMBHs in Section \ref{section: 3} and Section \ref{section: 4}. Next, we examine the potential for wind and collisional ejecta enrichment in our models in Section \ref{section: 5}. Finally, we summarize our results and conclude in Section \ref{section: 7}.

%%%%%%%%%%%%%%%%%%%%%%%%%%%%%%%%%%%%%%%%%%%%%%%%%%%%%%%%%%%%%%%%%%%%%%%%%%%

\section{Numerical methods and initial conditions}\label{section: 2}

\subsection{Numerical methods: the updated \bifrost{} code}

For the numerical simulations of this study we use the GPU accelerated direct summation \nbody{} code \bifrost{} \citep{Rantala2023} coupled with the rapid stellar population synthesis code \sevn{} \citep{Iorio2023}. \bifrost{} has been updated since the previous study of the FROST-CLUSTERS project \citep{Rantala2025b}. The key improvements include models for mass loss in stellar collisions, supermassive stars and the ability to perform simulations on AMD accelerated processing unit (APU) hardware. In the following we briefly describe these main code updates.

\subsubsection{Gravitational dynamics}

Gravitational dynamics in the \bifrost{} code is modelled using the hierarchical variant \citep{Rantala2021} of the fourth order forward symplectic integrator \citep{Chin1997,Chin2005,Chin2007,Dehnen2017a,Bernard2025}. The time integration technique is momentum conserving and efficient for multi scale dynamical systems with a large dynamical range. For calculating the so-called gradient acceleration term responsible for cancelling the leading second order error terms we use the displacement approximation of \cite{Omelyan2006} which simplifies the implementation of the algorithm especially on GPUs \citep{Rantala2024b}. For strongly interacting small scale few-body systems including binaries, fly-bys, short-lived triple interactions, hierarchical triples as well as small clusters around massive BHs we use both secular and regularised integration techniques \citep{Wang2020,Rantala2020,Rantala2023,Rantala2024b} including post-Newtonian (PN) equations of motion for BHs up to the order PN3.5.

\subsubsection{Hardware acceleration and algorithmic updates}

\bifrost{} performs the direct summation calculations for determining the particle accelerations and potential energies on GPUs, where the $\bigO{N^2}$ double loops over the particles are mapped onto the massively parallel grid-like architecture \citep{Rantala2021} using the CUDA programming standard for Nvidia GPUs. We have extended \bifrost{} with support for AMD GPUs based on the HIP programming model and the ROCm software stack. To avoid code duplication, the original CUDA based implementation is kept as the single source code, only a special C header file redefines the CUDA calls into HIP calls on-the-fly when compiling for a HIP target, a technique referred to as \emph{hipifly}. Hardware-specific parameters such as the configuration of the computational grid are defined at compile-time via preprocessor macros. Moreover, to reduce the number of copies of the particle ensemble, \bifrost{} makes use of node-local MPI shared memory windows which enable multiple MPI ranks to access the same single copy of the ensemble on a shared memory compute node, reducing the volume of the network communication and the number of messages exchanged.

\subsubsection{Stellar evolution}

\bifrost{} is coupled with the fast stellar population synthesis code \sevn{} \citep{Iorio2023, Mapelli2020, Spera2015, Spera2017}. In \sevn{} we use the PARSEC-based (\citealt{Bressan2012, Chen2015, Costa2021, Nguyen2022, Costa2025}) stellar tracks \texttt{SEVNtracks\_parsec\_ov04\_AGB} with the overshooting parameter $\lambda = 0.4$ ranging in stellar mass from $\msol{2.2} \leq m_\star \leq \msol{600}$. For stripped pure He stars we select the \texttt{SEVNtracks\_parsec\_pureHe36} tracks. In this study we consider nine different logarithmically spaced stellar metallicities from $Z=0.0002=\zsol{0.01}$ to $Z=0.02=\zsol{1.0}$.

Due to their large number ($\sim 1400$), the simulations for this study were performed during the first half of 2025 using the most stable development branch of \bifrost{} at that time. This code branch did not include the \sevn{} binary stellar evolution coupling of \bifrost{} which we have used in our recent studies \citep{Rantala2025a,Rantala2025b}. Thus, binary stars in the simulations of this study evolve as single stars. We still include initial (primordial) binary star populations in our models as they significantly enhance the stellar collision rates in star clusters \citep{Fregeau2004,Gaburov2008,Rantala2025b} due to their larger collision cross sections compared to single stars. Typically only a single star can considerably grow via collisions in isolated clusters \citep{Baumgardt2011,Fujii2013,Rantala2024b,Rantala2025b}. For this rapid runaway stellar collisional channel (e.g. \citealt{Greene2020} and references therein) the growing stars rapidly ($\lesssim 1$--$2$ Myr) reach masses above $m_\star \gtrsim \msol{600}$, above the maximum stellar mass in the current \sevn{} stellar tracks. For these extremely massive stars and their collisions we use simplified evolution routines outlined in the following sections.

\subsubsection{Collisions involving compact objects}

Stars with masses $m_\star$ and radii $R_\star$ within the tidal disruption radius $R_\mathrm{t}$ \citep{Kochanek1992} defined as
\begin{equation}
    R_\mathrm{t} = 1.3 \left( \frac{M_\bullet}{m_\star} \right)^\mathrm{1/3} R_\star
\end{equation}
from any BH in the simulation will tidally disrupt. The BH accretes $50\%$ of the stellar material while the other half is instantaneously removed from the simulation. Gravitational wave driven BH-BH mergers occur at the separation when the innermost stable circular orbits of the BHs overlap. For the BHs in our simulations this separation is always very small, less than $<\Rsol{1}$. At the moment of the BH-BH merger, the remnant experiences relativistic gravitational wave mass loss, and receives a gravitational wave recoil kick based on fitting functions to numerical relativity \citep{Zlochower2015}. For additional details of the compact object collision procedures see \cite{Rantala2024b} and \cite{Rantala2025b}.

\subsection{Stellar collisions and collision products}

\subsubsection{Mass loss in stellar collisions}

Due to its simplicity, mass conservation in stellar collisions remains a common but not necessarily a well motivated assumption in N-body simulations (e.g. \citealt{Lombardi2002,Glebbeek2008,Glebbeek2009,Glebbeek2013}). Hydrodynamical stellar collision simulations of \cite{Glebbeek2013} suggest that the collisional mass loss can be parametrized as 
\begin{equation}\label{eq: phidef}
    f_\mathrm{loss} = \frac{C q}{(q+1)^2}
\end{equation}
in which $q=m_\mathrm{2}/m_\mathrm{1}<1$ and $C$ is a constant that may depend on the structure of the stars. We use a mass loss prescription that is simplified and only depends on the mass ratio of the merging stars. In this work we set $C=0.3$ \citep{Glebbeek2013} yielding $f_\mathrm{loss,max} = 7.5\%$ for equal mass stellar mergers. In the limit of $q\rightarrow0$ there is no mass loss. The final mass of the collision product is thus $m_\mathrm{remnant} = (m_\mathrm{1}+m_\mathrm{2})(1-f_\mathrm{loss}) = m_\mathrm{1}+m_\mathrm{2} - f_\mathrm{loss} m_\mathrm{1} - f_\mathrm{loss} m_\mathrm{2}$. The lost mass $m_\mathrm{loss} = (m_\mathrm{1}+m_\mathrm{2})f_\mathrm{loss}$ is instantly removed from the N-body simulation. The mass loss recipe always satisfies $m_\mathrm{1}+m_\mathrm{2}-m_\mathrm{lost} \geq m_\mathrm{1}$. We note that recent work on extremely massive stellar collisions (e.g. \citealt{RamirezGaleano2025, RomanGarza2026}) suggests that this assumption may not be always valid and the primary stars may suffer from catastrophic mass loss especially if they have  marginally bound extended envelopes.

We note that in \nbody{} simulations it is common to ignore the mass loss $f_\mathrm{loss} m_\mathrm{1}$ from the primary star. As we include this term, the total lost mass in $N_\mathrm{coll}$ collisions when building up a massive star may result in a cumulative mass loss larger than what would be possible for a single collision, i.e. $f_\mathrm{loss,cumu} > f_\mathrm{loss,max} = 7.5\%$. This is because each collision repeatedly removes mass from the primary star. We elaborate this process in detail in Appendix \ref{appendix: cumuloss} and show that when doubling an initial mass of $m_\mathrm{1,init}$ via collisions with $q\ll1$ using the mass loss recipe of Eq. \eqref{eq: phidef} and $C=0.3$, we have $f_\mathrm{loss,cumu} \sim 0.214 \sim 2.85 \times f_\mathrm{loss,max}$. Such efficient cumulative collisional mass loss cannot occur in models that only include mass loss from the secondary star.

\subsubsection{Massive stellar collision product radii}\label{section: radii}

The radii of the very and extremely massive stars in our models exceeding $m_\star\gtrsim \msol{600}$ are modelled by extrapolating the zero age main sequence (ZAMS) radii of the massive stars from the \texttt{PARSEC} stellar tracks \citep{Costa2025} including the metallicity dependence of the radii. We fit power-law relations of the form
\begin{equation}\label{eq: our-stellar-masssize}
    \frac{R_\star}{\Rsol{1}} = a(Z) \left( \frac{m_\star}{\msol{100}} \right)^\mathrm{\delta(Z)}
\end{equation}
to ZAMS stellar radii in the mass range of $\msol{400} \lesssim m \lesssim \msol{600}$, and obtain the coefficients $a(Z)$ and $\delta(Z)$ for nine different metallicities between $\zsol{0.01}$ and $\zsol{1.0}$. In \bifrost{}, given $m$ and $Z$ we linearly interpolate $a(Z)$ and $\delta(Z)$ from tabulated values to calculate $R_\star$. The coefficients $a(Z)$ and $\delta(Z)$ are listed in Table \ref{tab: mass-radius}. Stars with higher metallicities have initially larger radii. However, their higher wind mass loss rates especially above $Z \gtrsim \zsol{0.5}$ result in rapid mass loss and consequently decreased radii at later evolutionary stages.

\begin{table}
    \centering
    \begin{tabular}{cccc}
        \hline
        $Z/\zsol{}$ & $a(Z)$  & $\delta(Z)$ & $R_\star$ for a $\msol{10^3}$ star [$\Rsol{}$]\\
        \hline
        $0.010$ & $9.588$ & $0.562$ & 35.0\\
        $0.018$ & $9.878$ & $0.571$ & 36.8\\
        $0.032$ & $10.276$ & $0.584$ & 39.4\\
        $0.056$ & $10.571$ & $0.606$ & 42.7\\
        $0.100$ & $10.154$ & $0.686$ & 49.2\\
        $0.178$ & $9.940$ & $0.765$ & 57.8\\
        $0.316$ & $8.689$ & $0.960$ & 79.3\\
        $0.562$ & $5.533$ & $1.458$ & 158.9\\
        $1.000$ & $5.793$ & $1.814$ & 377.0\\
        \hline
    \end{tabular}
    \caption{The coefficients for the ZAMS mass size relation of Eq. \eqref{eq: our-stellar-masssize} for stars exceeding $\geq \msol{600}$ in mass. The radii of an example star of $\msol{10^3}$ are provided at nine different metallicities.}
    \label{tab: mass-radius}
\end{table}

We note that our massive stars have relatively small radii compared to the proposed protostar like models of extremely massive and supermassive stars. These stars can have extended radii depending on their gas accretion rates $\dot{m}_\mathrm{acc}$, potentially reaching several orders of magnitude larger sizes compared to Eq. \eqref{eq: our-stellar-masssize} for high $\dot{m}_\mathrm{acc}$ (e.g. \citealt{Haemmerle2018, RamirezGaleano2025, RomanGarza2026}). Our relatively compact EMS and SMS sizes are consistent with low gas accretion rates of $\dot{m}_\mathrm{acc} \lesssim \ratesol{10^{-3}}$ \citep{Hosokawa2012}, and by a factor of $\sim3$ smaller compared to \cite{Gieles2018}. At higher accretion rates, the increased sizes of the stars enhance their collision cross sections, but on the other hand may facilitate collisional mass loss from their diffuse envelopes (e.g. \citealt{Reinoso2023}). For additional discussion of the massive stellar radii see \cite{Rantala2025b}.

\subsubsection{Massive stellar collision product wind loss rates}\label{section: windrates}

Stellar wind mass loss by line driven winds \citep{Castor1975,Abbott1982,Garmany1985} plays a fundamental role in the evolution of massive stars \citep{Puls2008,Vink2011}. The wind mass loss rates $dm_\mathrm{wind}/dt$ depend on the mass, radius, temperature, luminosity and metallicity of the stars (e.g. \citealt{Niuwenhuijzen1990}). The metallicity dependence of the wind loss rates of massive stars is commonly expressed as
\begin{equation}
\derfrac{m_\mathrm{wind}}{t} \propto \left( \frac{Z}{\zsol{}} \right)^p, 
\end{equation}
in which the power law index in within the range of $0.5 \lesssim p \lesssim 0.94$ \citep{Vink2001}. For this study use the wind loss recipe of \cite{Vink2018}
\begin{equation}\label{eq: vink2018}
\log\left( \derfrac{m_\mathrm{wind}}{t} \middle/ \frac{M_\odot}{\mathrm{yr}} \right) = -9.13 + 2.1 \log(m_\star/M_\odot) + 0.74 \log(Z/Z_\odot),
\end{equation}
which assumes relatively cool ($T_\mathrm{eff} = 15000$ K) inflated very massive stars with large Eddington factors in the mass and metallicity ranges of $\msol{100} \lesssim m_\star \lesssim \msol{900}$ and $0.01 \lesssim Z/\zsol{} \lesssim 1.0$. We note that Eq. \eqref{eq: vink2018} is used in our models beyond $m_\star \gtrsim \msol{900}$. The wind model or very similar recipes have been recently used both in analytic models \citep{Gieles2018} and hydrodynamical simulations including collisional \nbody{} dynamics \citep{Fujii2024}. The wind loss rates for extremely massive stars in this study are higher than in our previous studies (\citealt{Rantala2024b,Rantala2025a,Rantala2025b}) where we assumed a simple power-law scaling from the most massive PARSEC stellar tracks. In addition, together with the included mass loss in stellar collisions, we expect somewhat lower maximum stellar and IMBH masses at comparable star cluster densities compared to our earlier models.

We note that the \cite{Vink2018} wind recipe we have adopted is not the only commonly used wind loss rate formulation used in \nbody{} and cluster Monte Carlo simulations of runaway stellar collisional IMBH formation. In general, the adopted wind recipes in \nbody{} simulations make up a substantial difference in the resultant wind mass loss rates especially for sub-solar metallicities \citep{Banerjee2020}. In the updated level C main sequence stellar evolution formulation of the \texttt{NBODY6++GPU} code, \cite{Kamlah2022} and \cite{Vergara2025} assume the wind models of \cite{Vink2001} for optically thin, line-driven outflows for hot and massive O and B type stars. Their massive and extremely luminous stars beyond the Humphreys-Davidson limit \citep{Humphreys1994} ($L_\star/L_\mathrm{\odot} > 6\times10^5$ and $10^{-5}\times R_\star/R_\odot (L/L_\mathrm{\odot})^\mathrm{1/2}>1$) undergo luminous blue variable (LBV) like mass loss with wind loss rates of $\dot{m}_\mathrm{LBV} = f_\mathrm{LBV} \times \ratesol{10^{-4}}$ \citep{Belczynski2010} with $f_\mathrm{LBV} \sim 1.5$. This LBV like wind loss rate effectively behaves as a maximum wind rate in the model. For our wind rate recipe $\dot{m} = 1.5\times\ratesol{10^{-4}}$ is reached at $\sim \msol{1700}$ for $Z=\zsol{0.01}$ and $\sim \msol{760}$ at $Z=\zsol{0.10}$. As such, the wind rate estimations of the models in \texttt{NBODY6++GPU} and \bifrost{} considerably differ for extremely massive and supermassive stars. This has important implications for the final IMBH masses in the simulations and for the amount of enriched material produced by the extremely massive stars during their lifetimes.

\subsubsection{Massive collision product ages and lifetimes}

Stellar mergers\footnote{In this numerical work we use terms stellar collision and stellar merger interchangeably. To be precise, in the real Universe, a stellar collision can lead to either a common envelope episode, or a collision in which the two stars merge and mix into one.} may provide a fresh supply of fuel into the centres of evolved stars (e.g. \citealt{Leonard1989,Lombardi2002}). This stellar rejuvenation process is commonly treated in \nbody{} simulations using parametrized age assignment formulas for the collision products depending on the types, ages and the masses of the merging stars (e.g. \citealt{Hurley2002}). Recently, \cite{Vergara2025} demonstrated that widely used rejuvenation formulas can result in excessive rejuvenation when low mass stars collide with extremely massive and supermassive stars. We instead follow the rejuvenation prescription of \cite{Mapelli2016} which cannot lead to such an excessive rejuvenation. We introduce a small modification to the prescription to take into account the mass loss in the collisions. For a primary star with a mass $m_\mathrm{1}$ merging with a secondary star with a mass of $m_\mathrm{2}$ ($m_\mathrm{1} \geq m_\mathrm{2}$) and a collisional mass loss of $m_\mathrm{loss}$, the age of the collision product is $A(m_\mathrm{1}+m_\mathrm{2}-m_\mathrm{loss}) = f_\mathrm{rej}(m_\mathrm{1},m_\mathrm{2},m_\mathrm{loss}) A(m_\mathrm{1})$. Here $A$ is the age of the star in its current evolutionary stage (main sequence or evolved phase) and $f_\mathrm{rej}$ is the rejuvenation factor. For the age assignment procedure we group together the \sevn{} stellar evolution phases $1$--$3$ (core hydrogen burning phases) and $4$--$6$ (core helium burning phases). We define the rejuvenation factor as
\begin{equation}
    f_\mathrm{rej}(m_\mathrm{1},m_\mathrm{2},m_\mathrm{loss}) = \frac{m_\mathrm{1}}{m_\mathrm{1}+m_\mathrm{2}-m_\mathrm{loss}} \frac{T(m_\mathrm{1}+m_\mathrm{2}-m_\mathrm{loss})}{T(m_\mathrm{1})}
\end{equation}
which adds the collisional mass loss $m_\mathrm{loss}$ into the recipe of \cite{Mapelli2016}. Here $T$ is the total duration of the current stellar evolutionary stage. Note that in the limit of $q=m_\mathrm{2} / m_\mathrm{1} \rightarrow 0$ the rejuvenation factor $f_\mathrm{rej} \rightarrow 1$, so multiple collisions with low mass secondary stars do not excessively rejuvenate the collision product.

We determine the lifetimes of collisionally formed stars after which above $m_\star \gtrsim \msol{600}$ they collapse into IMBHs using extrapolated PARSEC stellar tracks similarly to our extremely massive star radius determination procedure. Together with the rejuvenation procedure, the lifetimes $T_\mathrm{life}$ of our extremely massive stars are typically in the range of $2.5$ Myr $\lesssim T_\mathrm{life} \lesssim 5.0$ Myr, consistent with \nbody{} and Cluster Monte Carlo models in the literature (e.g. \citealt{Vergara2025}).

\subsection{Initial conditions}

\begin{figure}
\includegraphics[width=0.9\columnwidth]{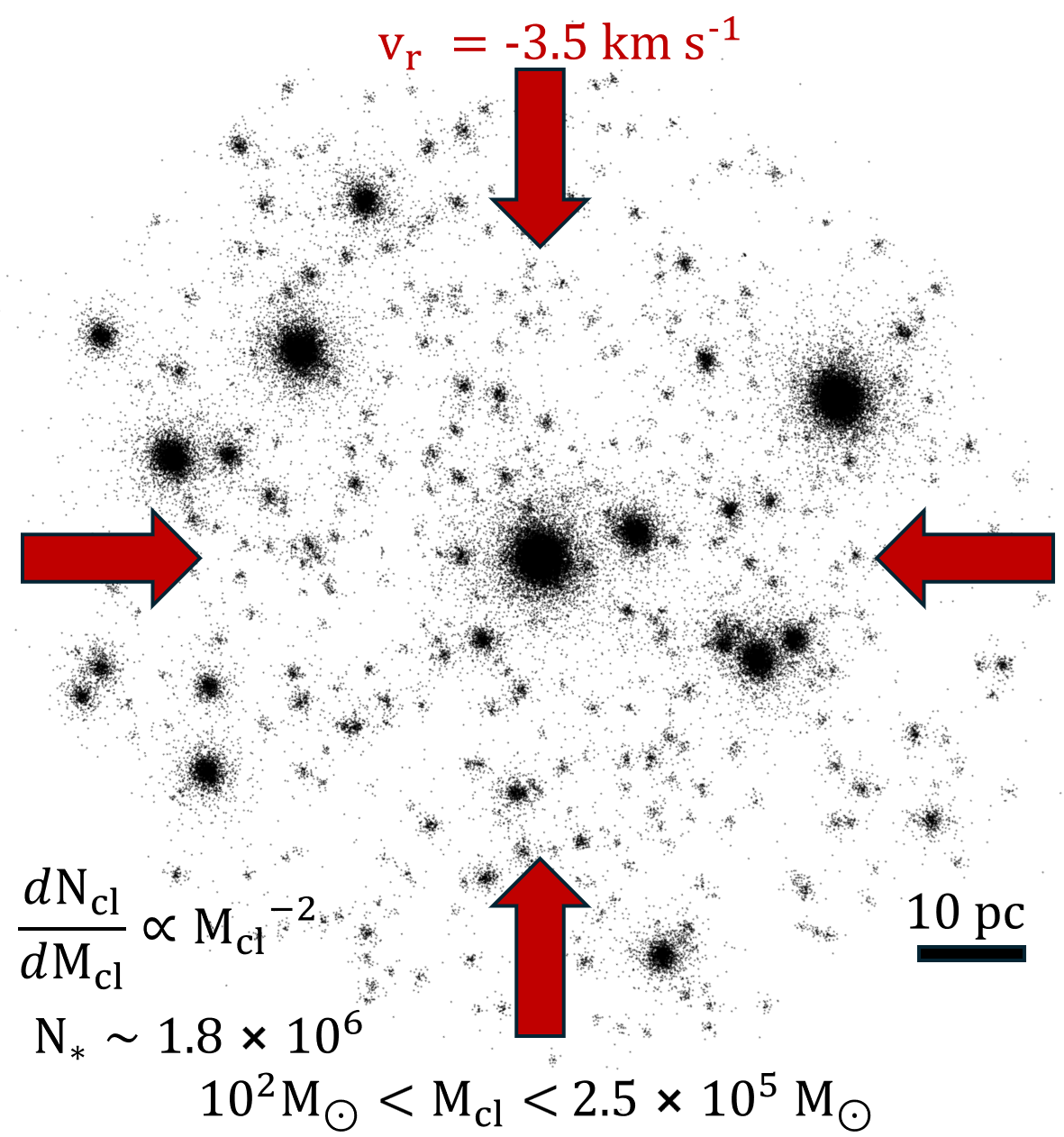}
\caption{A schematic illustration of the initial conditions for the hierarchically assembling cluster setups.}
\label{fig: ic-0}
\end{figure}

\subsubsection{Isolated models}\label{section: IC-isolated}

\begin{table*}
    \centering
    \begin{tabular}{cccccccccc}
        \hline
        Isolated & $N$ & $M_\mathrm{cl}$ & $m_\mathrm{max,0}$ & $f_\mathrm{h}$ & $r_\mathrm{h}$ & $\rho_\mathrm{c}$ & $\Sigma_\mathrm{h}$ & $N_\mathrm{metal}$ & $N_\mathrm{random}$\\
        models & & $[\msol{}]$ & $[\msol{}]$ & & [pc] & [$\rhosol{10^5}$] & [$\Sigmasol{10^4}$] & &\\
        \hline
        IM1D1Z[1--9] & $3.9\times10^4$ & $2.3\times10^4$ & $123.1$ & $0.580$ & $1.23$ & $0.07$ & $0.41$ & $9$ & $10$\\
        IM1D2Z[1--9] & $3.9\times10^4$ & $2.3\times10^4$ & $123.1$ & $0.395$ & $0.84$ & $0.21$ & $0.89$ & $9$ & $10$\\
        IM1D3Z[1--9] & $3.9\times10^4$ & $2.3\times10^4$ & $123.1$ & $0.269$ & $0.57$ & $0.65$ & $1.91$ & $9$ & $10$\\
        IM1D4Z[1--9] & $3.9\times10^4$ & $2.3\times10^4$ & $123.1$ & $0.184$ & $0.39$ & $2.07$ & $4.11$ & $9$ & $10$\\
        IM1D5Z[1--9] & $3.9\times10^4$ & $2.3\times10^4$ & $123.1$ & $0.125$ & $0.26$ & $6.54$ & $8.86$ & $9$ & $10$\\
        \hline
        IM2D1Z[1--9] & $1.0\times10^5$ & $5.9\times10^4$ & $150.0$ & $0.580$ & $1.45$ & $0.10$ & $0.75$ & $9$ & $10$\\
        IM2D2Z[1--9] & $1.0\times10^5$ & $5.9\times10^4$ & $150.0$ & $0.395$ & $0.99$ & $0.32$ & $1.62$ & $9$ & $10$\\
        IM2D3Z[1--9] & $1.0\times10^5$ & $5.9\times10^4$ & $150.0$ & $0.269$ & $0.67$ & $1.01$ & $3.49$ & $9$ & $10$\\
        IM2D4Z[1--9] & $1.0\times10^5$ & $5.9\times10^4$ & $150.0$ & $0.184$ & $0.46$ & $3.19$ & $7.52$ & $9$ & $10$\\
        IM2D5Z[1--9] & $1.0\times10^5$ & $5.9\times10^4$ & $150.0$ & $0.125$ & $0.31$ & $10.09$ & $16.20$ & $9$ & $10$\\
        \hline
        IM3D1Z[1--9] & $2.5\times10^5$ & $1.5\times10^5$ & $150.0$ & $0.580$ & $1.72$ & $0.16$ & $1.37$ & $9$ & $10$\\
        IM3D2Z[1--9] & $2.5\times10^5$ & $1.5\times10^5$ & $150.0$ & $0.395$ & $1.17$ & $0.49$ & $2.94$ & $9$ & $10$\\
        IM3D3Z[1--9] & $2.5\times10^5$ & $1.5\times10^5$ & $150.0$ & $0.269$ & $0.80$ & $1.55$ & $6.34$ & $9$ & $10$\\
        IM3D4Z[1--9] & $2.5\times10^5$ & $1.5\times10^5$ & $150.0$ & $0.184$ & $0.54$ & $4.90$ & $13.66$ & $9$ & $10$\\
        IM3D5Z[1--9] & $2.5\times10^5$ & $1.5\times10^5$ & $150.0$ & $0.125$ & $0.37$ & $15.50$ & $29.43$ & $9$ & $10$\\
         \hline
    \end{tabular}
    \caption{The sample of $1350$ \nbody{} simulations of isolated models of this study (label I) including three different star cluster masses $M_\mathrm{cl}$ (labels M$1$--M$3$), five different normalizations of the cluster mass radius relation $f_\mathrm{h}$ i.e. different cluster densities (labels D$1$--D$5$), and nine different metallicities (labels Z$1$--Z$9$). Ten random realizations of each cluster model are performed. In the table, we also list the number of stars in the clusters $N$, their initial IMF cut-off mass $m_\mathrm{max,0}$, their 3D half-mass radii $r_\mathrm{h}$ as well as their 3D central densities $\rho_\mathrm{c}$ and 2D half-mass surface densities $\Sigma_\mathrm{h}$.}
    \label{tab: isolated}
\end{table*}

We construct idealised, isolated, spherically symmetric star cluster models following the \cite{Plummer1911} density profile. For the isolated models we use three different cluster masses of $M_\mathrm{cl}=\msol{2.3\times10^4}$, $M_\mathrm{cl}=\msol{5.9\times10^4}$ and $M_\mathrm{cl}=\msol{1.5\times10^5}$ corresponding to $3.9\times10^4 \lesssim N \lesssim 2.5\times10^5$ individual stars. As in our previous FROST-CLUSTERS studies, we parametrise the initial mass-size relation of the star clusters as
\begin{equation}\label{eq: mass-size-isolated}
    \frac{r_\mathrm{h}}{\mathrm{pc}} = \frac{f_\mathrm{h} R_\mathrm{4}}{1.3} \left( \frac{M_\mathrm{\star}}{10^4M_\odot} \right)^\beta,
\end{equation}
in which $r_\mathrm{h}$ is the 3D half mass radius. We set $\beta=0.180\pm0.028$ and $R_\mathrm{4}=2.365\pm0.106$ following \cite{Brown2021}. In \cite{Rantala2024b} and \cite{Rantala2025b}, we used the normalisation parameter $f_\mathrm{h} = 0.125$ to capture the small observed birth radii of embedded clusters (e.g. \citealt{Marks2012}). For young local clusters \citep{Brown2021} $f_\mathrm{h} = 1.0$, and for $z\sim10$ \textit{JWST} clusters \citep{Adamo2024} $f_\mathrm{h} \sim 0.15$--$0.20$, however, we note that these most likely differ from the birth radii of the clusters. For the isolated simulation sample we use five different normalisations of the cluster mass size relation in the range of $0.125 \leq f_\mathrm{h} \leq 0.580$. For a fixed cluster mass $M_\mathrm{cl}$ this corresponds to a span of two orders of magnitude in the central Plummer model density $\rho_\mathrm{c}$. With the three different cluster masses and five different densities, our isolated cluster models span a range of $\sim7$ in cluster mass $M_\mathrm{cl}$, a factor of $\sim70$ in half mass surface density up to $\Sigma_\mathrm{h} = \Sigmasol{2.9\times10^5}$ and a factor of $\sim240$ in central stellar density up to $\rho_\mathrm{c} = \rhosol{1.6\times10^6}$.

Besides the star cluster densities, we also extend our previous studies in metallicity. In \cite{Rantala2024b} and \cite{Rantala2025b} we assumed a fixed stellar metallicity of $Z=0.0002 = \zsol{0.01}$. For the isolated simulation sample we explore nine different logarithmically spaced metallicities from $Z=\zsol{0.01}$ to $\zsol{1.0}$. Finally, we run $N_\mathrm{random}=10$ random realisations of each isolated cluster model, and our full isolated cluster sample consists of in total $1350$ \nbody{} models. The main isolated cluster setups of this study with their key physical properties are listed in Table \ref{tab: isolated}. Each isolated simulation (I) is labelled according to their mass (label M; $1$--$3$), density (D; $1$--$5$) and metallicity (Z; $1$--$9$). We do not indicate the individual random realizations of the model in the labels.

\subsubsection{Stellar populations}

We assume coeval zero age main sequence stellar populations following the \cite{Kroupa2001} initial mass function. The maximum initial stellar mass $m_\mathrm{max,0}$ depends on the cluster mass $M_\mathrm{cl}$ following \cite{Weidner2006} and \cite{Yan2023} with the upper limit $m_\mathrm{max,0} = \msol{150}$ for massive clusters \citep{Rantala2025b}. The binary star populations are initialised as described in \cite{Rantala2025b} with the binary population properties observed by \cite{Moe2017}, \cite{Winters2019} and \cite{Offner2023}. The model results in an initial binary fraction of $f_\mathrm{b}\sim0.29$ with most of the massive stars initially in binaries. Unlike in \cite{Rantala2025b}, in this work we do not consider an initial triple star population.

Each isolated model is run until $t=7.5$ Myr at which point all the massive BH progenitor stars have already ended their lives. We use \bifrost{} accuracy parameters of $\eta=0.2$ for free-fall, fly-by and gradient time-step criteria \citep{Rantala2023}, few-body subsystem radii of $r_\mathrm{ngb} = 5$ mpc as well as tolerance parameters $\eta_\mathrm{GBS}=10^{-8}$ and $\eta_\mathrm{endtime}=10^{-3}$ for algorithmically regularised few-body integration.

\subsubsection{Hierarchically assembling cluster models}\label{section: IC-hierarchical}

The initial conditions for our hierarchical star cluster assembly regions are set up as in \cite{Rantala2024b} and \cite{Rantala2025a}. A schematic illustration of a representative system is provided in Fig. \ref{fig: ic-0}. Individual star cluster masses up to $M_\mathrm{cl} \sim \msol{2.5\times10^5}$ are sampled from the universal power-law cluster mass function with a slope of $-2$ \citep{Elmegreen1996,Zhang1999,Adamo2020,Lahen2020}. The total mass of the hierarchical assembly region is $M_\mathrm{region} = \msol{10^6}$ corresponding to $N\sim1.8\times10^6$ individual stars in the clustered setups. We use the scatter of \cite{Brown2021} in Eq. \eqref{eq: mass-size-isolated} for our initial cluster mass size relation in the hierarchical setups leading to somewhat different initial cluster densities in different random realizations of the models.

Motivated by the structure of hierarchical star cluster formation regions in the solar mass resolution hydrodynamical star burst simulations of \cite{Lahen2020}, the centre-of-masses of the individual star clusters ($N_\mathrm{cl} \leq 960$) are sampled within a uniform sphere with a radius of $r_\mathrm{region} = 50$ pc. The hierarchical region is collapsing with a radial velocity component of $v_\mathrm{r} = -3.5$ km/s and a random component of the same magnitude. The most massive sub-cluster is placed at the origin with a zero initial velocity. We further ensure that all sub-clusters are initially gravitationally bound to the hierarchically clustered region. For additional details of the structure of the hierarchical region setup see \cite{Rantala2024b}.

In this work we extend the study of hierarchically assembling star clusters to previously unexplored low density and extremely dense models beyond our fiducial star cluster mass radius relation normalization of $f_\mathrm{h}=0.125$. In total, we model nine different normalizations of the initial star cluster mass size relation. The maximum initial half mass densities of individual sub-clusters reach $\rho_\mathrm{h} \sim \rhosol{6.5\times10^7}$ corresponding to maximum half mass surface densities of $\Sigma_\mathrm{h}\sim\Sigmasol{3.9\times10^6}$. These densities considerably exceed the observed densities of YMCs in the local Universe \citep{Krumholz2019}, and even the densities of the $z\sim10$ \textit{JWST} proto GCs with $R_\mathrm{e} \sim 1$ pc and $M_\mathrm{cl}\sim \msol{10^6}$ \citep{Adamo2024}. However, these are not necessarily the birth densities of the clusters, and it has been shown that dense star clusters can rapidly expand by a factor up to $\sim10$ in their size (e.g. \citealt{ArcaSedda-DRAGON2a,Lahen2025a}) in their early evolution. For the Cosmic Gems clusters this argument yields maximum birth surface densities of $\Sigma_\mathrm{h}\sim\Sigmasol{1.6\times10^7}$, a factor of $\sim4$ higher than our densest model. Thus, our sub-cluster densities cover most of the plausible densities expected for forming star clusters in the Universe, except potentially the very extreme end. Each hierarchical model is simulated at $Z=\zsol{0.1}$ while we additionally run the densest models also at $Z=\zsol{0.01}$. Three random realizations of each model are generated resulting in total of $27$ models at $Z=\zsol{0.1}$ and three extremely dense models with $Z=\zsol{0.01}$. We summarize the key properties of the central sub-clusters of the models with $M_\mathrm{cl}=\msol{2.5\times10^5}$ in Table \ref{tab: hierarchical-models}. Similarly to the isolated models, each hierarchical setup (H) is labelled according to their density (D; $1$--$9$) and metallicity (Z; $1$--$2$). Each hierarchical model is run until $t=7.5$ Myr just as the isolated models with the same user given accuracy parameters.

\begin{table*}
    \centering
    \begin{tabular}{llllccc}
\hline   
Hierarchical model & $f_\mathrm{h}$ & min $r_\mathrm{h}$ & max $r_\mathrm{h}$ & min--max $\rho_\mathrm{c}$ & min--max $\rho_\mathrm{h}$ & min--max $ \Sigma_\mathrm{h}$\\
central cluster & & pc & pc & [$\rhosol{10^5}$] & [$\rhosol{10^5}$] & [$\Sigmasol{10^5}$]\\
\hline   
HD1 (3 models) & $0.580$ & $1.67$ & $2.16$ & $0.03$-$0.06$ & $0.03$--$0.06$ & $0.05$--$0.08$ \\ 
HD2 (3 models) & $0.395$ & $1.14$ & $1.47$ & $0.08$-$0.18$ & $0.09$--$0.20$ & $0.11$--$0.18$ \\ 
HD3 (3 models) & $0.269$ & $0.78$ & $1.00$ & $0.27$-$0.58$ & $0.30$--$0.64$ & $0.23$--$0.39$ \\ 
HD4 (3 models) & $0.184$ & $0.53$ & $0.68$ & $0.85$-$1.84$ & $0.94$--$2.02$ & $0.51$--$0.84$ \\ 
HD5 (3 models) & $0.125$ & $0.36$ & $0.46$ & $2.71$-$5.84$ & $2.97$--$6.41$ & $1.09$--$1.82$ \\ 
HD6 (3 models) & $0.067$ & $0.24$ & $0.32$ & $8.58$-$18.51$ & $9.42$--$20.33$ & $2.35$--$3.93$ \\ 
HD7 (3 models) & $0.034$ & $0.17$ & $0.22$ & $27.19$-$58.67$ & $29.87$--$64.45$ & $5.08$--$8.48$ \\ 
HD8 (3 models) & $0.014$ & $0.11$ & $0.15$ & $86.19$-$186.00$ & $94.68$--$204.32$ & $10.95$--$18.29$ \\ 
HD9 (3 models) & $0.004$ & $0.08$ & $0.10$ & $273.23$-$589.61$ & $300.14$--$647.69$ & $23.63$--$39.47$ \\ 
\hline   
    \end{tabular}
    \caption{The densities of the central sub-clusters of our hierarchical cluster assembly models. Each central sub-cluster has a mass of $M_\mathbf{cl}=\msol{2.5\times10^5}$ and consists of $N\sim4\times10^5$ individual stars. The star cluster densities cover a range of four orders of magnitude in half-mass and central densities $\rho_\mathrm{h}$ and $\rho_\mathrm{c}$ as well as a factor of $\sim500$ in half mass surface densities $\Sigma_\mathrm{h}$. The total stellar mass within the hierarchically assembling regions is $M_\star\sim \msol{10^6}$ in $N=1.8\times10^6$ individual stars.}
    \label{tab: hierarchical-models}
\end{table*}

\subsection{Model assumptions and uncertainties}

We briefly outline alternative model assumptions not explored in this study and their likely effect on the runaway collision cascades and the final IMBH masses. While this study mainly focuses on the effects of metallicity and cluster density, we note that our previous studies in the FROST-CLUSTERS project (I; \citealt{Rantala2024b}, II; \citealt{Rantala2025b}) we examined the consequences of different modelling assumptions such as collision treatment, star cluster density profiles and an extended IMF (sections 2.2.3 and 2.3 of \citealt{Rantala2025b}). While the details of the initial density profile (for a fixed $r_h$) seems to have a negligible effect on the IMBH masses, including a collision mass loss prescription lowers the IMBH masses by a factor of $\lesssim2$. Whether initial binary stars are included in the models has a large effect on the first collisions and in models with high velocity dispersions the initial binaries are crucial for the collision cascades to occur in the first place \citep{Rantala2024b, Rantala2025b}. In our observationally motivated initial models the binary fraction is always high for massive stars ($f_\mathrm{bin}>0.9$ for $m_\star > \msol{10}$). While we have not systematically varied $f_\mathrm{bin}$ in our models, the exact massive star binary fraction likely has a relatively weak effect on the IMBH masses, at least in relatively brief collision cascades \citep{GonzalezPrieto2024}. The effect of the stellar binary population properties (mass ratio, semi-major axis and eccentricity distributions) on the IMBH masses from runaway stellar collisions in cluster-scale $\textit{N}$-body simulations remains largely unexplored due to the computational costs of extended model parameter space.

For IMBH mass growth by TDEs we have selected a model in which $50\%$ of the stellar material is accreted into the BH. This value may be somewhat optimistic, and it has been shown (e.g. \citealt{Rizzuto2021}) that for low ($10\%$) accretion fractions the TDE driven growth channel becomes inefficient. Moreover, we have not included a model for stellar tides in the star BH interactions (e.g. \citealt{Press1977, Samsing2018}). While stellar tidal drag forces may increase the TDE rates via tidal captures, the fraction of TDEs strongly affected by the stellar tides is expected to be low ($\lesssim15\%$; \citealt{Rizzuto2023}). Within the short ($\lesssim5$ Myr) timescales for IMBH formation the galactic tidal field, which we have not modelled, likely plays a negligible role \citep{ArcaSedda-DRAGON2a}. However, we note that the galactic tidal field will have a major effect on the long term evolution ($>100$ Myr--$1$ Gyr) of the IMBH host clusters (e.g. \citealt{Lutzgendorf2013b}).

%%%%%%%%%%%%%%%%%%%%%%%%%%%%%%%%%%%%%%%%%%%%%%%%
\section{Runaway stellar collisions across star cluster metallicities and densities: isolated models}\label{section: 3}

\begin{figure*}
\includegraphics[width=0.85\textwidth]{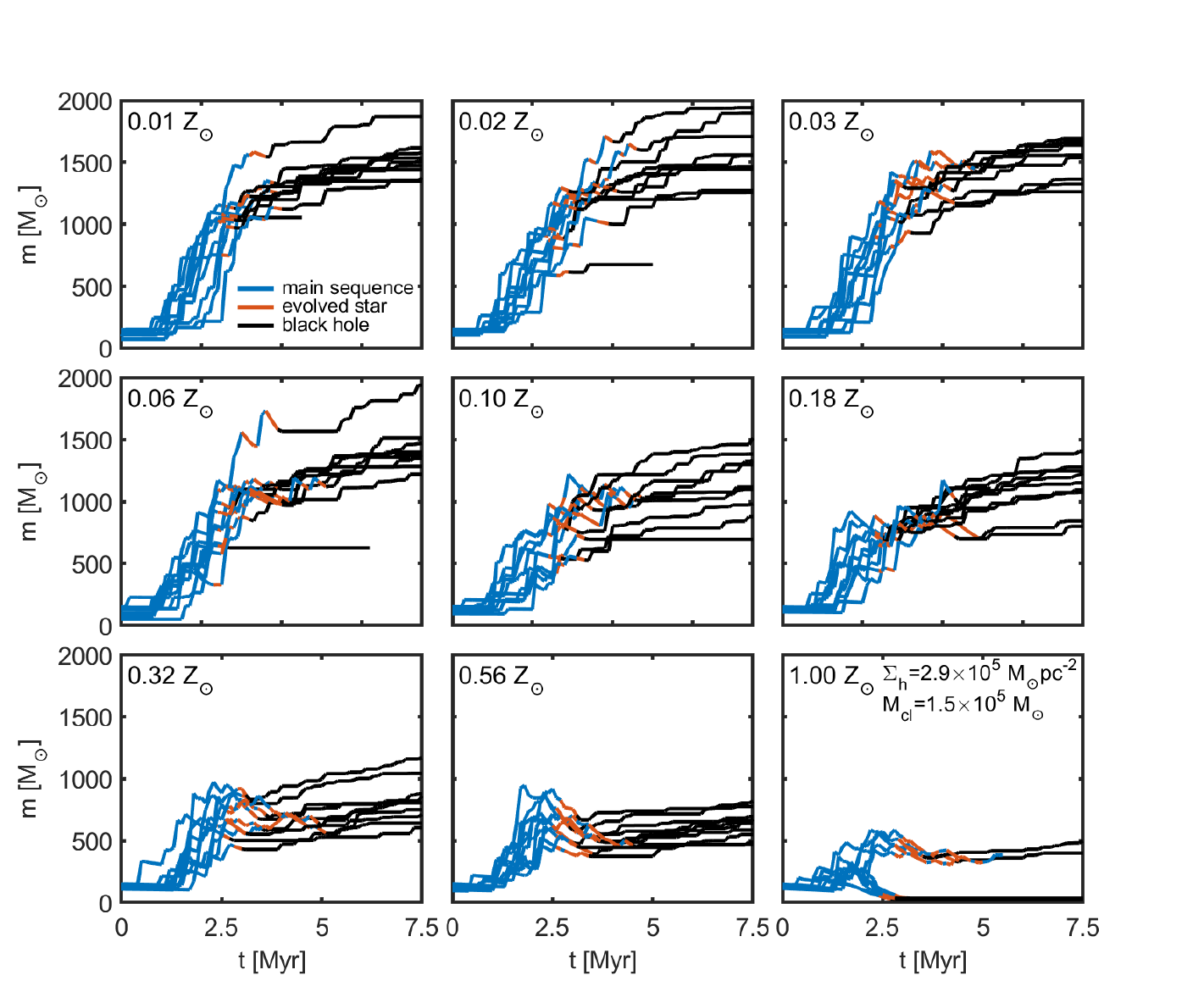}
\caption{The mass growth histories of selected massive stars and IMBHs in the isolated dense star cluster models IM3D5Z1--9 with $M_\mathrm{cl} = 1.5\times10^5$ and $\Sigma_\mathrm{h} = \Sigmasol{2.9\times10^5}$. Each line shows the growth history of a star that reached the highest mass in its host cluster during the simulation before collapsing into an IMBH. In a number of models the massive star is disrupted by a stellar BH or an IMBH, or the IMBH is ejected from its host cluster after merging with a stellar BH due to a GW recoil kick before the end of the simulations at $t=7.5$ Myr. The metallicity increases in each panel starting from $Z=\zsol{0.01}$ on the top left. When increasing the metallicity from $Z=\zsol{0.01}$ to $Z=\zsol{0.10}$, the overall picture of the stellar mass growth remains similar while the maximum IMBH mass decreases from $M_\bullet \sim \msol{1900}$ to $\msol{1500}$ in the models due to increased wind losses. Above $Z\gtrsim\zsol{0.2}$ the picture qualitatively changes as the stronger wind losses can quench the collision cascades, and the massive stars lose substantial amounts of mass before the ends of their lives. At solar metallicity, only few IMBHs can form.}
\label{fig: mass-evolution}
\end{figure*}

\subsection{Runaway collisions and star cluster metallicity}

We present the mass growth histories of selected stars that considerably grew by stellar collisions in the isolated simulations models IM3D5Z1--9 with $M_\mathrm{cl} = \msol{1.5\times10^5}$ and $\Sigma_\mathrm{h}= \Sigmasol{2.9\times10^5}$ in Fig. \ref{fig: mass-evolution}. Each growth history represents a massive star that obtained the highest mass in its simulation as a star before collapsing into an IMBH. In a small number of simulations the IMBH is ejected before $t=7.5$ Myr after merging with a stellar BH due to a relativistic GW recoil kick. Furthermore, in a number of models the most massive star strongly interacts with a stellar BH for which we focus in a separate study \citep{Rantala2026b}. Still, the low metallicity models IM3D5Z1 behave qualitatively similarly as our isolated setups in \cite{Rantala2024b} and \cite{Rantala2025b} despite the somewhat higher stellar wind mass loss rates for $\geq \msol{600}$ stars and the now included mass loss in collisions. At $Z=\zsol{0.01}$, the collision cascades commence between $t\sim0.7$--$2.4$ Myr. The growth of the stars by collisions is almost monotonic, and results in the collisional build-up of stars with masses in the range of $\msol{982} \lesssim m_\star \lesssim \msol{1583}$. After the collapse of the stars into IMBHs their growth proceeds via TDEs, and to a minor extent, mergers with stellar BHs. By $t=7.5$ Myr, the IMBHs have grown into the mass range of $\msol{1055} \lesssim M_\bullet \lesssim \msol{1870}$, up to $\sim 20\%$ from their mass just after their formation. We note that the IMBH growth via TDEs can be artificially boosted in simplified, isolated models of young massive star clusters. In more realistic hierarchically assembling models, the clusters increase in size and decrease in mean density due to cluster mergers \citep{Rantala2025b} which can be understood in terms of virial arguments (e.g. \citealt{Naab2009}). Thus, the $20\%$ IMBH mass growth by TDEs in $\lesssim5$ Myr should be regarded as an optimistic upper limit. Between $t=3$--$4$ Myr and $t=7.5$ Myr the mass growth via IMBH-BH mergers is minor, and IMBH-IMBH mergers are rare in isolated monolithic cluster setups \citep{Rantala2025a}. We note that in the long-term evolution ($>10$--$100$ Myr) of the IMBHs mergers with stellar BHs may become an important source of mass growth if the IMBH is retained in the cluster \citep{Antonini2019, Fragione2020, Rizzuto2022, ArcaSedda2023, ArcaSedda-DRAGON2b}.

At somewhat higher metallicities between $\zsol{0.02} \lesssim Z \lesssim \zsol{0.10}$ the overall picture remains qualitatively similar to $Z=\zsol{0.01}$. After the collision cascades begin, the stellar mass growth is almost monotonic. Most cluster models form an IMBH which further grow via TDEs. However, due to higher wind mass loss rates of extremely massive stars, the peak collisional stellar masses are somewhat lower. At $Z=\zsol{0.10}$ the maximum stellar masses are in the range of $\msol{558} \lesssim m_\star \lesssim \msol{1221}$, $23$--$43\%$ lower compared to the models with $Z=\zsol{0.01}$. Nevertheless, our results demonstrate that IMBH formation through stellar collisions remains viable even at moderate metallicities of $Z=\zsol{0.10}$ at fixed initial cluster mass and half-mass density. The final IMBH masses at $Z=\zsol{0.10}$ are in range between $\msol{694} \lesssim M_\bullet \lesssim \msol{1513}$.

At higher metallicities above $Z>\zsol{0.1}$ the picture described above qualitatively changes. High wind mass loss rates at metallicities above $Z \gtrsim \zsol{0.2}$ efficiently prevent stellar mass growth, even though the runaway collision cascades still proceed. This results in non-monotonic growth histories for the stars. Instead, the mass growth histories of the stars $m_{\star}(t)$ show a maximum followed by a decline, a feature which becomes more prominent towards higher metallicities. Before forming IMBHs at the end of their lives, the high wind mass loss rates of the stars cannot be balanced with collisional mass gain, and the massive stars can substantially lose mass before the end of their lives (e.g. \citealt{Mapelli2016}). Still, almost every dense cluster model forms an IMBH above a few hundred solar masses up to $Z\sim \zsol{0.5}$. At $Z=\zsol{0.56}$, the maximum collisional stellar masses reach $\msol{529} \lesssim m_\star \lesssim \msol{946}$ while the formed IMBHs have masses of $\msol{376} \lesssim m_\mathrm{\bullet} \lesssim \msol{705}$ with up to $\sim \msol{250}$ of stellar material lost via strong winds.

Finally, at solar metallicity, IMBHs rarely form. With $Z=\zsol{1.0} $ only two random realizations out of $10$ produce an IMBH in the models IM3D5Z9. The maximum collisional stellar masses achieved are $\msol{200} \lesssim m_\mathrm{\star} \lesssim \msol{588}$ as high stellar wind rates quench the mass growth. Thus, in most of the solar metallicity models the maximum final BH mass is in the stellar mass BH range. We note that IMBHs cannot from at $Z=\zsol{1.0}$ from stars with ZAMS masses of $m_\mathrm{\star} \lesssim \msol{600}$ through isolated single stellar evolution.

\subsection{Runaway collisions and star cluster density}

\subsubsection{The faction of clusters that form an IMBH}

\begin{figure}
\includegraphics[width=0.9\columnwidth]{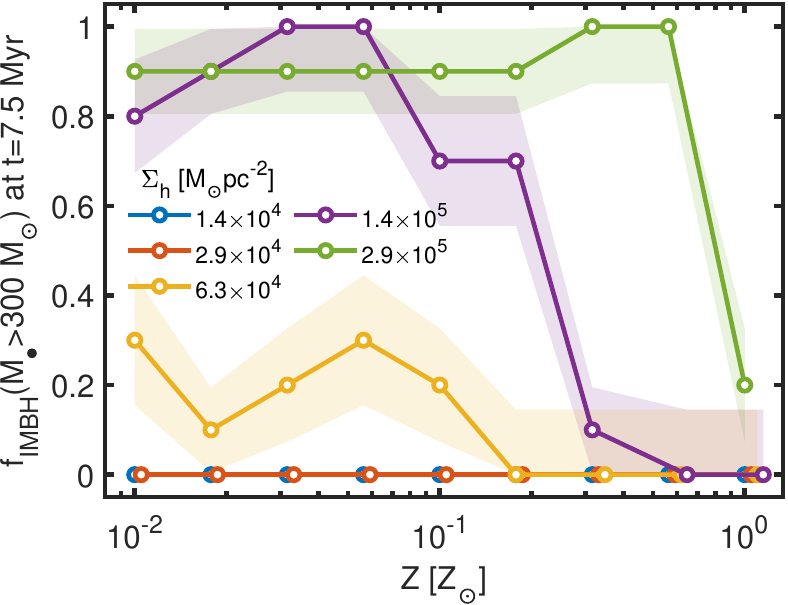}
\caption{The fraction $f_\mathrm{IMBH}$ of star clusters forming an IMBH with $M_\bullet > \msol{300}$ via runaway collisions in the isolated setups and retaining it until $t=7.5$ Myr with different densities and metallicities. In the models $M_\mathrm{cl}=\msol{1.5\times10^5}$. The IMBH fraction sensitively depends on both cluster density and metallicity.}
\label{fig: prob}
\end{figure}

Next, we explore the effect of the star cluster density on the fraction of the clusters that form an IMBH, and the masses of the IMBHs that form. As in \cite{Rantala2024b}, the mass segregation timescale of our isolated star clusters is shorter than the lifetimes of the most massive stars. However, the cluster core collapse timescale is longer, i.e. $t_\mathrm{seg} < t_\mathrm{life} \lesssim t_\mathrm{cc}$. As such, the collision cascades are driven by the high cluster densities \citep{Paiella2025}. The IMBH formation fraction is calculated as a fraction of random realizations that formed an IMBH out of $N_\mathrm{random}=10$ of each star cluster model. For this analysis we adopt an IMBH mass threshold of $M_\bullet> 2\times m_\mathrm{max,0} = \msol{300}$ for runaway collisional IMBHs, and required that the IMBH is retained in the cluster until $t=7.5$ Myr. The adopted runaway IMBH mass threshold excludes low mass IMBHs formed through a single merger of the components of a massive binary system. We will discuss these low mass binary merger IMBHs more in detail below. We focus on the isolated cluster models IM3D[1--5]Z[1--9] which all have $M_\mathrm{cl}=\msol{1.5\times10^5}$. The half-mass surface densities of the models lie in the range of $\Sigmasol{1.4\times10^4} \leq \Sigma_\mathrm{h} \leq \Sigmasol{2.9\times10^5}$ while their metallicities are $\zsol{0.01} \leq Z \leq \zsol{1.0}$. We present the IMBH fraction $f_\mathrm{IMBH}$ via a collision cascade for each model in Fig. \ref{fig: prob}. In the models IM3D[1--2]Z[1--9] with half-mass surface densities below $\Sigma_\mathrm{h} \lesssim \Sigmasol{2.9\times10^4}$ no runaway IMBHs form at any metallicity as very few collisions occur in the low density models. Increasing the cluster densities, the models IM3D3Z[1--9] with $\Sigma_\mathrm{h} = \Sigmasol{6.3\times10^4}$ are dense enough for forming runaway IMBHs with $f_\mathrm{IMBH}\sim0.1$--$0.4$ (mean $f_\mathrm{IMBH}\sim0.22$) when $Z \leq \zsol{0.10}$, but not at higher metallicities. Towards increasingly high cluster densities, the fraction of clusters with the runaway IMBH formation rapidly increases, and most high density cluster models almost always form a runaway IMBH. Above $\Sigma_\mathrm{h} \gtrsim \Sigmasol{1.4\times10^5}$, $f_\mathrm{IMBH} \gtrsim 0.8$ for models with a relatively low metallicity in the range of $Z \lesssim \zsol{0.06}$. At higher metallicities the IMBH formation fraction sharply decreases to $f_\mathrm{IMBH}\sim0.0$--$0.2$ after a critical metallicity that depends on the cluster (surface) density. For models with $\Sigma_\mathrm{h} = \Sigmasol{1.4\times10^5}$ the critical metallicity threshold is between $\zsol{0.10} \lesssim Z \lesssim \zsol{0.32}$ while for the most dense isolated models the threshold occurs above $Z \gtrsim \zsol{0.56}$.

\subsubsection{The masses of the formed IMBHs}

\begin{figure}
\includegraphics[width=1.0\columnwidth]{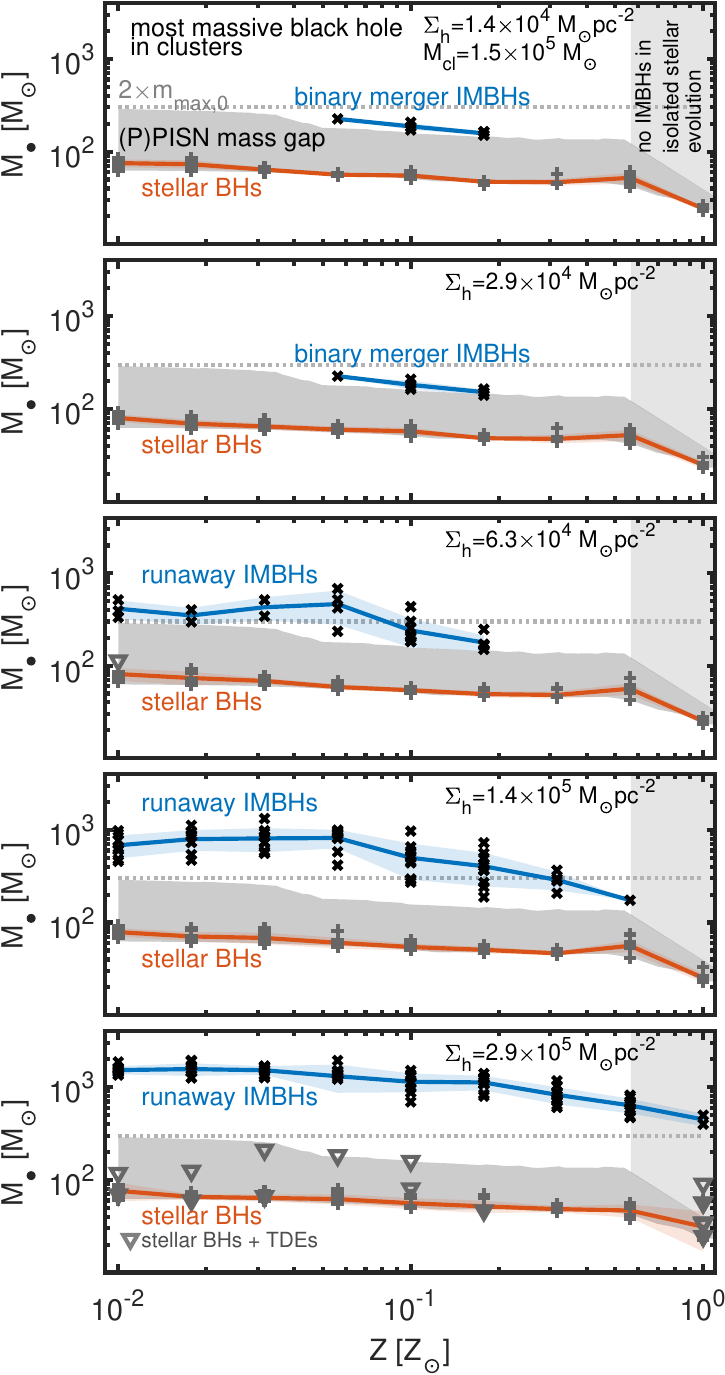}
\caption{The masses $M_\bullet$ of the most massive stellar BHs and IMBHs in models with $M_\mathrm{cl}=\msol{1.5\times10^5}$. From the top panel down the host cluster surface density increases from $\Sigma_\mathrm{h}=\msol{1.4\times10^4}$ to $\Sigma_\mathrm{h}=\msol{2.9\times10^5}$. In the two models with the lowest densities, no runaway collisions occur and IMBHs rarely form via a single massive binary merger. At $\Sigma_\mathrm{h} \gtrsim \Sigmasol{6.3\times10^4}$ IMBHs form through stellar collision cascades with $M_\bullet$ increasing with increasing $\Sigma_\mathrm{h}$ especially at low metallicities $Z\lesssim\zsol{0.10}$. At higher metallicities the IMBH formation and their masses are suppressed due to the stellar wind mass losses. In the models with highest density, individual stellar BHs may grow into the (P)PISN mass gap via micro-TDEs.}
\label{fig: Z-Mbh1}
\end{figure}

\begin{figure}
\includegraphics[width=1.0\columnwidth]{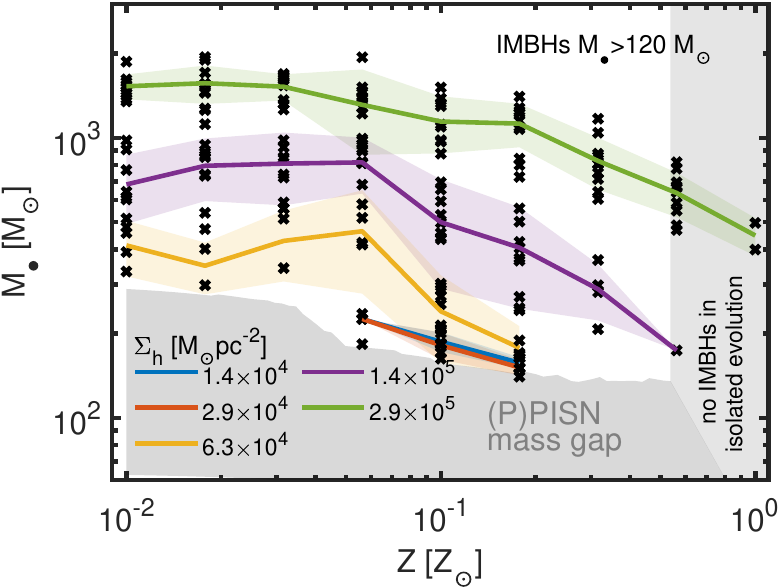}
\caption{The maximum IMBH masses $M_\bullet$ in the simulations with $M_\mathrm{cl}=\msol{1.5\times10^5}$ from Fig. \ref{fig: Z-Mbh1}. Above the critical surface density $\Sigma_\mathrm{h} \gtrsim \Sigmasol{6.3\times10^4}$ IMBHs can form via runaway collision cascades, exceeding $M_\bullet = \msol{10^3}$ in the models with $\Sigma_\mathrm{h} = \Sigmasol{2.9\times10^5}$. At low metallicities $Z \lesssim \zsol{0.10}$ the stellar wind losses in the models are relatively weak and do not affect the final IMBH masses. However, above a critical metallicity threshold (e.g. $Z \sim \zsol{0.06}$ for $\Sigmasol{6.3\times10^4} \lesssim \Sigma_\mathrm{h} \lesssim \Sigmasol{1.4\times10^5}$) the masses of the formed IMBH decline due to the wind mass losses of their progenitor stars before the ends of their lives.}
\label{fig: Z-Mbh2}
\end{figure}

We present the masses of the most massive stellar BHs and IMBHs in all the random realizations of the models IM3D[1--5]Z[1--9] in Fig. \ref{fig: Z-Mbh1}. Furthermore, we display the IMBH masses from models with different metallicities and surface densities in Fig. \ref{fig: Z-Mbh2} to allow for a straightforward comparison of the IMBH masses at different cluster surface densities. We note that while Fig. \ref{fig: mass-evolution} focused on massive stars and IMBHs selected by their maximum mass in the stellar phase, for Fig. \ref{fig: Z-Mbh1} and Fig. \ref{fig: Z-Mbh2} we have selected the most massive (IM)BHs still present in their host clusters at $t=7.5$ Myr. The figures also include the location of the metallicity dependent (P)PISN mass gap for BH masses assuming isolated single stellar evolution obtained from the \texttt{PARSEC} stellar tracks provided by the \sevn{} code. For metallicities above $Z \gtrsim \zsol{0.6}$, the high mass boundary of the (P)PISN mass gap is not captured by the available models limited to ZAMS masses below $Z \lesssim \msol{600}$. Thus, the exact extent of the (P)PISN mass gap and its upper edge at metallicities close to $Z=\zsol{1.0}$ is uncertain. Due to this and uncertainties in the wind loss rates (see Section \ref{section: windrates}) we emphasize that any formed IMBHs in this metallicity range $Z \gtrsim \zsol{0.6}$ should be regarded with caution. Finally, we note that Fig. \ref{fig: Z-Mbh1} and Fig. \ref{fig: Z-Mbh2} do not indicate how frequently the IMBHs with given masses form, but instead simply show the mass distribution of the most massive BHs and IMBHs in the models that formed IMBHs.

In the top two panels of Fig. \ref{fig: Z-Mbh1}, at low cluster surface densities below $\Sigma_\mathrm{h} \lesssim \Sigmasol{3\times10^4}$ IMBHs rarely form. The only IMBHs at the low surface densities originate from individual collisions between components of massive initial binary systems which predominantly occur at $t \gtrsim 2$ Myr in the simulations. By construction, $M_\bullet \lesssim 2\times m_\mathrm{max,0} = \msol{300}$ for these relatively rare binary merger IMBHs, and the most massive stellar BHs lie near the lower edge of the (P)PISN mass gap. A number of stellar BHs in the setups grow via micro-TDEs (stellar BH TDEs; see e.g. \citealt{Kremer2019,Rastello2025}). Our assumed TDE accretion factor of $0.5$ is relatively high for micro-TDEs and thus the masses of BHs grown via this channel should be considered as upper limits.

This low density picture changes above surface densities $\Sigma_\mathrm{h} \gtrsim \Sigmasol{6.3\times10^4}$. First, at low metallicities below $Z\lesssim \zsol{0.2}$, relatively short ($N_\mathrm{coll}>2$) collision sequences occur leading to the formation of IMBHs in the mass range of $\msol{300} \lesssim M_\bullet \lesssim \msol{600}$ from massive progenitor stars above the (P)PISN mass gap. These results are consistent with \cite{Mapelli2016} who found remnant masses up to $\sim \msol{250}$ at somewhat lower initial cluster densities. Above the critical metallicity of $Z \gtrsim \zsol{0.2}$ no IMBHs form. The collision rates in the runaway cascades increase with increasing cluster surface density, and the maximum metallicity in which any IMBHs can form also increases accordingly. At $\Sigma_\mathrm{h}=\Sigmasol{1.4\times10^4}$, the most massive IMBHs reach masses exceeding $M_\bullet \sim \msol{10^3}$ at $Z<\zsol{0.10}$, well above the (P)PISN mass gap at metallicities below $Z \lesssim \zsol{0.2}$. Finally, the bottom panel of Fig. \ref{fig: Z-Mbh1} shows our models with the highest initial densities. As already shown in Fig. \ref{fig: mass-evolution}, the maximum IMBH masses of the models M3D5Z[1--6] with $Z \lesssim \zsol{0.2}$ are in the range of $\msol{1.5\times10^3} \lesssim M_\bullet \lesssim \msol{1.9\times10^3}$. At higher metallicities ($Z \gtrsim \zsol{0.2}$) the IMBH masses considerably decrease due to the strong stellar winds. 

Fig. \ref{fig: Z-Mbh2} reveals two clear trends for the IMBH masses as a function of their host cluster metallicity and density. First, at low metallicities below $Z \lesssim \zsol{0.06}$, the IMBH masses steadily increase as a function of the host cluster mass above $\Sigma_\mathrm{h} \gtrsim \Sigmasol{6.3\times10^4}$. This is due to the combination of increasingly lengthy collision cascades and relatively weak stellar winds at low metallicities. Second, the IMBH masses strongly decline towards higher metallicities in all dense cluster models. Most importantly, the metallicity threshold after which the IMBH masses strongly decline depends on the cluster surface density. For models with $\Sigma_\mathrm{h} = \Sigmasol{6.3\times10^4}$--$\Sigmasol{1.4\times10^5}$ the IMBH masses begin to steadily decrease after $Z \gtrsim \zsol{0.06}$, and for the highest density models with  $\Sigma_\mathrm{h} = \Sigmasol{2.9\times10^5}$ at $Z \gtrsim \zsol{0.2}$. Finally, we note that the final IMBH masses are affected by the uncertainties in the extremely massive star wind mass loss rates. We briefly explore this in Appendix \ref{appedix: wind}.

%%%%%%%%%%%%%%%%%%%%%%%%%
\subsection{The combined effect of cluster mass, metallicity and surface density on the IMBH masses}

\begin{figure*}
\includegraphics[width=0.8\textwidth]{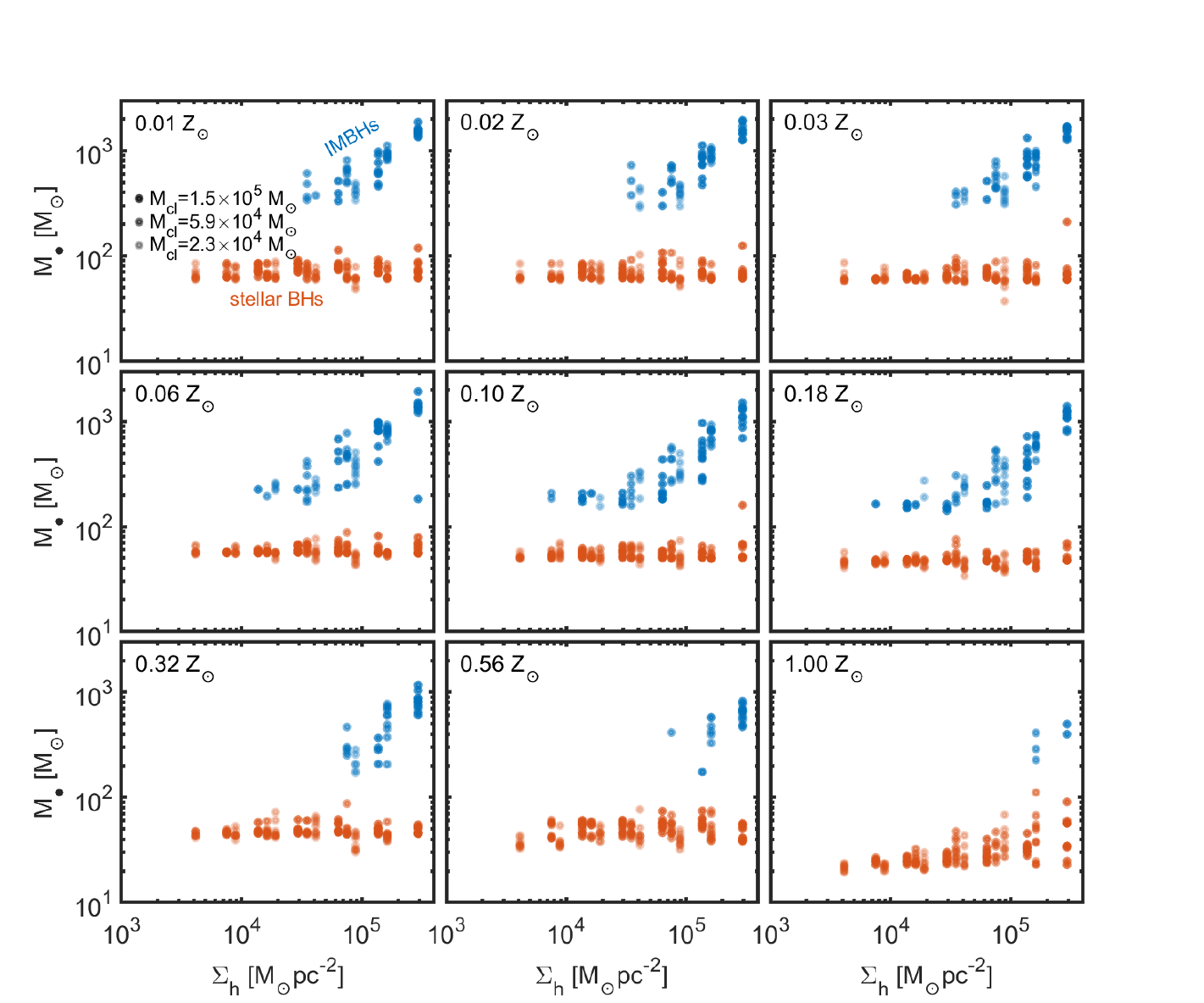}
\caption{The most massive stellar BHs and IMBHs in the isolated simulations at $t=7.5$ Myr at different metallicities as a function of the cluster half mass surface density $\Sigma_\mathrm{h}$. The cluster masses $\msol{2.3\times10^4} \lesssim M_\bullet \lesssim \msol{1.5\times10^5}$ are indicated by the increasingly dark shaded symbols. The maximum stellar BH masses typically lie close to the lower edge of the (P)PISN mass gap at each different metallicity. At high surface densities individual stellar BHs may grow via micro-TDEs as discussed in the main text. Especially at metallicities $Z\lesssim\zsol{0.32}$ there is a clear correlation between the IMBH masses and the host cluster half mass surface densities. For a fixed cluster half mass surface density $\Sigma_\mathrm{h}$ the cluster mass $M_\mathrm{cl}$ has little effect on $M_\mathrm{\bullet}$.}
\label{fig: Mbh-combined}
\end{figure*}

\subsubsection{Above a critical cluster mass threshold, $\Sigma_\mathrm{h}$ and $Z$ determine the maximum $M_\bullet$}

In the previous sections we have focused on our isolated star cluster models with $M_\mathrm{cl}=\msol{1.5\times10^5}$. We now extend these results to our lower cluster masses of $M_\mathrm{cl}=\msol{2.3\times10^4}$ and $M_\mathrm{cl}=\msol{5.9\times10^4}$. We note that all the examined cluster setups are massive enough to potentially form IMBHs: our previous simulations \citep{Rantala2024b, Rantala2025b} have demonstrated that dense but low mass ($M_\mathrm{cl}\lesssim\msol{10^4}$) clusters rarely produce an IMBH. As shown in Table \ref{tab: isolated}, the surface density ranges for isolated cluster models of different masses largely overlap. We present the masses of the most massive stellar BHs and IMBHs formed in all our $1350$ isolated star cluster models in Fig. \ref{fig: Mbh-combined}. For stellar BHs the results of the lower mass cluster models agree with the $M_\mathrm{cl}=\msol{1.5\times10^5}$ results presented in the previous section. The lowest cluster mass models with $M_\mathrm{cl}=\msol{2.3\times10^4}$ produce somewhat less massive stellar BHs because of their lower IMF cut-off mass $m_\mathrm{max,0}$ in the initial conditions as indicated in Table \ref{tab: isolated}. At metallicities below $Z \lesssim \zsol{0.2}$, IMBHs form in the models with different masses with a clear correlation between the cluster surface densities and the resulting IMBH masses. For a fixed metallicity $Z$ and surface density $\Sigma_\mathrm{h}$ the cluster mass $M_\mathrm{cl}$ has a little effect on the final IMBH masses $M_\bullet$. Thus, it seems that only two parameters,  $Z$ and $\Sigma_\mathrm{h}$, are required to determine the maximum $M_\mathrm{\bullet}$ in isolated setups. This is not unexpected as for our models $\Sigma_\mathrm{h}$ and $M_\mathrm{cl}$ are not completely independent but related through the shallow fixed power-law slope initial mass-radius relation of Eq. \eqref{eq: mass-size-isolated} of the clusters. Towards higher metallicities, the critical surface density threshold for IMBH formation increases, and above $Z\gtrsim \zsol{0.3}$ fewer IMBHs form, especially at solar metallicity.

\subsubsection{A LOESS smoothed model for $M_\bullet$ as a function of host cluster properties}

\begin{figure*}
\includegraphics[width=0.75\textwidth]{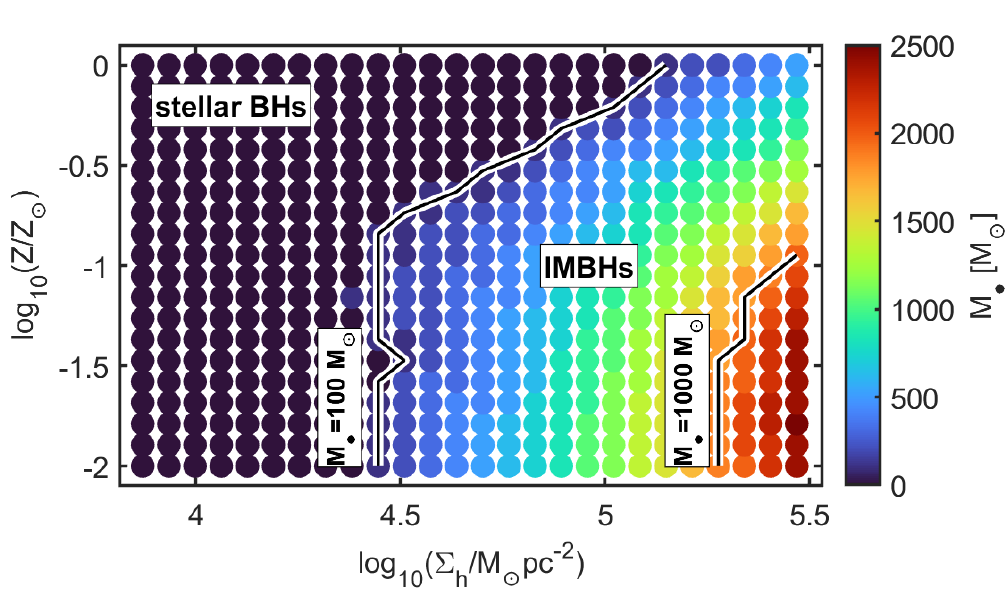}
\caption{The LOESS smoothed map presenting the maximum IMBH mass as a function of their host star cluster half mass surface density $\Sigma_\mathrm{h}$ and metallicity $Z$. The contours for $M_\bullet = \msol{100}$ and $M_\bullet = \msol{1000}$ show very regular angular shapes. Stellar BHs occupy the parameter space towards lower surface densities and higher metallicities from the critical contour of $M_\bullet = \msol{100}$ while increasingly massive IMBHs are found towards the bottom right corner with the highest cluster densities and lowest metallicities.}
\label{fig: loess}
\end{figure*}

Motivated by the evident correlations between the maximum IMBH masses and $\Sigma_\mathrm{h}$ at each cluster metallicity in Fig. \ref{fig: Mbh-combined}, we proceed to build a model for the formed IMBH masses as a function of their host cluster surface density and metallicity, i.e. $M_\bullet = M_\bullet(\Sigma_\mathrm{h},Z)$. However, even with our isolated simulation sample of $1350$ models, our models do not uniformly sample the two-dimensional $(\Sigma_\mathrm{h},Z)$ parameter space. Therefore, we produce a two-dimensional smoothed map $M_\bullet(\Sigma_\mathrm{h},Z)$ employing the publicly available Locally Weighted Regression (LOESS) software package\footnote{\url{https://users.physics.ox.ac.uk/~cappellari/software/}} of \cite{Cappellari2013}. LOESS is an adaptive smoothing technique designed to recover mean trends from scattered data points with noise in one or two dimensions \citep{Cleveland1988}. The resulting smoothed map $M_\bullet(\Sigma_\mathrm{h},Z)$ is presented in Fig. \ref{fig: loess}. We emphasize that the figure only displays the maximum masses of the IMBHs that can form in the clusters, not how frequently they form. As shown in Fig. \ref{fig: prob}, IMBHs rarely form at metallicities close to solar even in dense clusters.

Two separate regions are apparent in the smoothed map $M_\bullet(\Sigma_\mathrm{h},Z)$ of Fig. \ref{fig: loess}: stellar BHs at low densities and high metallicities, and IMBHs at high densities and low metallicities. At low surface densities the most massive BHs are always in the stellar mass regime with $M_\bullet \lesssim \msol{100}$. Above a critical metallicity, IMBHs can only form in sufficiently dense systems separated from the stellar BH regime by the surface density dependent critical metallicity. A metallicity threshold contour line $Z_\mathrm{\msol{100}}=Z(M_\bullet=\msol{100})$ is presented in Fig. \ref{fig: loess} and approximately follows the equation
\begin{equation}\label{eq: Zcrit}
    \log_\mathrm{10}\left( \frac{Z_\mathrm{\msol{100}}}{\zsol{}} \right) \approx \begin{cases}
    \leq \log_\mathrm{10}\left( \frac{Z_\mathrm{crit}}{\zsol{}} \right) \hspace{2.15cm} \text{if $\Sigma_\mathrm{h} = \Sigma_\mathrm{crit}$}\\
    k_\mathrm{crit} \log_\mathrm{10}\left( \frac{\Sigma_\mathrm{h}}{\Sigma_\mathrm{crit}} \right) + \log_\mathrm{10}\left( \frac{Z_\mathrm{crit}}{\zsol{}} \right) \hspace{0.0cm} \text{if $\Sigma_\mathrm{crit} < \Sigma_\mathrm{h}$ }
    \end{cases}
\end{equation}
in which $\Sigma_\mathrm{crit} \approx \Sigmasol{2.79\times10^4}$, $Z_\mathrm{crit}\approx \zsol{0.14}$ and $k_\mathrm{crit}\approx1.20$. The contour line for $M_\bullet = \msol{1000}$ displayed in Fig. \ref{fig: loess} shows a very similar angular shape as the contour for $Z_\mathrm{\msol{100}}$. Even at low metallicities, star cluster surface densities above $\Sigma_\mathrm{h}\gtrsim \Sigmasol{2\times10^5}$ are required to produce a $M_\bullet=\msol{10^3}$ IMBH.

\subsection{A parametrised fit model for estimating star cluster IMBH masses}

\begin{figure*}
\includegraphics[width=0.75\textwidth]{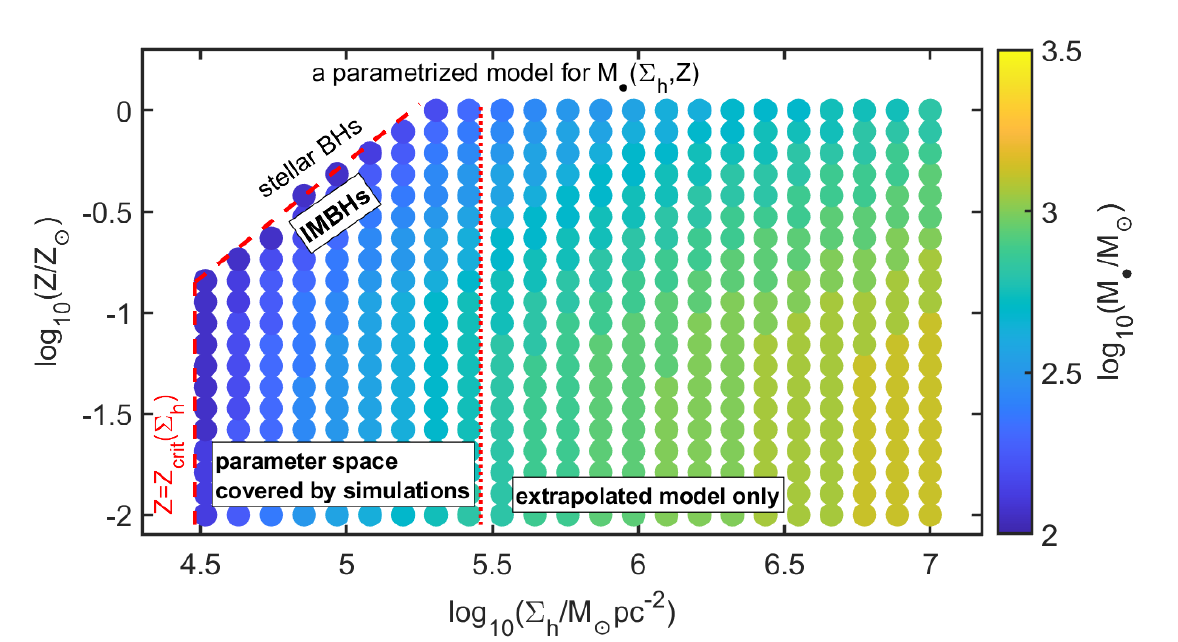}
\caption{The parametrized model for IMBH masses as a function of their host cluster half mass surface densities $\Sigma_\mathrm{h}$ and metallicities $Z$. For the illustration we use Eq. \eqref{eq: 2d-fit-inter} below $\log_\mathrm{10}(\Sigma_\mathrm{h}/\Sigmasol{})<5.45$, and Eq. \eqref{eq: 2d-fit-extra} above it.}
\label{fig: fit}
\end{figure*}

The very regular shapes of the critical metallicity contour lines for IMBH masses $M_\bullet=\msol{100}$ and $M_\bullet=\msol{1000}$ in Fig. \ref{fig: loess} and Eq. \eqref{eq: Zcrit} suggest that it should be possible to construct a parametrized model for the IMBH masses $M_\bullet(\Sigma_\mathrm{h},Z)$ that follows a relatively simple functional form. We show that this is indeed the case, but special care is required to ensure that the parametrized model remains well-behaved outside the parameter space of the performed simulations especially towards increasingly high cluster surface densities.

In order to find a suitable parametrized fitting formula for the LOESS smoothed map $M_\bullet(\Sigma_\mathrm{h},Z)$ in Fig. \ref{fig: loess} we first have to omit the stellar BH masses from the fitting process and focus on the IMBH regime on the right of the $M_\bullet=\msol{100}$ contour. This is because for each metallicity $Z$, $M_\bullet(\Sigma_\mathrm{h})$ has a sudden step function like jump from the lower edge of the (P)PISN mass gap into the IMBH mass regime at the critical surface density $\Sigma_\mathrm{crit}$ at which runaway stellar collisions first begin to occur. Such a step function feature is difficult to fit using continuous and well-behaved functions, and we instead choose to focus on the IMBH mass regime for the parametrization of $M_\bullet(\Sigma_\mathrm{h},Z)$. The stellar BH masses at low cluster surface densities and high metallicities can be obtained with any suitable fast stellar population synthesis model \citep{Hurley2000,Hurley2002,Iorio2023} instead.

Our aim is to find a relatively simple parametrized functional form for $M_\bullet(\Sigma_\mathrm{h},Z)$ which remains well behaved when extrapolated to higher star cluster surface densities beyond $\gtrsim \Sigmasol{2.9\times10^5}$. We provide two different parametrized functional forms: a simple formula that behaves well at high extrapolated cluster surface densities, and a somewhat more complex formula for the $(\Sigma_\mathrm{h},Z)$ parameter space of our simulations. For both formulas, we first fit $M_\bullet(\Sigma_\mathrm{h})$ for each fixed metallicity $Z$, and then find a second parametric model to describe the fit parameters as a function of metallicity. After experimenting with various functional forms to be used within the $(\Sigma_\mathrm{h},Z)$ parameter space of our simulations we arrive at
\begin{equation}\label{eq: 2d-fit-inter}
\begin{split}
    \frac{M_\bullet}{\msol{}} &= \theta_\mathrm{H}\Bigg( \Sigma_\mathrm{h} - \Sigma_\mathrm{crit}(Z) \Bigg) \Bigg\{ A(Z) \log_\mathrm{10}\left( \frac{\Sigma_\mathrm{h}}{\Sigmasol{}} \right)\\ &+ B(Z) \left[ \log_\mathrm{10}\left( \frac{\Sigma_\mathrm{h}}{\Sigmasol{}} \right) \right]^\mathrm{2} + C(Z) \Bigg\} 
\end{split}
\end{equation}
in which $\theta_\mathrm{H}(x)$ is the Heaviside step function, and the coefficients $\Sigma_\mathrm{crit}(Z), A(Z)$, $B(Z)$ and $C(Z)$ are metallicity dependent. For example, the coefficient $A(Z)$ can be represented using a piecewise defined function of the form $A(Z) = A_\mathrm{1} \log_\mathrm{10}(Z/\zsol{}) + A_\mathrm{2}$ in three metallicity intervals: $Z/\zsol{}<0.126$, $0.126\leq Z/\zsol{} < 0.398$ and $Z>0.398$. The coefficients $B(Z)$, $C(Z)$ and $\Sigma_\mathrm{crit}(Z)$ follow similar piecewise relations. We provide the relevant metallicity dependent coefficients $A_\mathrm{1}$, $A_\mathrm{2}$, $B_\mathrm{1}$, $B_\mathrm{2}$, $C_\mathrm{1}$, $C_\mathrm{2}$, $\Sigma_\mathrm{1}$ and $\Sigma_\mathrm{2}$ for the parametrized formulas in Table \ref{table: fit-coeff-inter}.

\begin{table}
    \centering
    \begin{tabular}{clll}
        \hline
         Coeff. & $Z/\zsol{}<0.126$ & $0.126 \leq Z/\zsol{} < 0.398$  & $Z/\zsol{} > 0.398$\\
         \hline
$A_\mathrm{1}$ & -1790.07 & +9707.84 & +1147.68 \\ 
$A_\mathrm{2}$ & -7392.65 & +2627.71 & -721.18 \\ 
$B_\mathrm{1}$ & +162.46 & -1015.72 & -166.92 \\ 
$B_\mathrm{2}$ & +829.42 & -211.38 & +126.15 \\ 
$C_\mathrm{1}$ & +4734.11 & -23585.20 & -2002.25 \\ 
$C_\mathrm{2}$ & +16556.82 & -7814.04 & +471.59 \\ 
$\Sigma_\mathrm{1}$ & +0.0386 & +0.91 & +0.91 \\ 
$\Sigma_\mathrm{2}$ & +4.53 & +5.22 & +5.22 \\ 
         \hline
    \end{tabular}
    \caption{The parametrized coefficients for constructing $A(Z)$, $B(Z)$, $C(Z)$ and $\Sigma(Z)$ in Eq. \eqref{eq: 2d-fit-inter}. For example, $A(Z) = A_\mathrm{1} \log_\mathrm{10}(Z/\zsol{}) + A_\mathrm{2}$.}
    \label{table: fit-coeff-inter}
\end{table}

An alternative fit parametrization for high star cluster surface densities above $\log_\mathrm{10}(\Sigma_\mathrm{high}/\Sigmasol{})=5.2$ is
\begin{equation}\label{eq: 2d-fit-extra}
\begin{split}
    \frac{M_\bullet}{\msol{}} &= \theta_\mathrm{H}\Bigg( \Sigma_\mathrm{h} - \Sigma_\mathrm{high} \Bigg) \Bigg\{ D(Z) \log_\mathrm{10}\left( \frac{\Sigma_\mathrm{h}}{\Sigmasol{}} \right) + E(Z) \Bigg\} 
\end{split}
\end{equation}
in which the parameters $D(Z)$ and $E(Z)$ are again metallicity dependent. The coefficients $D_\mathrm{1}$, $D_\mathrm{2}$, $E_\mathrm{1}$ and $E_\mathrm{2}$ to construct e.g. $D(Z)$ via $D(Z) = D_\mathrm{1} \log_\mathrm{10}(Z/\zsol{}) + D_\mathrm{2}$ can be found in Table \ref{table: fit-coeff-extra}. This second parametrized model for $M_\bullet$ remains well behaved also at high star cluster surface densities. We present the model in Fig. \ref{fig: fit} extrapolating the results for $M_\bullet$ up to high surface densities of $\Sigma_\mathrm{h} = \Sigmasol{3.2\times10^6}$, approximately an order of magnitude higher than our most dense models.

\begin{table}
    \centering
    \begin{tabular}{cccc}
        \hline
         Coeff. & $Z/\zsol{}<0.079$ & $0.079 \leq Z/\zsol{} < 0.316$  & $Z/\zsol{} > 0.316$\\
         \hline
$D_\mathrm{1}$ & -37.37 & -922.15 & -611.27 \\ 
$D_\mathrm{2}$ & +1452.33 & +466.71 & +628.40 \\ 
$E_\mathrm{1}$ & +81.66 & +4242.58 & +2620.25 \\ 
$E_\mathrm{2}$ & -6892.57 & -2280.02 & -3137.93 \\
         \hline
    \end{tabular}
    \caption{The parametrized coefficients for constructing $D(Z)$ and $E(Z)$ in Eq. \eqref{eq: 2d-fit-extra}. For $\Sigma_\mathrm{high}$ we use $\log_\mathrm{10}(\Sigma_\mathrm{high}/\Sigmasol{})=5.2$}
    \label{table: fit-coeff-extra}
\end{table}

For using the models of Eq. \eqref{eq: 2d-fit-inter} and Eq. \eqref{eq: 2d-fit-extra} for seeding IMBHs into star clusters in semi-analytic galaxy evolution frameworks (e.g. \texttt{L-Galaxies}; \citealt{Spinoso2023,Hoyer2025}, \texttt{CAT}; \citealt{Valiante2011, Trinca2022}, \texttt{Delphi}; \citealt{Dayal2019}) or high mass resolution ($\lesssim \msol{10}$) hydrodynamical simulations, the 3D half mass density $\rho_\mathrm{h}$ may be required instead of the projected $\Sigma_\mathrm{h}$. For the Plummer model the 2D and 3D half mass densities are related as \begin{equation}
\frac{\Sigma_\mathrm{h}(\rho_\mathrm{h})}{\Sigmasol{6.99\times10^4}} = \left( \frac{M_\mathrm{cl}}{\msol{10^5}} \right)^\mathrm{1/3} \left( \frac{\rho_\mathrm{h}}{\rhosol{10^5}} \right)^\mathrm{2/3}.
\end{equation}
For simulation models that can track star cluster masses but not their densities (e.g. gravitationally softened hydrodynamical models with mass resolutions above $m_\star>\msol{100})$, one should assume the normalization of the cluster birth mass size relation $f_\mathrm{h}$ from Eq. \eqref{eq: mass-size-isolated} to use the IMBH mass fitting formulas.

%%%%%%%%%%%%%%%%%%%%%%%%%%%%%%%%%%%%%%%%%%%%%%%%%%%%%%%%%%%%%%%%%%%
%%%%%%%%%%%%%%%%%%%%%%%%%%%%%%%%%%%%%
\section{Hierarchical star cluster assembly and IMBH formation at $Z=\zsol{0.01}$ and $Z=\zsol{0.1}$}\label{section: 4}

\begin{figure*}
\includegraphics[width=1.0\textwidth]{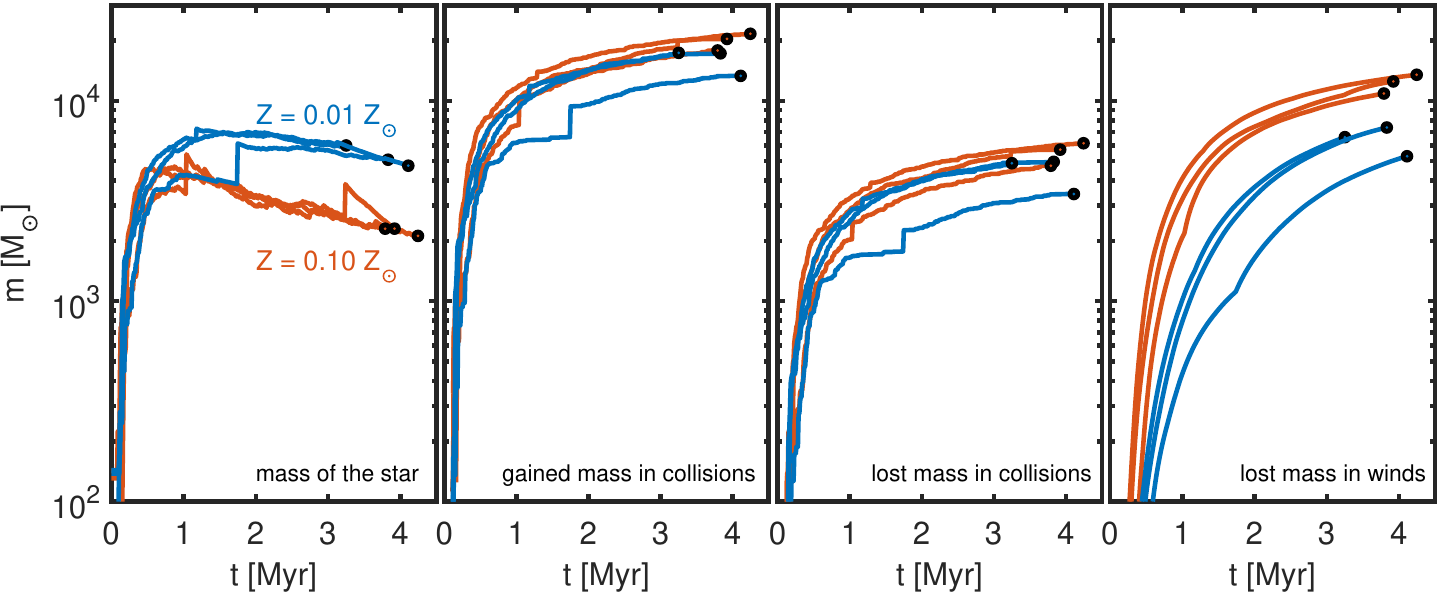}
\caption{The mass growth and loss histories of six most massive stars formed in our densest hierarchical models HD9Z1 ($Z=\zsol{0.01}$) and HD9Z2 ($Z=\zsol{0.10}$). First panel: the masses of the stars as function of time. Second panel: the cumulative gained mass by the stars i.e. the sum of secondary impacting stars $m_\mathrm{2}$. Third and fourth panels: the cumulative mass lost in collisions and stellar winds. While the gained masses are comparable in models with different metallicities, the higher wind loss rates at $Z=\zsol{0.10}$ lead to earlier quenching of the runaway mass growth. Note that the runaway collisions still proceed after $t\gtrsim1$ Myr even though the mass growth of the stars plateaus or turns into decline. No star in our densest models reaches $m_\star = \msol{10000}$ while the total gained mass in stellar collisions can exceed $m_\mathrm{gain} \gtrsim \msol{21600}$.}
\label{fig: massgrowth-hierarchical}
\end{figure*}

\subsection{IMBH formation in extremely dense hierarchically assembling star clusters}\label{section: hierarchical-imbh-4.1}

After our exploration of idealised, isolated cluster models initially in equilibrium we proceed to simulate hierarchically assembling massive star clusters. While the isolated setups can be regarded as useful numerical experiments to study IMBH formation in dense systems, the hierarchical models represent more realistic physically motivated initial setups for star cluster formation. We note that even the hierarchical \nbody{} models have a degree of idealization as they do not take into account the extended star formation history a star cluster would have in reality. We perform in total $30$ hierarchical star cluster assembly simulations as listed in Table \ref{tab: hierarchical-models} to explore the collisional massive star and IMBH formation as well as chemical enrichment by such massive stars in different metallicities and especially cluster densities. In total $27$ models are simulated at $Z=\zsol{0.10}$ (models HD[1-9]Z2) while the three densest models are also run with $Z=\zsol{0.01}$ (models HD9Z1). The central clusters of the densest models initially have half mass densities up to $\rho_\mathrm{h} \sim \rhosol{6.5\times10^7}$ corresponding to half mass surface densities of $\Sigma_\mathrm{h} \sim \Sigmasol{3.9\times10^6}$. Our previous studies of the hierarchical cluster assembly \citep{Rantala2024b, Rantala2025a, Rantala2025b} have focused on clusters at $Z=\zsol{0.01}$ with somewhat lower central densities up to $\rho_\mathrm{c} \sim$1--2$\times\rhosol{10^6}$. The central sub-clusters of our intermediate density models HD5Z[1-2] closely resemble the simulated massive young star cluster of \cite{Lahen2025b} formed in a dwarf galaxy merger starburst.

We present the mass growth histories of the most massive star in our densest cluster models HD9Z1 and HD9Z2 in the left panel of Fig. \ref{fig: massgrowth-hierarchical}. In all extremely dense models the cascade of runaway stellar collisions commences after $t\sim0.1$ Myr and the stars rapidly reach twice the maximum initial IMF cut-off mass of $2\times m_\mathrm{max,0}=\msol{300}$ by $t=0.12$---$0.17$ Myr. The stars increase their mass via collisions and reach $m_\star = \msol{1000}$ by $t=0.15$---$0.29$ Myr and $m_\star = \msol{4000}$ before $t=1.04$ Myr. However, no massive star reaches $m_\mathrm{\star}=\msol{10000}$ in our simulation sample despite the extremely high initial star cluster densities. This is due to the stellar wind mass losses being higher or comparable to the collisional mass gains of the extremely massive stars after $t>1$ Myr. At $Z=\zsol{0.10}$, the maximum stellar masses are in the range of $\msol{4496} \lesssim m_\mathrm{max} \lesssim \msol{5413}$, and for $Z=\zsol{0.01}$, $\msol{6131} \lesssim m_\mathrm{max} \lesssim \msol{7263}$. The times for the peak collisional stellar masses occur between $t=0.70$--$1.04$ Myr for $Z=\zsol{0.10}$ and somewhat later between $t=1.18$--$1.99$ Myr in the models with $Z=\zsol{0.01}$. After the moment of peak collisional stellar mass, the masses of the stars either remain relatively constant until the end of the stellar lifetimes at $t=3$--$4$ Myr ($Z=\zsol{0.01}$) or steadily decline ($Z=\zsol{0.10}$). The masses of the IMBHs at their formation are $\msol{2118} \lesssim M_\mathrm{\bullet} \lesssim \msol{2302}$ ($Z=\zsol{0.10}$)
and $\msol{4747} \lesssim M_\mathrm{\bullet} \lesssim \msol{5988}$ ($Z=\zsol{0.01}$), respectively.

\begin{figure}
\includegraphics[width=1.0\columnwidth]{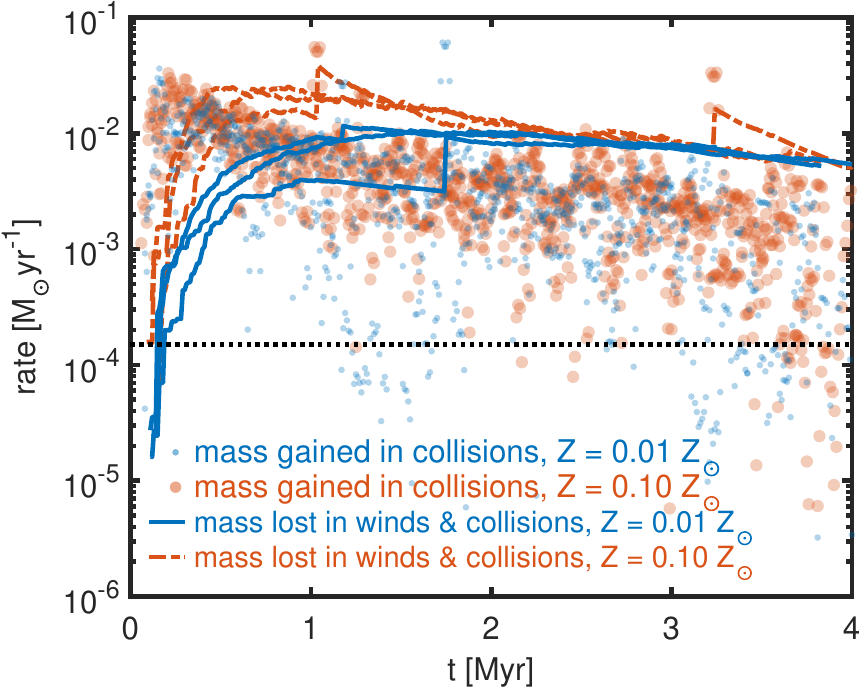}
\caption{The rates for gained (dots) and total lost mass (lines) from the most massive stars in our densest hierarchical models with $Z=\zsol{0.10}$ (HD9Z2) and $Z=\zsol{0.01}$ (HD9Z1). While the collisional mass gain rates at different metallicities are comparable, the total mass loss rates peak and exceed the mass gain rates earlier, and are in general higher in the models with $Z=\zsol{0.10}$. This is mainly due to the higher wind loss rates at higher metallicities.}
\label{fig: gain-loss-comparison}
\end{figure}

We examine the mass budget of the collisionally grown massive stars in our densest models in panels two to four of Fig. \ref{fig: massgrowth-hierarchical}. At lower metallicities $Z=\zsol{0.01}$, the masses of the extremely massive stars remain relatively constant after reaching their peak masses. This balance between the mass gained in the stellar collisions (the second panel) and the losses via collisions (the third panel) and winds (the final fourth panel) is maintained for $2$--$3$ Myr until the stars end their lives. The most massive star at $Z=\zsol{0.01}$ gains $m_\mathrm{gain} = \msol{17367}$ of material during its lifetime, and loses $m_\mathrm{coll} = \msol{4883}$ and $m_\mathrm{wind} = \msol{6590}$ via collisions and winds, respectively. The considerable amount of collisional mass loss is consistent with our estimates in Appendix \ref{appendix: cumuloss}. Both the collisional and wind mass losses are comparable to the mass of the formed IMBH ($M_\bullet = \msol{5988}$). In Fig. \ref{fig: gain-loss-comparison} we compare the total rates for stellar mass gain (via collisions) and mass loss (via collisions and winds). The collisional mass gain rates are calculated over time bins of $\Delta t=0.01$ Myr. The collisional mass gain rate peaks early with $\dot{m}_\mathrm{gain} \sim \ratesol{2\times10^{-2}}$, between $0.1$ Myr $\lesssim t \lesssim 0.2$ Myr after which it steadily declines. Collisions with other collisionally grown very massive stars are seen as brief peaks in the mass gain rate, although these collisions are relatively rare. For $Z=\zsol{0.01}$ the combined mass loss rate balances the mass gain rate around $t\sim0.8$--$1.7$ Myr, consistent with the plateau in the mass growth history of the stars in the left panel of Fig. \ref{fig: massgrowth-hierarchical}.

The most massive star at $Z=\zsol{0.10}$ gains $m_\mathrm{gain} = \msol{21621}$ of material during its lifetime, and loses $m_\mathrm{coll} = \msol{6144}$ and $m_\mathrm{wind} = \msol{13500}$ via collisions and winds for an IMBH mass of $M_\bullet = \msol{2302}$. In Fig. \ref{fig: gain-loss-comparison} we show that the collisional mass growth rates are comparable at different metallicities despite the metallicity dependence of our stellar models, including their radii and wind rates. Due to the higher wind loss rates at higher metallicities, the combined collision and wind loss rates reach and exceed the mass gain rates already around $0.27$ Myr $\lesssim t \lesssim 0.51$ Myr. This explains why the masses of the collisionally grown stars with $Z=\zsol{0.1}$ reach their peak masses earlier than their counterparts with $Z=\zsol{0.01}$ in the left panel of Fig. \ref{fig: massgrowth-hierarchical} and result in lower overall IMBH masses at the end of the simulations.

\subsection{Comparison to the dense model of \citet{Vergara2025}: the effect of wind loss rates}

\begin{figure}
\includegraphics[width=1.0\columnwidth]{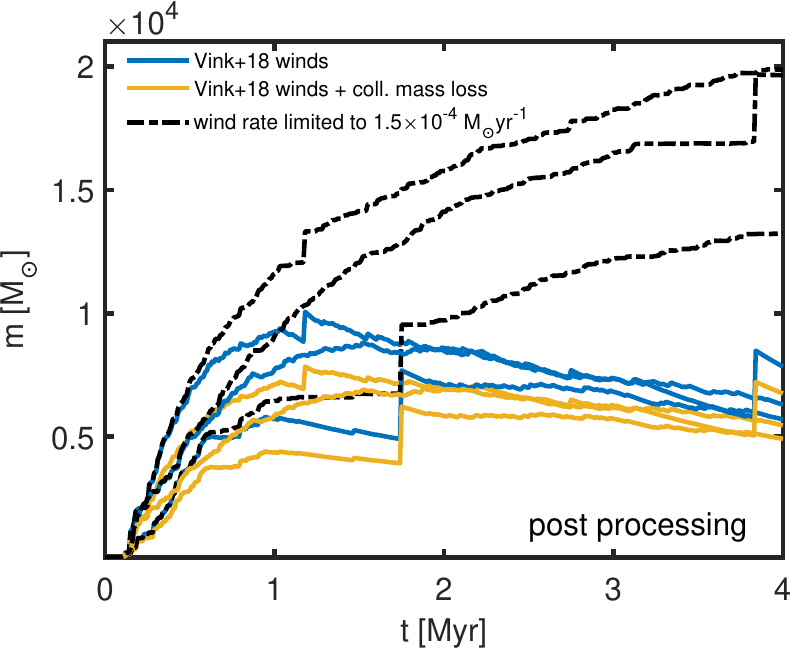}
\caption{The post processed stellar mass evolution models from our densest simulations exploring the effect of the upper wind mass loss limit on the maximum stellar masses. The solid lines indicate simple models with collisions with and without collisional mass loss including \citet{Vink2018} winds, capturing the overall behaviour of our full \nbody{} models. Placing an upper limit for the stellar wind loss rate (dashed line) results in monotonic stellar mass growth beyond $\msol{20000}$.}
\label{fig: postprocess}
\end{figure}

Recently, \cite{Vergara2025,Vergara2025b} modelled the formation of a collisionally grown massive star of $\gtrsim\msol{50000}$ and an IMBH in an extremely dense isolated star cluster using \texttt{NBODY6++GPU} and \texttt{MOCCA}. Even though their cluster mass of $M_\mathrm{cl} = \msol{5.86\times10^5}$ is by a factor $\sim2.3$ higher compared to the masses of central sub-cluster of our densest models, the half mass density of their model $\rho_\mathrm{h} = \rhosol{6.99\times10^7}$ is less than $10\%$ higher compared to ours. The initial stellar mass function of the two models is the same while our models initially also include stars in primordial binary systems. Despite the similar initial stellar densities of our models, \cite{Vergara2025} reach by a factor of $\sim7$ higher peak stellar masses compared to our simulations. The growth history of their massive star is closely comparable to our models until $t\sim0.3$--$0.4$ Myr. While the mass growth histories of our most massive stars plateau or turn into decline soon after this, the star of \cite{Vergara2025} continues to monotonically grow until the end of its life at $t>4$ Myr. Such growth history considerably differs from our models. 

We identify three differences in the extremely massive star models currently employed in \texttt{NBODY6++GPU}, \texttt{MOCCA} and \bifrost{} to explain the differences of the maximum stellar masses in the comparably dense cluster models. These are: 1) winds, 2) collisional mass loss from the primary stars 3) the extremely massive star radii at late evolutionary stages. The assumed wind loss rates are by far the dominant driver of the differences. In Fig. \ref{fig: massgrowth-hierarchical} and Fig. \ref{fig: gain-loss-comparison} we showed that the mass growth of our most massive stars is quenched due to collisional and especially wind losses even though the runaway collisions continue after $t>1$ Myr. In their level C stellar evolution for massive main sequence stars (see \citealt{Kamlah2022}) \cite{Vergara2025} employ the hot O and B type line driven winds of \cite{Vink2001}. Most importantly, their massive and extremely luminous stars beyond the Humphreys-Davidson limit \citep{Humphreys1994} ($L_\star/L_\mathrm{\odot} > 6\times10^5$ and $10^{-5}\times R_\star/R_\odot (L/L_\mathrm{\odot})^\mathrm{1/2}>1$) undergo luminous blue variable (LBV) like mass loss with wind loss rates of $\dot{m}_\mathrm{LBV} = \ratesol{1.5\times10^{-4}}$ \citep{Belczynski2010}, independent of stellar mass and metallicity. This effective ceiling sets the maximum main sequence wind mass loss rate in their models. In the \cite{Vink2018} wind model used in our study there is no such upper limit, and increasingly massive stars have increasingly high wind loss rates in the model.

We show that applying such a upper wind rate limit of  $\dot{m}_\mathrm{LBV} = 1.5 \times \ratesol{10^{-4}}$ in our models results in a qualitatively similar stellar mass growth as in \cite{Vergara2025}. In Fig. \ref{fig: postprocess} we perform a post processing analysis of the stars of our densest models with $Z=\zsol{0.01}$. Starting from the initial stars, we add infalling secondary star masses $m_\mathrm{2}$ at the times they collided with the massive primary in the simulations. In the basic post processed model we only apply mass gain by collisions, and wind losses according to \cite{Vink2018}. The stars are evolved using the simple model until $t=4$ Myr. In Fig. \ref{fig: postprocess} this model leads to a qualitatively similar behaviour compared to our \nbody{} simulations: the masses of the stars reach a plateau and decrease before the end of their lives. Next, we employ a wind rate limit of $\dot{m}_\mathrm{max} = 1.5 \times \ratesol{10^{-4}}$ in the simple model. For $Z=\zsol{0.01}$, the limiting wind rate is reached in our models around $m_\star \sim \msol{1700}$. Applying the wind rate limit we cover a qualitatively similar behaviour as seen in the model of \cite{Vergara2025}: the stars monotonically grow until the end of the lives, the most massive one reaching $\msol{20000}$. 

We note that our post-processed model with $m_\mathrm{max} = \msol{20000}$ only represents a lower limit for the maximum stellar mass. In a full \nbody{} simulation the increased mass of the star (compared to models in which the mass growth is quenched) would increase both the radii and gravitational focusing cross section of the star, further enhancing its growth via collisions. To verify this, we perform three isolated very dense cluster simulations in Appendix \ref{appendix: wind-limit} and Fig. \ref{fig: appendix-windrate-limit}. The most massive star of the three models reaches $m_\star>\msol{25000}$. These test simulations strongly support our findings in the post-processed model: including the maximum wind rate limit enables monotonic collisional stellar mass growth well into the SMS regime above $\gtrsim \msol{20000}$.

The second difference between the massive star prescriptions in \texttt{NBODY6++GPU}, \texttt{MOCCA} and \bifrost{} is the mass loss in collisions. \cite{Vergara2025} only apply collisional mass loss for the secondary star, and only for hyperbolic collisions. Even though their model takes the velocity of the hyperbolic collision into account while our simple model does not, their model never applies mass loss for the primary star of the collision. We have shown in Appendix \ref{appendix: cumuloss} that even $<10\%$ mass loss per collision can result in a substantial cumulative collisional mass loss during the lifetime of an extremely massive star, if collisional losses from the primary are also included in the models.

In our most dense models the collisional losses can become almost comparable to wind losses at $Z=\zsol{0.01}$. In the models without collisional mass loss, higher wind loss rates immediately after collisions compensate for the absence of the collisional ejecta. In Fig. \ref{fig: postprocess}, we further include the primary mass loss in stellar collisions in the simple post-processed model. The inclusion of collisional mass loss from the primary in the models only somewhat decreases the final stellar masses at $t=4$ Myr: without the collisional mass loss the final masses are $15$-$16\%$ higher.

Finally, the radii of extremely massive stars in the models included in \texttt{NBODY6++GPU}, \texttt{MOCCA} and \bifrost{} considerably differ late in the evolution of the stars. However, this occurs only at \mbox{$t>2$ Myr} and thus does not affect the differences of mass growth between $0.3$ Myr $\lesssim t \lesssim 2$ Myr discussed above. The peak stellar masses in our models are always reached before $t<2$ Myr. At this point the radii of the extremely massive stars have their maximum values of $R_\mathrm{\star} \sim \Rsol{107}$, consistent with the star of \cite{Vergara2025} that has a radius of $R_\mathrm{\star} \sim \Rsol{100}$ at $t=1.60$ Myr. However, after this point the radius of the star begins to increase in \texttt{NBODY6++GPU} and \texttt{MOCCA}, reaching $\Rsol{200}$, $\Rsol{1000}$ and $\Rsol{10000}$ at $t=2.36$ Myr, $t=3.25$ Myr and $t=3.91$ Myr, respectively. The maximum radius of the $\msol{50000}$ star is $\Rsol{2\times10^5}$, approximately by a factor $100$-$1000$ of larger than in our models (see Section \ref{section: radii}). This leads to a very large difference between the collisional cross sections (by a factor of $>10^4$) late in the evolution of the extremely massive stars between \texttt{NBODY6++GPU} and \bifrost, and contributes to the sustained high collisional mass gains of the extremely massive stars of \cite{Vergara2025}.

\begin{table}
    \centering
    \begin{tabular}{ccccc}
       \hline
        Hierarchical & $\Sigma_\mathrm{c}$ & $\rho_\mathrm{c}$ & $N_\mathrm{IMBH}$ &\\
        model & [$\Sigmasol{10^5}$] & [$\rhosol{10^5}$] & (max) & \\
       \hline
        HD1Z2 & $0.11$ & $0.21$ & 0 &\\
        HD2Z2 & $0.17$ & $2.04$ & 0  &\\
        HD3Z2 & $0.38$ & $3.88$ & 0--1 &\\
        HD4Z2 & $0.54$ & $4.90$ & 3--5 &\\
        HD5Z2 & $0.93$ & $9.16$ & 4--6 &\\
        HD6Z2 & $1.12$ & $13.47$ & 4--7 &\\
        HD7Z2 & $1.41$ & $14.01$ & 5--7 &\\
        HD8Z2 & $1.96$ & $26.9$ & 6--9 &\\
        HD9Z2 & $3.15$ & $121.1$ & 8--9 &\\
        \hline
    \end{tabular}
    \caption{The central stellar surface densities $\Sigma_\mathrm{c}$ and central 3D stellar densities $\rho_\mathrm{c}$ of the hierarchically assembled star cluster models with $Z=\zsol{0.10}$ at $t=7.5$ Myr. We also show the maximum number of IMBHs with masses exceeding $M_\bullet \gtrsim \msol{300}$ (at any moment of time in the simulations) as $N_\mathrm{IMBH}$. As expected, denser models result in more efficient IMBH formation. The densest low metallicity comparison simulations with $Z=\zsol{0.01}$ formed $9$--$12$ IMBHs.}
    \label{table: assembled}
\end{table}

\subsection{Final cluster central densities}

\begin{figure}
\includegraphics[width=\columnwidth]{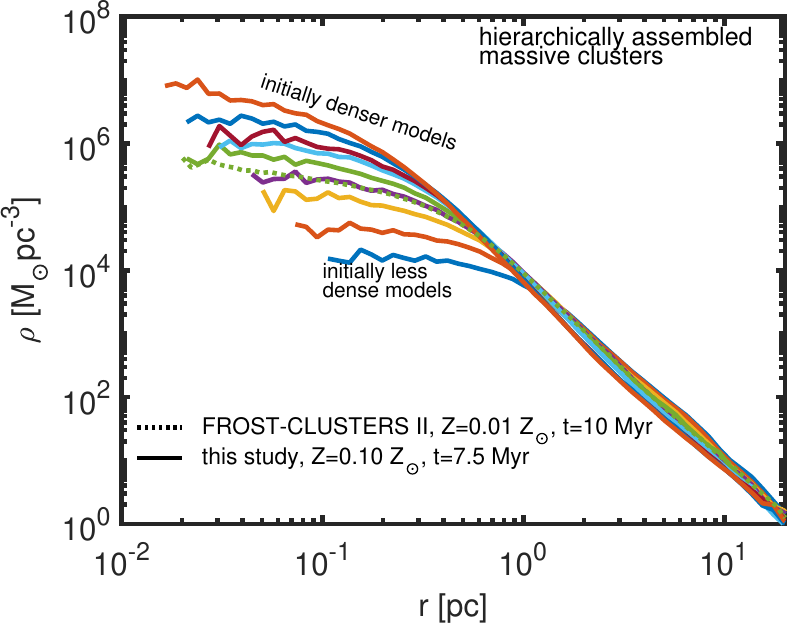}
\caption{The 3D stellar density profiles (solid lines) of the assembling massive central star cluster at $t=7.5$ Myr. We show for comparison the medium density low metallicity model HB-150 at $t=10$ Myr from our previous study \citep{Rantala2025b} as the dotted line. Initially denser models result in higher final central densities, as expected, however the central densities decrease in our densest models during the simulations due to strong two body relaxation effects.}
\label{fig: densityprofile}
\end{figure}

Next, we examine the densities of the hierarchically assembled star clusters. At $Z=\zsol{0.10}$, the hierarchical star cluster assembly proceeds as at lower metallicities (see e.g. figure 3 of \citealt{Rantala2025a}). At the end of the simulations at $t=7.5$ Myr all the sub-clusters have not merged yet with the central growing massive cluster especially in dense models in which the sub-clusters are less vulnerable for disruption. As the assembled clusters still have sub-structure and are out of equilibrium, the measured effective radii using the same recipe as in our previous study \citep{Rantala2025b} does not reliably correspond the final sizes of the assembled clusters. As the merging sub-clusters mainly deposit their stars in the outskirts of the massive central cluster later in its evolution, we can instead focus on the central stellar profiles of the assembled clusters which are less vulnerable to change compared to the outer parts.

We show the stellar density profiles of the massive assembling clusters at $t=7.5$ Myr in Fig. \ref{fig: densityprofile}. The illustrated profiles are averages of three random realizations of each hierarchical model. Our medium density model HD5Z2 ($Z=\zsol{0.10}, t=7.5$ Myr) has initial conditions closely resembling our previous model HB-150 in \cite{Rantala2025b}. We show this profile from \cite{Rantala2025b} at $Z=\zsol{0.01}, t=10$ Myr in  Fig. \ref{fig: densityprofile} to illustrate the profile evolution of the assembling clusters from $t=7.5$ Myr to $t=10$ Myr. In $2.5$ Myr, the central density in the models HD5Z2 decreases (see figure 5 of \citealt{Rantala2024b}) due to relaxation effects and stellar evolution mass losses while the remaining substructure is erased in the outer parts of the cluster. The densest assembled models reach high central densities of $\rho_\mathrm{c} \gtrsim \rhosol{10^7}$ measured within the central $\sim0.03$ pc. We list the central stellar densities $\rho_\mathrm{c}$ and the central surface densities $\Sigma_\mathrm{c}$ in Table \ref{table: assembled}. Initially denser models have larger final central densities, as expected. The central densities of the initially dense models above $\rho_\mathrm{c,init} \gtrsim \rhosol{8\times10^5}$ decrease during the cluster assembly. For extremely dense clusters the central density initially decreases due to strong two-body relaxation effects \citep{Binney2008} at high stellar densities. This rapid evolutionary phase lasts until $t\sim1.5$--$1.75$ Myr after which the central densities more gradually decline due to relaxation and stellar evolution effects, star cluster mergers and IMBH interactions in their aftermath. The final central densities of the initially densest clusters still remain high after the cluster assembly which allows sustained TDE rates after the IMBH formation. We will discuss the TDE rates in Section \ref{section: tde}.

\subsection{IMBH formation and GW mergers}

We list the total maximum number of IMBHs formed in each hierarchical simulation in Table \ref{table: assembled}. As expected, low density models up to HD3Z2 produce $\lesssim1$ IMBHs per simulation. At higher densities IMBHs form more efficiently: our densest models produce $N_\mathrm{IMBH}=8$--$9$ ($Z=\zsol{0.10}$) or $N_\mathrm{IMBH}=9$--$12$ ($Z=\zsol{0.01}$) IMBHs depending on metallicity. The mild metallicity dependence is consistent with our results from the isolated star cluster parameter study. Even in the densest models low mass clusters below $M_\mathrm{cl} \lesssim \msol{6\times10^3}$ do not form IMBHs. This highlights the critical role of massive stars ($m_\star \gtrsim \msol{100}$), which the low mass cluster models lack, in initiating the runaway collision cascades. However, above this critical mass threshold the IMBH masses are well described only by the cluster surface density and metallicity as seen in our isolated parameter study and \cite{Rantala2025b}.

Before $t=7.5$ Myr, in total eight GW driven mergers occur in our hierarchical simulations. These all occur in the setups with the two highest densities: H8D9Z2, H8D9Z1 and H8D9Z2. In the models HD8Z2 there is a single IMBH-IMBH merger with masses $M_\bullet=\msol{744}$ and $M_\bullet=\msol{562}$ ($q=0.76$). The rest of the seven GW driven mergers occur between stellar BHs and IMBHs with primary IMBH masses in the range of $\msol{743} \lesssim M_\bullet \lesssim \msol{3384}$ with mass ratios $0.02 \lesssim q \lesssim 0.06$. However, the IMBH interaction phase and merger phase is just beginning at $t=7.5$ Myr. In our previous studies (e.g. figures 6--8 in \citealt{Rantala2024b}) a large fraction of black hole interactions and GW driven mergers occurred between $7.5$ Myr $\leq t \leq 50$ Myr in the simulations.

At $t=7.5$ Myr, the centres of the low density $Z=\zsol{0.10}$ setups HD1Z2--HD3Z2 do not contain any IMBHs. For the intermediate density models HD4--HD7 the typical cluster centre contains an IMBH-IMBH binary ($58\%$ of the models), a single IMBH ($25\%$) or no IMBHs ($17\%$). In the highest density models central IMBH binaries and triples are the most common central IMBH configurations. Triple systems can occur more frequently in denser models as they produce the largest number of IMBHs, and dense stellar environments enable short sinking timescales in the aftermath of star cluster mergers. Rapid GW driven mergers in these densest models are expected as especially the triples are efficient in driving IMBH-IMBH mergers (e.g. \citealt{Liu2024,Rantala2025a}). Moreover, the high central stellar densities up to $\rho_\mathrm{c} \gtrsim \rhosol{10^7}$ enable faster stellar dynamical hardening of IMBH-IMBH binaries by stellar three body ejections compared to lower density clusters \citep{Souvaitzis2025}.

\subsection{Destruction of massive stellar binaries and supernova progenitors}

\begin{figure}
\includegraphics[width=\columnwidth]{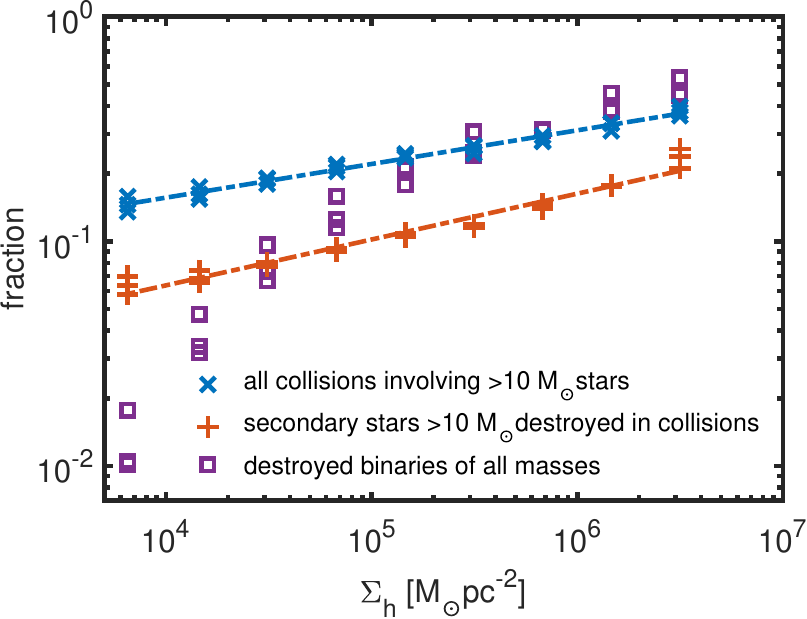}
\caption{The fraction of destroyed binary systems (square symbols) and potential supernova progenitors (plus symbols) as a function of the host star cluster surface density. Up to $\sim40\%$ of all supernova progenitors (cross symbols) can experience a stellar collision in the densest clusters.}
\label{fig: destruction}
\end{figure}

The extremely dense models of massive star clusters are particularly hostile environments for the longevity of massive initial (primordial) stellar binaries. In our initial conditions we use the same stellar binary population properties regardless the star cluster density. The higher destruction rate of more massive stellar binaries in denser models originates from the shorter mass segregation and dynamical friction timescales in denser environments. In our models, $>90\%$ of stars with $m_\mathrm{\star} \geq \msol{10}$ are initially members of binary systems. At the centres of the clusters, the frequent interactions with stars and later stellar BHs and IMBHs can unbind binary components, or drive them into merging with each other. The surviving massive stellar binaries evolve into BH-BH binaries which are potential future gravitational wave sources in their host clusters. Whereas our lowest density models HD1Z2 contain $81$ stellar BH-BH binaries at $t=7.5$ Myr, the corresponding highest density models HD9Z2 only have $2$--$3$. Moreover, stellar binaries with both components in a BH progenitor mass range of $\gtrsim \msol{19}$ are $\sim5$ times more numerous in the models HD1Z2 compared to HD9Z2. A similar trend is observable in the overall stellar binary content of the assembled star clusters. We illustrate this is in Fig. \ref{fig: destruction}. In our densest models $53\%$ of the binaries are destroyed during the first $t=7.5$ Myr, especially in the inner parts of the assembled clusters (see figure 10 of \citealt{Rantala2025b}). Meanwhile, the least dense models only have $1$--$2$\% of their binaries destroyed.

We note that stellar mergers in stellar mass binaries or with the extremely massive stars can destroy a substantial number of the massive stars ($\geq \msol{10}$) initially present in the clusters. This can lead to depletion of single stellar BH \citep{Paiella2025} and core collapse supernova progenitors in the clusters. We show in Fig. \ref{fig: destruction} both the fraction of $\geq \msol{10}$ stars that participated in any collisions, and $\geq \msol{10}$ stars which were destroyed as the secondary stars of the stellar mergers. As our models contain close initial binaries, the fraction of destroyed $\geq \msol{10}$ stars is $6$--$7\%$ for our low density models with $\Sigma_\mathrm{h} \lesssim \Sigmasol{10^4}$. However, the fraction of destroyed massive stars rapidly increases as a power-law function of increasing cluster surface density, and $21$--$26\%$ of the $\geq \msol{10}$ secondary stars are destroyed in our densest models. In total, up to $\sim40\%$ of stars more massive than $\geq \msol{10}$ participate in a collision as a primary or a secondary star. The destruction of $>\msol{10}$ stars is compensated (up to $\sim50\%$ in the densest models) by the generation of new $>\msol{10}$ stars in the collisions of lower mass stars.\\

The overall stellar wind and collisional ejecta losses are enhanced by the stellar collisions in dense environments while simultaneously the supernova energy budget may be decreased by $20$--$40\%$ in the densest systems. Furthermore, in addition to being suppressed, the release of supernova energy is on average delayed in high density clusters. Between $t=3$ Myr and $5$ Myr, the number of stars that reach the end of their lives is only $10\%$--$40\%$ in the dense models HD9Z2 compared to the low density models HD1Z2. Similarly, the early radiation feedback from massive interacting binary stars is prevented in the densest environments as the binaries merge or become unbound. Overall, the enhanced enrichment together with the suppressed and delayed feedback may help to produce enriched stellar populations in dense environments. However, we note that the suppression and delay of the supernova energy release and binary star radiation feedback is accompanied by the extreme radiation field of $>\msol{100}$--$\msol{1000}$ produced in the collisions. The cumulative wind and collisional ejecta budgets of the models are examined in Section \ref{section: 5}.

\subsection{Rates for stellar collisions and TDEs}\label{section: tde}

\begin{figure}
\includegraphics[width=\columnwidth]{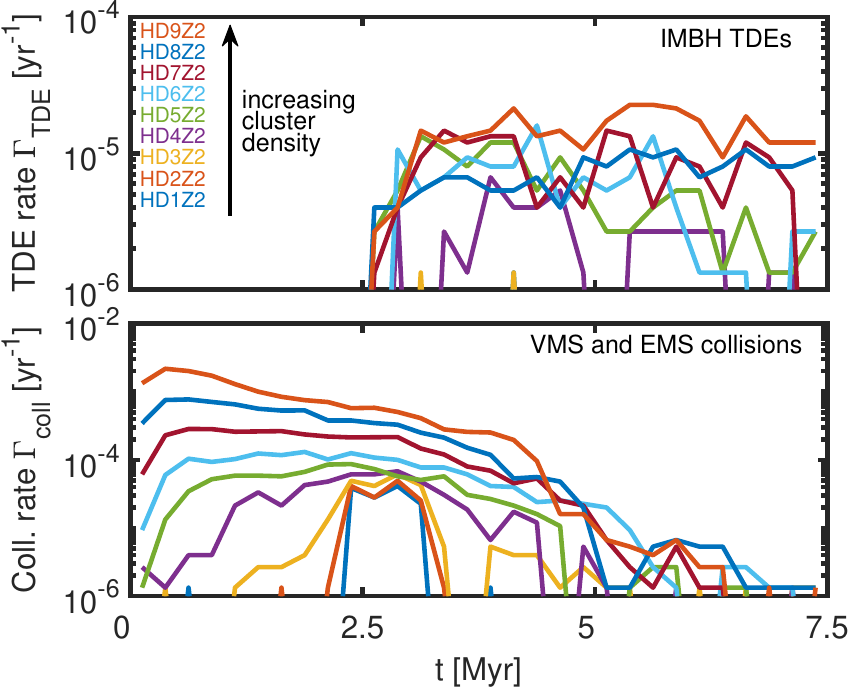}
\caption{Top panel: the IMBH TDE rates $\Gamma_\mathrm{TDE}$ in the hierarchical models HD1Z2--HD9Z2. In the densest models the TDE rates $\Gamma_\mathrm{TDE}>10^{-5}$ yr$^{-1}$ can be sustained from the IMBH formation until the end of the simulations at $t=7.5$ Myr. Bottom panel: the rates of VMS and EMS collisions in the hierarchical models. In the densest models the maximum collision rates reach $\Gamma_\mathrm{coll}\sim2\times10^{-3}$ yr$^{-1}$.}
\label{fig: tde}
\end{figure}

We show the very and extremely massive star ($m_\mathrm{\star,1}+m_\mathrm{\star,2}>\msol{150}$) collision rates $\Gamma_\mathrm{coll}$ and the IMBH ($M_\bullet>\msol{100}$) TDE rates $\Gamma_\mathrm{TDE}$ as a function of time in Fig. \ref{fig: tde}. At comparable cluster densities of $\Sigma_\mathrm{h} \sim 1$--$\Sigmasol{2\times10^5}$, both the maximum very massive star collision rates and the TDE rates are by a factor of few lower in our models HD5Z2 compared to our previous studies (models HB-150 in \citealt{Rantala2025b}). This is mostly due to the somewhat lower IMBH masses in the current runs as we now assume stronger winds and collisional mass loss at $Z=\zsol{0.10}$ compared to our setups in \cite{Rantala2025a} and \cite{Rantala2025b}. 

In the hierarchical low density models (HD1Z2--HD2Z2) the only collisions beyond the mass limit of $\msol{150}$ occur after $t>2.25$ Myr when individual binaries merge late in their evolution. In models with intermediate densities (HD3Z2--HD4Z2) the peak VMS and EMS collision rate occur between $\sim1.5$ Myr $\lesssim t \lesssim 3$ Myr as in our previous studies. However, in our densest models (HD6Z2 and denser) the collision rates are already initially high due to the extremely high stellar densities of the models. The peak VMS collision rate in our densest hierarchical models HD9Z2 reach $ \Gamma_\mathrm{coll}\sim\collratesol{2.2\times10^{-3}}$ per assembling cluster. The VMS collision rates rapidly decline after $t\gtrsim4$--$5$ Myr after the most massive stars end their lives. Similarly, IMBH TDEs commence after $t\gtrsim2.5$ Myr after the IMBH formation commences. The maximum collision and TDE rates by IMBHs in models HD9Z2 is are $\Gamma_\mathrm{coll}\sim\collratesol{2.2\times10^{-3}}$ and $\Gamma_\mathrm{TDE} \sim \collratesol{2.3\times10^{-5}}$ per assembling cluster. While in \cite{Rantala2025b} our peak TDE rates rapidly declined after $1$--$2$ Myr of the IMBH formation, the TDE rates remain relatively constant after the IMBH formation in our densest models until the end of the simulations. The sustained TDE rates compared to our previous studies are enabled by up to an order of magnitude higher maximum central stellar densities in this study. The central ($\leq 0.1$ pc) cluster densities of $\rho_\mathrm{c} \gtrsim \rhosol{10^7}$ and velocity dispersions of $\gtrsim 30$ km s$^{-1}$ may be high enough to sustain long term runaway IMBH growth via TDEs \citep{Rizzuto2023, Stone2017}, enabling TDE driven IMBH growth beyond $M_\bullet > \msol{10^4}$ (see e.g. \citealt{Tal2017b}).

%%%%%%%%%%%%%%%%%%%%%%%%%%%%%%%%%%%%%%%%%%%%%%%%%%%%%%%%%%%%%%%
\section{Wind and collisional ejecta mass loss budgets}\label{section: 5}

\subsection{Cumulative rates of winds and collisional ejecta}

\subsubsection{Winds}

\begin{figure*}
\includegraphics[width=\textwidth]{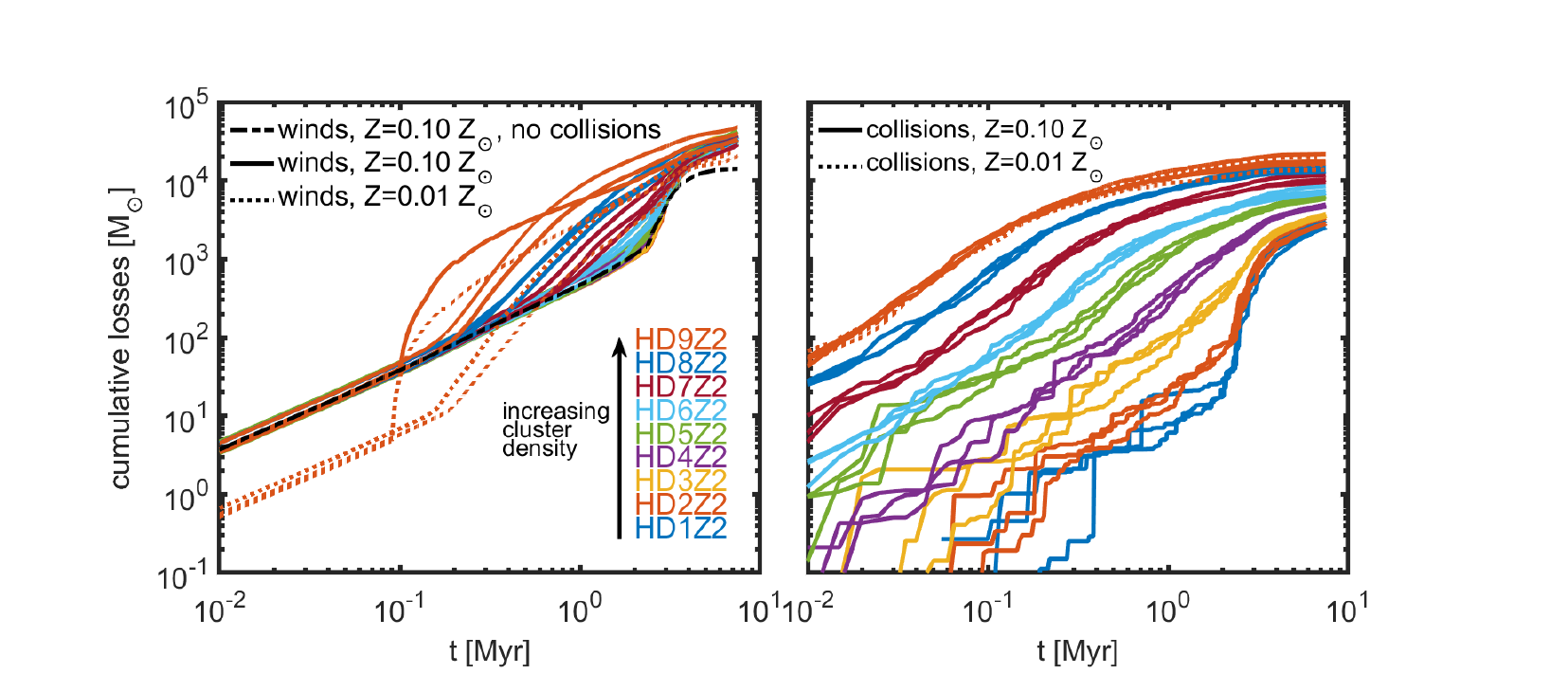}
\caption{The cumulative mass loss rates from the stars through winds (left panel) and by collisional ejecta (right panel) in the hierarchical setups of different initial densities. The nine models with $Z=\zsol{0.10}$ are displayed using the solid line while the densest comparison setup with $Z=\zsol{0.01}$ is shown as the dotted line. At $\zsol{0.10}$ the total wind losses in the models are very similar at different densities until the setup HD6Z2 after which individual extremely massive stars begin to dominate over the mass loss budget of the stellar population. At the low metallicity the wind losses are weaker, as expected. The cumulative amount of collisional ejecta strongly depends on the cluster density, and at early times and in the low metallicity model exceeds the total mass of the wind material. The collisional losses are very similar in the models with $Z=\zsol{0.10}$ and $\zsol{0.01}$.}
\label{fig: enrich-1}
\end{figure*}

\begin{figure}
\includegraphics[width=\columnwidth]{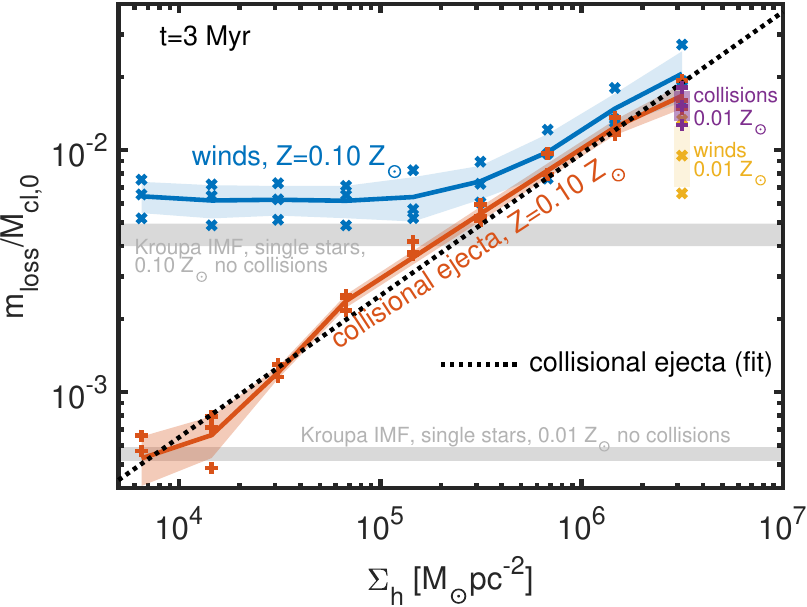}
\caption{The total mass losses in winds and stellar collisions per initial stellar mass $m_\mathrm{loss}/M_\mathrm{cl,0}$ as a function of the central cluster surface density $\Sigma_\mathrm{h}$ in the models with $Z=\zsol{0.10}$. In low density models up to $\Sigma_\mathrm{h}\sim1$--$\Sigmasol{2\times10^5}$ the wind losses are relatively constant $(m_\mathrm{loss}/M_\mathrm{cl,0}\sim7\times10^{-3})$ after which the total losses rise due to extremely massive star formation at high cluster densities. The collisional losses scale with the cluster densities following a power-law relation. In the densest models with $Z=\zsol{0.01}$ (top right corner) the amount of collisional ejecta exceeds the wind losses. We also show the expected amount of mass loss for a single star population without any collisions as the horizontal shaded regions.
}
\label{fig: enrich-2}
\end{figure}

\begin{figure}
\includegraphics[width=\columnwidth]{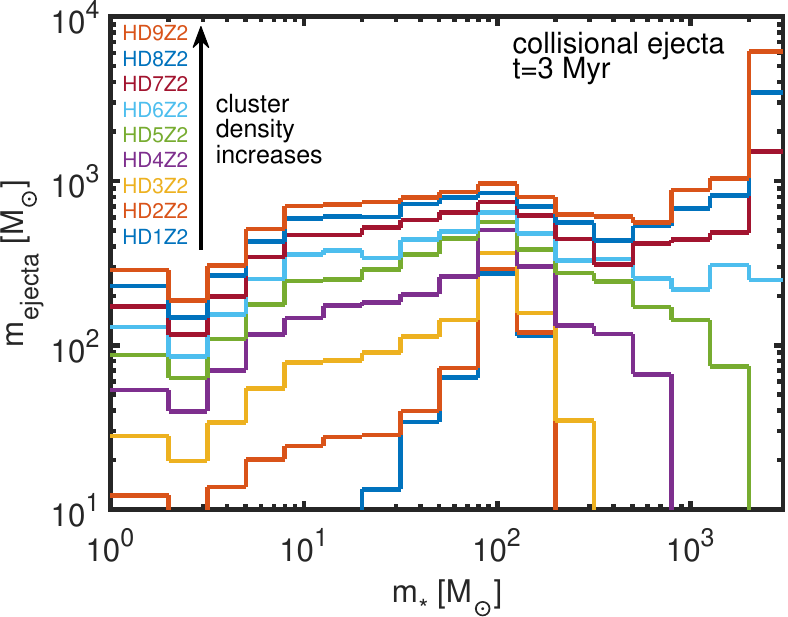}
\caption{The amount of collisional ejecta originating from stars of different (ZAMS) masses. The shown lines are averages of three random realizations of each modelled cluster density. The amount of collisional ejecta steadily increases with increasing density in each stellar mass bin, starting from $\sim \msol{100}$ stars, extending to low mass stars first and after the model HD4Z2 towards the high mass stars. Most notably, in very dense models (HD6Z2 and above), the largest contribution to the collisional ejecta budget comes from extremely massive stars above $\gtrsim \msol{2000}$.}
\label{fig: enrich-3}
\end{figure}

As shown in Section \ref{section: hierarchical-imbh-4.1}, the very massive stars built up by stellar collisions may lose up $\sim \msol{20000}$ via winds and collisional ejecta during their lifetimes, the total lost masses exceeding the final IMBH masses by a factor of few. We now examine the total budget of mass loss from the entire stellar population of the assembling clusters in more detail. 

We present the total cumulative losses in winds and collisions in the hierarchical cluster assembly models in Fig. \ref{fig: enrich-1}. There are in total 27 clustered setups at $Z=\zsol{0.10}$ with nine different initial densities, three random realizations each. The densest model is also simulated using a lower metallicity of $Z=\zsol{0.01}$. In the left panel of Fig. \ref{fig: enrich-1}, at low and intermediate cluster densities (models HD1Z2--HD5Z2 with $Z=\zsol{0.10}$) the total cumulative wind losses are very similar. The cumulative wind losses gradually increase until $t\sim2.4$ Myr after which the wind rates rapidly rise as massive stars reach the final stages of their lives. In a hierarchical system of $M_\star = \msol{10^6}$, the total wind losses at $t=7.5$ Myr are $m_\mathrm{wind}\sim3$--$\msol{4\times10^4}$. Even though stars can grow by collisions and IMBHs form in the models HD3Z2--HD5Z2, the total wind output of these massive stars is still small comparable to the rest of the stellar population. In denser systems the situation changes. Starting from HD6Z2 a small number of individual extremely massive stars comprise a large fraction of the wind mass loss budget, and the cumulative loss histories of the clusters begin to deviate from the low density clusters when extremely massive stars first form and exceed $\gtrsim\msol{600}$. This behaviour is especially visible in the densest initial setups. The maximum cumulative wind mass reached by $t=7.5$ Myr is $m_\mathrm{wind} \sim \msol{4.8\times10^4}$ with a single extremely massive star contributing almost $30\%$ of this. In the dense low metallicity models (HD9Z1 with $Z=\zsol{0.01}$) the cumulative wind masses are initially by a factor of $\sim5$ lower compared to the $Z=\zsol{0.10}$ setups. By the end of the simulations the maximum wind losses can still reach $\sim\msol{3\times10^4}$, a factor of $\sim1.6$ lower compared to the higher metallicity models. The relatively high total wind output in the low metallicity model is due to the fact that at low metallicities the runaway collisions and total losses are balanced at higher stellar masses. The higher $m_\star$ compensates for the lower $Z$ in the wind loss rates.

Young massive star clusters are expected to expulse their remaining gas content around $t\sim3$ Myr due to the first supernovae exploding in the systems. We show the cumulative wind losses at $t=3.0$ Myr in Fig. \ref{fig: enrich-2} as a function of the central sub-cluster half mass surface density $\Sigma_\mathrm{h}$ for $Z=\zsol{0.10}$ and for the densest model also at $Z=\zsol{0.01}$. The results are normalized by the total initial stellar mass of the clusters in the region $M_\mathrm{cl,0} \sim \msol{10^6}$. For comparison, we display the expected total wind losses at $Z=\zsol{0.01}$ and $\zsol{0.10}$ in the case of no collisions, assuming the stellar masses in our initial conditions and single stellar evolution using the \texttt{PARSEC} stellar tracks in \sevn. At $Z=\zsol{0.01}$, $m_\mathrm{wind}/M_\mathrm{cl}$ for the single population without collisions is $\sim 5$--$6\times10^{-4}$ while at $Z=\zsol{0.10}$, $m_\mathrm{wind}/M_\mathrm{cl,0} \sim5\times10^{-3}$. In the actual simulations the total wind losses are somewhat higher ($m_\mathrm{wind}/M_\mathrm{cl,0} \sim7\times10^{-3}$) due to binary star mergers forming more massive stars in the models. Above $\Sigma_\mathrm{h} \gtrsim1$--$\msol{2\times10^5}$ the relative wind losses increase, reaching $m_\mathrm{wind}/M_\mathrm{cl,0} \sim 0.02$ by $t=3.0$ Myr. For models with $Z=\zsol{0.01}$ the relative wind losses at $t=3.0$ Myr are somewhat lower, $m_\mathrm{wind}/M_\mathrm{cl,0} \sim 9\times10^{-3}$.

\subsubsection{Collisional ejecta}

Substantial amounts of stellar material can be lost via ejecta in stellar collisions, decreasing the masses of the formed IMBHs but simultaneously providing enriched material processed by the stars. Whereas the total cumulative wind losses are very similar in low and intermediate density clusters the cumulative collisional ejecta rates show a strong dependence on the initial cluster densities. We illustrate this in the right panel of Fig. \ref{fig: enrich-1} first focusing on the models with $Z=\zsol{0.10}$. In our models with the lowest initial density, stellar collisions occur predominantly in evolved massive binaries late in their evolution with final cumulative collisional ejecta masses reaching $m_\mathrm{coll} \sim \msol{2500}$ at $t=7.5$ Myr. With increasing cluster densities the cumulative ejecta masses steadily increase and shift towards earlier times, reflecting the mass segregation timescales of the clusters. Whereas the wind output of the stellar populations increases after a few Myr as massive stars evolve, the situation is opposite for the collisional ejecta. The highest rates of collisional ejecta losses coincide with the peak collision rates that occur early before $t<2$ Myr (see Fig. \ref{fig: gain-loss-comparison}). By $t=7.5$ Myr, the densest models output in total $m_\mathrm{coll} \gtrsim \msol{21500}$. While the cumulative wind losses showed a substantial dependence on the cluster metallicity, the collisional ejecta losses in the models with $Z=\zsol{0.01}$ and $Z=\zsol{0.10}$ follow almost identical evolution.

We show the cluster surface density dependence of the total collisional ejecta masses at $t=3$ Myr normalized by the zero age stellar mass of the clustered region $M_\mathrm{cl,0}$ in Fig. \ref{fig: enrich-2}. At $Z=\zsol{0.10}$, the $m_\mathrm{coll}/M_\mathrm{cl,0}$ monotonically increases with the cluster surface density and can be well fitted with a power-law of 
\begin{equation}\label{eq: ejecta}
\log_\mathrm{10}\left( \frac{m_\mathrm{coll}}{M_\mathrm{cl,0}} \right) = 0.59 \log_\mathrm{10}\left( \frac{\Sigma_\mathrm{h}}{\Sigmasol{}} \right) -5.53.
\end{equation}
Below surface densities of $\Sigma_\mathrm{h} \lesssim 2$--$\Sigmasol{3\times10^5}$ at $Z=\zsol{0.10}$ the winds are always the most important enrichment channel. In the densest models the final relative ejecta losses are $m_\mathrm{coll}/M_\mathrm{cl,0}\sim0.016$, comparable to the total wind losses in the models. At $Z=\zsol{0.01}$ the results for the ejecta are very similar, however, the wind losses are by a factor of few lower in our low metallicity models. As the collisional ejecta rates show little metallicity dependence, we expect that our models with $Z=\zsol{0.01}$ closely follow Eq. \eqref{eq: ejecta} for the ejecta as well. Thus, at $t=3.0$ Myr at $Z=\zsol{0.01}$ the collisional ejecta is likely the dominant enrichment channel in all but the few lowest density models.

Finally, we note that only $\sim50\%$ of the initial stellar mass of the hierarchical cluster assembly region ends up in the central regions of the final assembled star clusters \citep{Rantala2025b}. As massive stars sink to the centres of their host sub-clusters, the wind losses and collisions occur in environments that most likely end up in the central assembling cluster as opposed to being stripped up to the diffuse outer envelope of the system. Thus, we expect that the total wind and collision losses normalized by the final cluster mass are approximately two times larger than $m_\mathrm{wind}/M_\mathrm{cl,0}$ and $m_\mathrm{coll}/M_\mathrm{cl,0}$, up to $\sim4$--$5\%$ in the densest clusters, together reaching up to $10\%$ at $Z=\zsol{0.10}$.

\subsection{Which stars pollute via collisional ejecta?}

In the previous section we analyzed the total amount of collisional ejecta from stellar collisions regardless from which stars the ejecta originates from. The masses of the stars responsible for the ejecta are important as their mass and age determines the central stellar temperature, the occurring nuclear fusion processes, and ultimately the surface composition of the stars. While a detailed element-by-element enrichment calculation is beyond the scope of this work, we next examine the mass distribution of the stars from which the collisional ejecta originates from. For the stellar mass label we always use the ZAMS mass of the star, fixed by the initial conditions or determined by the stellar collision routines via the \sevn{} track switcher. For the analysis we use $15$ logarithmically spaced stellar mass bins between $\msol{2} \lesssim m_\star \lesssim \msol{2000}$ and one bin below ($m_\mathrm{\star} \lesssim \msol{2}$) and above ($m_\mathrm{\star} \gtrsim \msol{2000}$) the mass range. 

The resulting histogram for the cumulative collisional ejecta losses is shown at $t=3.0$ Myr in Fig. \ref{fig: enrich-3}. At low densities (model HD1Z2) most of the ejecta originates from $m_\star\sim\msol{100}$ stars as the most massive stars. Towards higher densities, the distribution first extends towards lower stellar masses (models HD2Z2 and HD3Z2), and after model HD4Z2 also towards the extremely massive star range as they begin to form in numbers at sufficiently high densities. This picture changes around the model HD6Z2 with an initial threshold surface density of $\Sigma_\mathrm{h}\sim2$--$\Sigmasol{4\times10^5}$. After this threshold surface density the extremely massive stars dominate the ejecta mass budget, and most of the collisional ejecta originates from few stars colliding with a large number of low mass companions.

\subsection{Discussion: peculiar GC enrichment patterns}

The increased mass-loss rates enabled by the collisionally growing stars over just a few Myr may have important implications for chemical enrichment in the cluster-forming regions (e.g. \citealt{WangKroupa2020}). \citet{lahen2024} argued, based on an extrapolation of a hydrodynamical star-formation simulation, that wind enrichment on a level of a few per cent in a $M_\mathrm{cl}=2\times \msol{10^5}$ cluster could produce some notably chemically peculiar stars ($\Delta \mathrm{Na/Fe}\gtrsim0.2$--$0.3$) in the core region of the cluster even with normal VMS winds. For a single EMS or SMS the chemical enrichment would indeed occur very concentrated in the core of the cluster. The chemical output of convective EMSs or SMSs \citep{Denissenkov2014, Charbonnel2023} mixed with the pristine ISM would provide a good match to the range of chemical variations in multiple populations of massive star clusters. In our models such transition from mostly VMS wind pollution to the regime dominated by individual EMSs (and SMSs depending on the wind rate model) occurs at cluster surface densities around $\Sigma_\mathrm{h}\sim2$--$\Sigmasol{4\times10^5}$. Using the star-by-star distribution of wind mass fractions per cluster obtained in \citet{lahen2024}, assuming that 2\% (5\%) of the cluster mass is recycled, and that the most extreme abundance ratios for [Al/Fe] describe the SMS wind composition, we approximate that $4$\% (7\%) of the stars in such a cluster would be considered as second population with $\Delta \mathrm{Na/Fe}\gtrsim0.3$. The number of stars with large offsets from the pristine composition would still be sub-dominant unless the mass fraction of recycled material was even larger or a fraction of the pristine stars were removed e.g. due to dynamical evolution, to change the number ratio. On the other hand, such peculiar stars have been observed to be the minority in GCs of relatively low initial mass \citep{Gratton2019}. A more direct abundance analysis with long-term dynamical evolution model for the cluster would be needed to quantify this population ratio in more detail.

%%%%%%%%%%%%%%%%%%%%%%%%%%%%%%%%%%%%%%%%%%%%%%%%%%%%%%%%%%%%%%%%%%%
%%%%%%%%%%%%%%%%%%%%%%%%%%%%%%%%%%%%%
    
%%%%%%%%%%%%%%%%%%%%%%%%%%%%%%%%%%%%%
\section{Summary and conclusions}\label{section: 7}

We have performed a large sample of direct \nbody{} simulations using the \bifrost{} code including stellar evolution and post-Newtonian BH dynamics to explore the hierarchical assembly and the early evolution of massive star clusters. While the competition of stellar wind mass losses and stellar collisional growth has been previously studied (e.g. \citealt{Mapelli2016}), we systematically examine the effect of initial star cluster masses ($\msolrange{10^4}{M_\mathrm{cl}}{10^6}$), densities ($\Sigmasolrange{4\times10^3}{\Sigma_\mathrm{h}}{4\times10^6}$) and metallicities ($\zsolrange{0.01}{Z}{1.0}$) on the extremely massive star and IMBH formation through runaway stellar collisions in a set of more than $N_\mathrm{sim} \geq 1440$ direct simulations. Our key findings can be summarized as follows.

Our simulation results demonstrate that the formation of IMBHs with masses in the range of $\msolrange{300}{M_\bullet}{6000}$ via runaway stellar collisions is plausible in massive, dense clusters at metallicities up to $Z\sim0.2$--$0.3\zsol{}$. 
Below this critical threshold metallicity, the initial star cluster density, rather than the metallicity, determines the IMBH masses. At higher metallicities especially above $\gtrsim \zsol{0.5}$ the IMBH formation becomes increasingly improbable, and the masses of the IMBHs that form are systematically lower than at low metallicities because of the higher stellar wind mass loss rates. Thus, our simulations favour IMBH formation in environments resembling the $z\sim10$ \textit{JWST} proto globular clusters \citep{Adamo2024} while IMBH formation in the local Universe conditions (higher metallicities, lower cluster densities) is heavily suppressed. We find that for massive star clusters above $M_\mathrm{cl} \gtrsim \msol{10^4}$ it is possible to formulate relatively simple empirical models that describes the maximum IMBH masses $M_\bullet=M_\bullet(\Sigma_\mathrm{h},Z)$ as a function of the metallicity and surface density of their host clusters. We provide two such models that can be both used within the ($\Sigma_\mathrm{h},Z$) parameter space of our simulations, and also be used at higher initial star cluster surface densities beyond $\Sigma_\mathrm{h}\gtrsim \Sigmasol{10^6}$.

We have performed the densest models of hierarchically assembling star clusters to date up to initial sub-cluster half mass surface densities of $\Sigma_\mathrm{h} \sim \Sigmasol{4\times10^6}$ including models at $Z=\zsol{0.01}$ and $Z=\zsol{0.10}$. Increasingly dense cluster models produce increasingly massive collisionally grown stars and IMBHs as well as result in high rates of mass loss from the stars. The substantial amount of wind and collisional mass loss up to $\sim \msol{20000}$ per extremely massive star may also contribute to the extreme abundance variations possible in the future generation stars with total stellar mass loss budgets up to $\sim5$--$10\%$ of the final assembled cluster mass in the densest models. At $Z=\zsol{0.10}$, the collisional ejecta losses become comparable to the total wind losses at $\Sigma_\mathrm{h} \sim \Sigmasol{3\times10^5}$ while at low metallicity ($Z=\zsol{0.01}$) the collisional losses always dominate. We find that the stellar population mass loss budget transforms from being VMS driven into being dominated by individual extremely massive (or even supermassive) stars at a threshold surface density of $\Sigma_\mathrm{h}\sim2$--$\Sigmasol{4\times10^5}$. In the densest models, the central stellar densities up to $\rho_\mathrm{c} \gtrsim \rhosol{10^7}$ can support sustained IMBH TDE rates above $\Gamma_\mathrm{TDE} \gtrsim 10^{-5}$ yr$^{-1}$ per cluster close to the regime in which runaway IMBH growth via TDEs becomes possible \citep{Stone2017,Rizzuto2023}. Our densest hierarchical models also result in the destruction of both supernova progenitors (up to $\sim 26$--$40\%$) and binary stars (up to $\gtrsim 50\%$) during the first $t=7.5$ Myr of cluster evolution, potentially considerably affecting the stellar feedback budget of the stellar populations in extremely dense star clusters.

Despite the high stellar collision rates exceeding $\gtrsim 10^{-3}$ yr$^{-1}$, none of the collisionally grown stars reach a mass of $\msol{10000}$ or above with our standard adopted wind prescription \citep{Vink2018} due to the high wind mass loss rates. The uncertainties in the modelling assumptions for the extremely massive and supermassive stars translate into large uncertainties in the final IMBH masses. We highlight the effect of wind model and rate uncertainties, the recipes for collisional mass loss and the radii of collisionally grown stars. We show that different but widely used models for extremely massive star wind rates (\citealt{Humphreys1994,Vink2001,Belczynski2010} versus \citealt{Vink2018}) result in final IMBH masses differing by a factor of $\sim4$. In our idealised comparison simulation with the wind rates resembling the rates adopted by \cite{Vergara2025}, we reach high stellar and IMBH masses exceeding $\msol{25000}$. We emphasize that EMS stellar wind models are decisive in determining whether the runaway collisional channel can result in SMBHs seeds of $M_\bullet \gtrsim \msol{10^4}$ or above.

For improved runaway collisional IMBH mass predictions accurate information of the massive cluster birth radii and densities would be required. Unfortunately, well resolved birth radii of young ($\lesssim$ a few Myr) still embedded clusters are challenging to obtain even in the local Universe. Interestingly, local young massive star clusters rarely reach as compact sizes and high mean stellar densities as high-redshift proto-GCs (e.g. \citealt{Ryon2017, Levy2024}). Moreover, the current high resolution cosmological GC formation simulations still rely on gravitational softening, suppressing strong small-scale gravitational interactions, which would be essential to self-consistently model the initial formation phase of the clusters. Our results highlight the need for star-by-star resolution star cluster formation simulations at different metallicities including both hydrodynamics an non-softened stellar dynamics, preferentially in their galactic and cosmological environments for informed estimates of the initial cluster birth densities. We have recently taken the first steps towards such models in the galactic context \citep{Lahen2025a,Lahen2025b} by including the accurate small-scale dynamics module \ketju{} \citep{Rantala2017,Rantala2020,Mannerkoski2023} into the star-by-star resolution hydrodynamical galaxy formation framework \texttt{GRIFFIN} \citep{Lahen2020}.
    
\section*{Acknowledgments}
AR thanks Abbas Askar, Rainer Spurzem and Marcelo Vergara for useful discussions on \nbody{} model comparisons, and Angela Adamo, Adélaïde Claeyssens as well as Vasily Belokurov for motivating the plausible star cluster densities at low and high redshifts, as well as Daniel Schaerer and Miroslava Dessauges-Zavadsky for the encouraging discussion on the SMBH seed number densities. The numerical simulations were performed using facilities in Germany hosted by the Max Planck Computing and Data Facility (MPCDF) and the Leibniz Supercomputing Centre (LRZ) in Garching, and the JUWELS Booster of the Jülich supercomputing centre (GCS project 59949 frost-smbh-origins). TN acknowledges support from the Deutsche Forschungsgemeinschaft (DFG, German Research Foundation) under Germany's Excellence Strategy - EXC-2094 - 390783311 from the DFG Cluster of Excellence "ORIGINS". NL was supported by a Gliese Fellowship at the Zentrum f\"ur Astronomie, Universit\"at Heidelberg, Germany. BR acknowledges support by the European Research Council via ERC Consolidator grant KETJU (no. 818930). The development of the SEVN code was enabled by M. Mapelli’s ERC Consolidator grant DEMOBLACK by the European Research Council under contract no. 770017.

\section*{Data availability statement}
The data relevant to this article will be shared on reasonable request to the corresponding author.

%%%%%%%%%%%%%%%%%%%%%%%%%%%%%%%%%%%%%%%%%%%%%%%%%%
%%%%%%%%%%%%%%%%%%%% REFERENCES %%%%%%%%%%%%%%%%%%

\bibliographystyle{mnras}
\interlinepenalty=10000
\bibliography{manuscript}

@ARTICLE{2005ApJ...627..632E,
       author = {{Elmegreen}, Bruce G. and {Elmegreen}, Debra Meloy},
        title = "{Stellar Populations in 10 Clump-Cluster Galaxies of the Hubble Ultra Deep Field}",
      journal = {\apj},
         year = 2005,
        month = jul,
       volume = {627},
       number = {2},
        pages = {632-646},
          doi = {10.1086/430514},
archivePrefix = {arXiv},
       eprint = {astro-ph/0504032},
 primaryClass = {astro-ph},
       adsurl = {https://ui.adsabs.harvard.edu/abs/2005ApJ...627..632E}
}

@ARTICLE{2008ApJ...687...59G,
       author = {{Genzel}, R. and {Burkert}, A. and {Bouch{\'e}}, N. and {Cresci}, G. and {F{\"o}rster Schreiber}, N.~M. and {Shapley}, A. and {Shapiro}, K. and {Tacconi}, L.~J. and {Buschkamp}, P. and {Cimatti}, A. and {Daddi}, E. and {Davies}, R. and {Eisenhauer}, F. and {Erb}, D.~K. and {Genel}, S. and {Gerhard}, O. and {Hicks}, E. and {Lutz}, D. and {Naab}, T. and {Ott}, T. and {Rabien}, S. and {Renzini}, A. and {Steidel}, C.~C. and {Sternberg}, A. and {Lilly}, S.~J.},
        title = "{From Rings to Bulges: Evidence for Rapid Secular Galaxy Evolution at z \raisebox{-0.5ex}\textasciitilde 2 from Integral Field Spectroscopy in the SINS Survey}",
      journal = {\apj},
         year = 2008,
        month = nov,
       volume = {687},
       number = {1},
        pages = {59-77},
          doi = {10.1086/591840},
archivePrefix = {arXiv},
       eprint = {0807.1184},
 primaryClass = {astro-ph},
       adsurl = {https://ui.adsabs.harvard.edu/abs/2008ApJ...687...59G}
}

@ARTICLE{2009ApJ...692...12E,
       author = {{Elmegreen}, Bruce G. and {Elmegreen}, Debra Meloy and {Fernandez}, Maria Ximena and {Lemonias}, Jenna Jo},
        title = "{Bulge and Clump Evolution in Hubble Ultra Deep Field Clump Clusters, Chains and Spiral Galaxies}",
      journal = {\apj},
         year = 2009,
        month = feb,
       volume = {692},
       number = {1},
        pages = {12-31},
          doi = {10.1088/0004-637X/692/1/12},
archivePrefix = {arXiv},
       eprint = {0810.5404},
 primaryClass = {astro-ph},
       adsurl = {https://ui.adsabs.harvard.edu/abs/2009ApJ...692...12E}
}

@BOOK{Aarseth2003,
       author = {{Aarseth}, Sverre J.},
        title = "{Gravitational N-Body Simulations}",
         year = "2003",
        publisher = "Cambridge University Press",
       adsurl = {https://ui.adsabs.harvard.edu/abs/2003gnbs.book.....A}
}

@ARTICLE{Abac2025,
       author = {{Abac}, Adrian and {Abramo}, Raul and {Albanesi}, Simone and {Albertini}, Angelica and {Agapito}, Alessandro and {Agathos}, Michalis and {Albertus}, Conrado and {Andersson}, Nils and {Andrade}, Tomas and {Andreoni}, Igor and {Angeloni}, Federico and {Antonelli}, Marco and {Antoniadis}, John and {Antonini}, Fabio and {Arca Sedda}, Manuel and {Artale}, M. Celeste and {Ascenzi}, Stefano and {Auclair}, Pierre and {Bachetti}, Matteo and {Badger}, Charles and {Banerjee}, Biswajit and {Barba-Gonzalez}, David and {Barta}, Daniel and {Bartolo}, Nicola and {Bauswein}, Andreas and {Begnoni}, Andrea and {Beirnaert}, Freija and {Bejger}, Michal and {Belgacem}, Enis and {Bellomo}, Nicola and {Bernard}, Laura and {Grazia Bernardini}, Maria and {Bernuzzi}, Sebastiano and {Berry}, Christopher P.~L. and {Berti}, Emanuele and {Bertone}, Gianfranco and {Bettoni}, Dario and {Bezares}, Miguel and {Bhagwat}, Swetha and {Bisero}, Sofia and {Bizouard}, Marie Anne and {Blanco-Pillado}, Jose J. and {Blasi}, Simone and {Bonino}, Alice and {Borghese}, Alice and {Borhanian}, Ssohrab and {Bortolas}, Elisa and {Botticella}, Maria Teresa and {Branchesi}, Marica and {Breschi}, Matteo and {Brito}, Richard and {Brocato}, Enzo and {Broekgaarden}, Floor S. and {Bulik}, Tomasz and {Buonanno}, Alessandra and {Burgio}, Fiorella and {Burrows}, Adam and {Calcagni}, Gianluca and {Canevarolo}, Sofia and {Cappellaro}, Enrico and {Capurri}, Giulia and {Carbone}, Carmelita and {Casadio}, Roberto and {Cayuso}, Ramiro and {Cerda-Duran}, Pablo and {Char}, Prasanta and {Chaty}, Sylvain and {Chiarusi}, Tommaso and {Chruslinska}, Martyna and {Cireddu}, Francesco and {Cole}, Philippa and {Colombo}, Alberto and {Colpi}, Monica and {Compere}, Geoffrey and {Contaldi}, Carlo and {Corman}, Maxence and {Crescimbeni}, Francesco and {Cristallo}, Sergio and {Cuoco}, Elena and {Cusin}, Giulia and {Dal Canton}, Tito and {Dalya}, Gergely and {D'Avanzo}, Paolo and {Davari}, Nazanin and {De Luca}, Valerio and {De Renzis}, Viola and {Della Valle}, Massimo and {Del Pozzo}, Walter and {De Santi}, Federico and {Ludovico De Santis}, Alessio and {Dietrich}, Tim and {Dimastrogiovanni}, Ema and {Domenech}, Guillem and {Doneva}, Daniela and {Drago}, Marco and {Dupletsa}, Ulyana and {Duval}, Hannah and {Dvorkin}, Irina and {Elias-Rosa}, Nancy and {Fairhurst}, Stephen and {Fantina}, Anthea F. and {Fasiello}, Matteo and {Fays}, Maxime and {Fender}, Rob and {Fischer}, Tobias and {Foucart}, Francois and {Fragos}, Tassos and {Foffa}, Stefano and {Franciolini}, Gabriele and {Fumagalli}, Jacopo and {Gair}, Jonathan and {Gamba}, Rossella and {Garcia-Bellido}, Juan and {Garcia-Quiros}, Cecilio and {Arpad Gergely}, Laszlo and {Ghirlanda}, Giancarlo and {Ghosh}, Archisman and {Giacomazzo}, Bruno and {Gittins}, Fabian and {Giudice}, Ines Francesca and {Goncharov}, Boris and {Gonzalez}, Alejandra and {Goriely}, Stephane and {Graziani}, Luca and {Greco}, Giuseppe and {Gualtieri}, Leonardo and {Guidi}, Gianluca Maria and {Gupta}, Ish and {Haney}, Maria and {Hannam}, Mark and {Harms}, Jan and {Harutyunyan}, Arus and {Haskell}, Brynmor and {Haungs}, Andreas and {Hazra}, Nandini and {Hemming}, Gary and {Heng}, Ik Siong and {Hinderer}, Tanja and {van der Horst}, Alexander and {Hu}, Qian and {Husa}, Sascha and {Iacovelli}, Francesco and {Illuminati}, Giulia and {Inguglia}, Gianluca and {Izquierdo Villalba}, David and {Janquart}, Justin and {Janssens}, Kamiel and {Jenkins}, Alexander C. and {Jones}, Ian and {Kacskovics}, Balazs and {Klessen}, Ralf S. and {Kokkotas}, Kostas and {Kuan}, Hao-Jui and {Kumar}, Sumit and {Kuroyanagi}, Sachiko and {Laghi}, Danny and {Lamberts}, Astrid and {Lambiase}, Gaetano and {Larrouturou}, Francois and {Leaci}, Paola and {Lenzi}, Michele and {Levan}, Andrew and {Li}, T.~G.~F. and {Li}, Yufeng and {Liang}, Dicong and {Limongi}, Marco and {Liu}, Boyuan and {Llanes-Estrada}, Felipe J. and {Loffredo}, Eleonora and {Long}, Oliver and {Lope-Oter}, Eva and {Lukes-Gerakopoulos}, Georgios and {Maggio}, Elisa and {Maggiore}, Michele and {Mancarella}, Michele and {Mapelli}, Michela and {Marchant}, Pablo and {Margiotta}, Annarita and {Mariotti}, Alberto and {Marriott-Best}, Alisha and {Marsat}, Sylvain and {Martinez-Pinedo}, Gabriel and {Maselli}, Andrea and {Mastrogiovanni}, Simone and {Matos}, Isabela and {Melandri}, Andrea and {Mendes}, Raissa F.~P. and {Mendonca Soares de Souza}, Josiel and {Mentasti}, Giorgio and {Mezcua}, Mar and {Mosta}, Philipp and {Mondal}, Chiranjib and {Moresco}, Michele and {Mukherjee}, Tista and {Muttoni}, Niccolo and {Nagar}, Alessandro and {Narola}, Harsh and {Nava}, Lara and {Navarro Moreno}, Pablo and {Nelemans}, Gijs},
        title = "{The Science of the Einstein Telescope}",
      journal = {arXiv e-prints},
         year = 2025,
        month = mar,
          eid = {arXiv:2503.12263},
        pages = {arXiv:2503.12263},
          doi = {10.48550/arXiv.2503.12263},
archivePrefix = {arXiv},
       eprint = {2503.12263},
 primaryClass = {gr-qc},
       adsurl = {https://ui.adsabs.harvard.edu/abs/2025arXiv250312263A}
}

@ARTICLE{Abac2025b,
       author = {{Abac}, A.~G. and {Abouelfettouh}, I. and {Acernese}, F. and {Ackley}, K. and {Adamcewicz}, C. and {Adhicary}, S. and {Adhikari}, D. and {Adhikari}, N. and {Adhikari}, R.~X. and {Adkins}, V.~K. and {Afroz}, S. and {Agapito}, A. and {Agarwal}, D. and {Agathos}, M. and {Aggarwal}, N. and {Aggarwal}, S. and {Aguiar}, O.~D. and {Ahrend}, I.-L. and {Aiello}, L. and {Ain}, A. and {Ajith}, P. and {Akutsu}, T. and {Albanesi}, S. and {Ali}, W. and {Al-Kershi}, S. and {All{\'e}n{\'e}}, C. and {Allocca}, A. and {Al-Shammari}, S. and {Altin}, P.~A. and {Alvarez-Lopez}, S. and {Amar}, W. and {Amarasinghe}, O. and {Amato}, A. and {Amicucci}, F. and {Amra}, C. and {Ananyeva}, A. and {Anderson}, S.~B. and {Anderson}, W.~G. and {Andia}, M. and {Ando}, M. and {Andr{\'e}s-Carcasona}, M. and {Andri{\'c}}, T. and {Anglin}, J. and {Ansoldi}, S. and {Antelis}, J.~M. and {Antier}, S. and {Aoumi}, M. and {Appavuravther}, E.~Z. and {Appert}, S. and {Apple}, S.~K. and {Arai}, K. and {Alvarez}, C. Araujo and {Araya}, A. and {Araya}, M.~C. and {Sedda}, M. Arca and {Areeda}, J.~S. and {Aritomi}, N. and {Armato}, F. and {Armstrong}, S. and {Arnaud}, N. and {Arogeti}, M. and {Aronson}, S.~M. and {Arun}, K.~G. and {Ashton}, G. and {Aso}, Y. and {Asprea}, L. and {Assiduo}, M. and {Assis de Souza Melo}, S. and {Aston}, S.~M. and {Astone}, P. and {Attadio}, F. and {Aubin}, F. and {AultONeal}, K. and {Avallone}, G. and {Avila}, E.~A. and {Babak}, S. and {Badger}, C. and {Bae}, S. and {Bagnasco}, S. and {Baiotti}, L. and {Bajpai}, R. and {Baka}, T. and {Baker}, A.~M. and {Baker}, K.~A. and {Baker}, T. and {Baldi}, G. and {Baldicchi}, N. and {Ball}, M. and {Ballardin}, G. and {Ballmer}, S.~W. and {Banagiri}, S. and {Banerjee}, B. and {Bankar}, D. and {Baptiste}, T.~M. and {Baral}, P. and {Baratti}, M. and {Barayoga}, J.~C. and {Barish}, B.~C. and {Barker}, D. and {Barman}, N. and {Barneo}, P. and {Barone}, F. and {Barr}, B. and {Barsotti}, L. and {Barsuglia}, M. and {Barta}, D. and {Bartoletti}, A.~M. and {Barton}, M.~A. and {Bartos}, I. and {Basalaev}, A. and {Bassiri}, R. and {Basti}, A. and {Bawaj}, M. and {Baxi}, P. and {Bayley}, J.~C. and {Baylor}, A.~C. and {Baynard}, II, P.~A. and {Bazzan}, M. and {Bedakihale}, V.~M. and {Beirnaert}, F. and {Bejger}, M. and {Belardinelli}, D. and {Bell}, A.~S. and {Bellie}, D.~S. and {Bellizzi}, L. and {Benoit}, W. and {Bentara}, I. and {Bentley}, J.~D. and {Ben Yaala}, M. and {Bera}, S. and {Bergamin}, F. and {Berger}, B.~K. and {Bernuzzi}, S. and {Beroiz}, M. and {Berry}, C.~P.~L. and {Bersanetti}, D. and {Bertheas}, T. and {Bertolini}, A. and {Betzwieser}, J. and {Beveridge}, D. and {Bevilacqua}, G. and {Bevins}, N. and {Bhandare}, R. and {Bhatt}, R. and {Bhattacharjee}, D. and {Bhattacharyya}, S. and {Bhaumik}, S. and {Bhagwat}, S. and {Biancalana}, V. and {Bianchi}, A. and {Bilenko}, I.~A. and {Billingsley}, G. and {Binetti}, A. and {Bini}, S. and {Binu}, C. and {Biot}, S. and {Birnholtz}, O. and {Biscoveanu}, S. and {Bisht}, A. and {Bitossi}, M. and {Bizouard}, M.-A. and {Blaber}, S. and {Blackburn}, J.~K. and {Blagg}, L.~A. and {Blair}, C.~D. and {Blair}, D.~G. and {Bode}, N. and {Boettner}, N. and {Boileau}, G. and {Boldrini}, M. and {Bolingbroke}, G.~N. and {Bolliand}, A. and {Bonavena}, L.~D. and {Bondarescu}, R. and {Bondu}, F. and {Bonilla}, E. and {Bonilla}, M.~S. and {Bonino}, A. and {Bonnand}, R. and {Borchers}, A. and {Borhanian}, S. and {Boschi}, V. and {Bose}, S. and {Bossilkov}, V. and {Bothra}, Y. and {Boudon}, A. and {Bourg}, L. and {Boyle}, M. and {Bozzi}, A. and {Bradaschia}, C. and {Brady}, P.~R. and {Branch}, A. and {Branchesi}, M. and {Braun}, I. and {Briant}, T. and {Brillet}, A. and {Brinkmann}, M. and {Brockill}, P. and {Brockmueller}, E. and {Brooks}, A.~F.},
        title = "{GW231123: A Binary Black Hole Merger with Total Mass 190─265 M$_{{\ensuremath{\odot}}}$}",
      journal = {\apjl},
         year = 2025,
        month = nov,
       volume = {993},
       number = {1},
          eid = {L25},
        pages = {L25},
          doi = {10.3847/2041-8213/ae0c9c},
archivePrefix = {arXiv},
       eprint = {2507.08219},
 primaryClass = {astro-ph.HE},
       adsurl = {https://ui.adsabs.harvard.edu/abs/2025ApJ...993L..25A}
}

@ARTICLE{Abbott1982,
       author = {{Abbott}, D.~C.},
        title = "{The theory of radiatively driven stellar winds. II. The line acceleration.}",
      journal = {\apj},
         year = 1982,
        month = aug,
       volume = {259},
        pages = {282-301},
          doi = {10.1086/160166},
       adsurl = {https://ui.adsabs.harvard.edu/abs/1982ApJ...259..282A}
}

@ARTICLE{Abbott2017b,
       author = {{Abbott}, B.~P. and {Abbott}, R. and {Abbott}, T.~D. and {Acernese}, F. and {Ackley}, K. and {Adams}, C. and {Adams}, T. and {Addesso}, P. and {Adhikari}, R.~X. and {Adya}, V.~B. and {Affeldt}, C. and {Afrough}, M. and {Agarwal}, B. and {Agathos}, M. and {Agatsuma}, K. and {Aggarwal}, N. and {Aguiar}, O.~D. and {Aiello}, L. and {Ain}, A. and {Ajith}, P. and {Allen}, B. and {Allen}, G. and {Allocca}, A. and {Altin}, P.~A. and {Amato}, A. and {Ananyeva}, A. and {Anderson}, S.~B. and {Anderson}, W.~G. and {Antier}, S. and {Appert}, S. and {Arai}, K. and {Araya}, M.~C. and {Areeda}, J.~S. and {Arnaud}, N. and {Arun}, K.~G. and {Ascenzi}, S. and {Ashton}, G. and {Ast}, M. and {Aston}, S.~M. and {Astone}, P. and {Aufmuth}, P. and {Aulbert}, C. and {AultONeal}, K. and {Avila-Alvarez}, A. and {Babak}, S. and {Bacon}, P. and {Bader}, M.~K.~M. and {Bae}, S. and {Baker}, P.~T. and {Baldaccini}, F. and {Ballardin}, G. and {Ballmer}, S.~W. and {Banagiri}, S. and {Barayoga}, J.~C. and {Barclay}, S.~E. and {Barish}, B.~C. and {Barker}, D. and {Barone}, F. and {Barr}, B. and {Barsotti}, L. and {Barsuglia}, M. and {Barta}, D. and {Bartlett}, J. and {Bartos}, I. and {Bassiri}, R. and {Basti}, A. and {Batch}, J.~C. and {Baune}, C. and {Bawaj}, M. and {Bazzan}, M. and {B{\'e}csy}, B. and {Beer}, C. and {Bejger}, M. and {Belahcene}, I. and {Bell}, A.~S. and {Berger}, B.~K. and {Bergmann}, G. and {Berry}, C.~P.~L. and {Bersanetti}, D. and {Bertolini}, A. and {Betzwieser}, J. and {Bhagwat}, S. and {Bhandare}, R. and {Bilenko}, I.~A. and {Billingsley}, G. and {Billman}, C.~R. and {Birch}, J. and {Birney}, I.~A. and {Birnholtz}, O. and {Biscans}, S. and {Bisht}, A. and {Bitossi}, M. and {Biwer}, C. and {Bizouard}, M.~A. and {Blackburn}, J.~K. and {Blackman}, J. and {Blair}, C.~D. and {Blair}, D.~G. and {Blair}, R.~M. and {Bloemen}, S. and {Bock}, O. and {Bode}, N. and {Boer}, M. and {Bogaert}, G. and {Bohe}, A. and {Bondu}, F. and {Bonnand}, R. and {Boom}, B.~A. and {Bork}, R. and {Boschi}, V. and {Bose}, S. and {Bouffanais}, Y. and {Bozzi}, A. and {Bradaschia}, C. and {Brady}, P.~R. and {Braginsky}, V.~B. and {Branchesi}, M. and {Brau}, J.~E. and {Briant}, T. and {Brillet}, A. and {Brinkmann}, M. and {Brisson}, V. and {Brockill}, P. and {Broida}, J.~E. and {Brooks}, A.~F. and {Brown}, D.~A. and {Brown}, D.~D. and {Brown}, N.~M. and {Brunett}, S. and {Buchanan}, C.~C. and {Buikema}, A. and {Bulik}, T. and {Bulten}, H.~J. and {Buonanno}, A. and {Buskulic}, D. and {Buy}, C. and {Byer}, R.~L. and {Cabero}, M. and {Cadonati}, L. and {Cagnoli}, G. and {Cahillane}, C. and {Calder{\'o}n Bustillo}, J. and {Callister}, T.~A. and {Calloni}, E. and {Camp}, J.~B. and {Canepa}, M. and {Canizares}, P. and {Cannon}, K.~C. and {Cao}, H. and {Cao}, J. and {Capano}, C.~D. and {Capocasa}, E. and {Carbognani}, F. and {Caride}, S. and {Carney}, M.~F. and {Casanueva Diaz}, J. and {Casentini}, C. and {Caudill}, S. and {Cavagli{\`a}}, M. and {Cavalier}, F. and {Cavalieri}, R. and {Cella}, G. and {Cepeda}, C.~B. and {Cerboni Baiardi}, L. and {Cerretani}, G. and {Cesarini}, E. and {Chamberlin}, S.~J. and {Chan}, M. and {Chao}, S. and {Charlton}, P. and {Chassande-Mottin}, E. and {Chatterjee}, D. and {Chatziioannou}, K. and {Cheeseboro}, B.~D. and {Chen}, H.~Y. and {Chen}, Y. and {Cheng}, H.-P. and {Chincarini}, A. and {Chiummo}, A. and {Chmiel}, T. and {Cho}, H.~S. and {Cho}, M. and {Chow}, J.~H. and {Christensen}, N. and {Chu}, Q. and {Chua}, A.~J.~K. and {Chua}, S. and {Chung}, A.~K.~W. and {Chung}, S. and {Ciani}, G. and {Ciolfi}, R. and {Cirelli}, C.~E. and {Cirone}, A. and {Clara}, F. and {Clark}, J.~A. and {Cleva}, F. and {Cocchieri}, C. and {Coccia}, E. and {Cohadon}, P.-F. and {Colla}, A.},
        title = "{GW170104: Observation of a 50-Solar-Mass Binary Black Hole Coalescence at Redshift 0.2}",
      journal = {\prl},
         year = 2017,
        month = jun,
       volume = {118},
       number = {22},
          eid = {221101},
        pages = {221101},
          doi = {10.1103/PhysRevLett.118.221101},
archivePrefix = {arXiv},
       eprint = {1706.01812},
 primaryClass = {gr-qc},
       adsurl = {https://ui.adsabs.harvard.edu/abs/2017PhRvL.118v1101A}
}

@ARTICLE{Abdurrouf2025,
       author = {{Abdurro'uf} and {Coe}, Dan and {Resseguier}, Tom and {Murphy}, Calla and {Xu}, Xinfeng and {Adamo}, Angela and {Roy}, Namrata and {Henry}, Alaina and {Kokorev}, Vasily and {Brammer}, Gabriel and {Fujimoto}, Seiji and {Ferguson}, Henry C. and {Pagul}, Amanda and {Windhorst}, Rogier A. and {Heckman}, Timothy and {Diego}, Jose M. and {Akins}, Hollis B. and {Allingham}, Joseph and {Amor{\'\i}n}, Ricardo O. and {Berg}, Danielle A. and {Brada{\v{c}}}, Maru{\v{s}}a and {Bradley}, Larry D. and {Chen}, Wenlei and {Chisholm}, John and {Conselice}, Christopher J. and {Dayal}, Pratika and {Dessauges-Zavadsky}, Miroslava and {Faisst}, Andreas L. and {Finkelstein}, Steven L. and {Fudamoto}, Yoshinobu and {Furtak}, Lukas J. and {Harikane}, Yuichi and {Hsiao}, Tiger Yu-Yang and {Jimenez-Teja}, Yolanda and {Koekemoer}, Anton M. and {Larson}, Rebecca L. and {Lucas}, Ray A. and {Messa}, Matteo and {Mowla}, Lamiya and {Nakane}, Minami and {Noirot}, Ga{\"e}l and {Pan}, Richard and {Pascale}, Massimo and {Richard}, Johan and {Ricotti}, Massimo and {Robbins}, Luke and {Schaerer}, Daniel and {Sun}, Fengwu and {Vanzella}, Eros and {Welch}, Brian and {Willott}, Chris and {Zitrin}, Adi},
        title = "{Spatially Resolved Physical Properties of Young Star Clusters and Star-forming Clumps in the Brightest z>6 Galaxy, the Strongly Lensed Cosmic Spear at z=6.2}",
      journal = {arXiv e-prints},
         year = 2025,
        month = dec,
          eid = {arXiv:2512.08054},
        pages = {arXiv:2512.08054},
          doi = {10.48550/arXiv.2512.08054},
archivePrefix = {arXiv},
       eprint = {2512.08054},
 primaryClass = {astro-ph.GA},
       adsurl = {https://ui.adsabs.harvard.edu/abs/2025arXiv251208054A}
}

@ARTICLE{Adamo2020,
       author = {{Adamo}, Angela and {Zeidler}, Peter and {Kruijssen}, J.~M. Diederik and {Chevance}, M{\'e}lanie and {Gieles}, Mark and {Calzetti}, Daniela and {Charbonnel}, Corinne and {Zinnecker}, Hans and {Krause}, Martin G.~H.},
        title = "{Star Clusters Near and Far; Tracing Star Formation Across Cosmic Time}",
      journal = {\ssr},
         year = 2020,
        month = jun,
       volume = {216},
       number = {4},
          eid = {69},
        pages = {69},
          doi = {10.1007/s11214-020-00690-x},
archivePrefix = {arXiv},
       eprint = {2005.06188},
 primaryClass = {astro-ph.GA},
       adsurl = {https://ui.adsabs.harvard.edu/abs/2020SSRv..216...69A}
}

@ARTICLE{Adamo2024,
       author = {{Adamo}, Angela and {Bradley}, Larry D. and {Vanzella}, Eros and {Claeyssens}, Ad{\'e}la{\"\i}de and {Welch}, Brian and {Diego}, Jose M. and {Mahler}, Guillaume and {Oguri}, Masamune and {Sharon}, Keren and {Abdurro'uf} and {Hsiao}, Tiger Yu-Yang and {Xu}, Xinfeng and {Messa}, Matteo and {Lassen}, Augusto E. and {Zackrisson}, Erik and {Brammer}, Gabriel and {Coe}, Dan and {Kokorev}, Vasily and {Ricotti}, Massimo and {Zitrin}, Adi and {Fujimoto}, Seiji and {Inoue}, Akio K. and {Resseguier}, Tom and {Rigby}, Jane R. and {Jim{\'e}nez-Teja}, Yolanda and {Windhorst}, Rogier A. and {Hashimoto}, Takuya and {Tamura}, Yoichi},
        title = "{Bound star clusters observed in a lensed galaxy 460 Myr after the Big Bang}",
      journal = {\nat},
         year = 2024,
        month = aug,
       volume = {632},
       number = {8025},
        pages = {513-516},
          doi = {10.1038/s41586-024-07703-7},
archivePrefix = {arXiv},
       eprint = {2401.03224},
 primaryClass = {astro-ph.GA},
       adsurl = {https://ui.adsabs.harvard.edu/abs/2024Natur.632..513A}
}

@ARTICLE{Alexander2017,
       author = {{Alexander}, Tal and {Bar-Or}, Ben},
        title = "{A universal minimal mass scale for present-day central black holes}",
      journal = {Nature Astronomy},
         year = 2017,
        month = aug,
       volume = {1},
          eid = {0147},
        pages = {0147},
          doi = {10.1038/s41550-017-0147},
archivePrefix = {arXiv},
       eprint = {1701.00415},
 primaryClass = {astro-ph.GA},
       adsurl = {https://ui.adsabs.harvard.edu/abs/2017NatAs...1E.147A}
}

@ARTICLE{Amaro-Seoane2012,
       author = {{Amaro-Seoane}, Pau and {Aoudia}, Sofiane and {Babak}, Stanislav and {Bin{\'e}truy}, Pierre and {Berti}, Emanuele and {Boh{\'e}}, Alejandro and {Caprini}, Chiara and {Colpi}, Monica and {Cornish}, Neil J. and {Danzmann}, Karsten and {Dufaux}, Jean-Fran{\c{c}}ois and {Gair}, Jonathan and {Jennrich}, Oliver and {Jetzer}, Philippe and {Klein}, Antoine and {Lang}, Ryan N. and {Lobo}, Alberto and {Littenberg}, Tyson and {McWilliams}, Sean T. and {Nelemans}, Gijs and {Petiteau}, Antoine and {Porter}, Edward K. and {Schutz}, Bernard F. and {Sesana}, Alberto and {Stebbins}, Robin and {Sumner}, Tim and {Vallisneri}, Michele and {Vitale}, Stefano and {Volonteri}, Marta and {Ward}, Henry},
        title = "{Low-frequency gravitational-wave science with eLISA/NGO}",
      journal = {Classical and Quantum Gravity},
         year = 2012,
        month = jun,
       volume = {29},
       number = {12},
          eid = {124016},
        pages = {124016},
          doi = {10.1088/0264-9381/29/12/124016},
archivePrefix = {arXiv},
       eprint = {1202.0839},
 primaryClass = {gr-qc},
       adsurl = {https://ui.adsabs.harvard.edu/abs/2012CQGra..29l4016A}
}

@ARTICLE{Angus2024,
       author = {{Angus}, C.~R. and {Woosley}, S.~E. and {Foley}, R.~J. and {Nicholl}, M. and {Villar}, V.~A. and {Taggart}, K. and {Pursiainen}, M. and {Ramsden}, P. and {Srivastav}, S. and {Stevance}, H.~F. and {Moore}, T. and {Auchettl}, K. and {Hoogendam}, W.~B. and {Khetan}, N. and {Yadavalli}, S.~K. and {Dimitriadis}, G. and {Gagliano}, A. and {Siebert}, M.~R. and {Aamer}, A. and {de Boer}, T. and {Chambers}, K.~C. and {Clocchiatti}, A. and {Coulter}, D.~A. and {Drout}, M.~R. and {Farias}, D. and {Fulton}, M.~D. and {Gall}, C. and {Gao}, H. and {Izzo}, L. and {Jones}, D.~O. and {Lin}, C.-C. and {Magnier}, E.~A. and {Narayan}, G. and {Ramirez-Ruiz}, E. and {Ransome}, C.~L. and {Rest}, A. and {Smartt}, S.~J. and {Smith}, K.~W.},
        title = "{Double ``acct'': A Distinct Double-peaked Supernova Matching Pulsational Pair Instability Models}",
      journal = {\apjl},
         year = 2024,
        month = dec,
       volume = {977},
       number = {2},
          eid = {L41},
        pages = {L41},
          doi = {10.3847/2041-8213/ad9264},
archivePrefix = {arXiv},
       eprint = {2409.02174},
 primaryClass = {astro-ph.HE},
       adsurl = {https://ui.adsabs.harvard.edu/abs/2024ApJ...977L..41A}
}

@ARTICLE{Antonini2019,
       author = {{Antonini}, Fabio and {Gieles}, Mark and {Gualandris}, Alessia},
        title = "{Black hole growth through hierarchical black hole mergers in dense star clusters: implications for gravitational wave detections}",
      journal = {\mnras},
         year = 2019,
        month = jul,
       volume = {486},
       number = {4},
        pages = {5008-5021},
          doi = {10.1093/mnras/stz1149},
archivePrefix = {arXiv},
       eprint = {1811.03640},
 primaryClass = {astro-ph.HE},
       adsurl = {https://ui.adsabs.harvard.edu/abs/2019MNRAS.486.5008A}
}

@ARTICLE{ArcaSedda2023,
       author = {{Arca Sedda}, Manuel and {Mapelli}, Michela and {Benacquista}, Matthew and {Spera}, Mario},
        title = "{Isolated and dynamical black hole mergers with B-POP: the role of star formation and dynamics, star cluster evolution, natal kicks, mass and spins, and hierarchical mergers}",
      journal = {\mnras},
         year = 2023,
        month = apr,
       volume = {520},
       number = {4},
        pages = {5259-5282},
          doi = {10.1093/mnras/stad331},
archivePrefix = {arXiv},
       eprint = {2109.12119},
 primaryClass = {astro-ph.GA},
       adsurl = {https://ui.adsabs.harvard.edu/abs/2023MNRAS.520.5259A}
}

@ARTICLE{ArcaSedda-DRAGON2a,
       author = {{Arca Sedda}, Manuel and {Kamlah}, Albrecht W.~H. and {Spurzem}, Rainer and {Giersz}, Mirek and {Berczik}, Peter and {Rastello}, Sara and {Iorio}, Giuliano and {Mapelli}, Michela and {Gatto}, Massimiliano and {Grebel}, Eva K.},
        title = "{The DRAGON-II simulations - I. Evolution of single and binary compact objects in star clusters with up to 1 million stars}",
      journal = {\mnras},
         year = 2024,
        month = mar,
       volume = {528},
       number = {3},
        pages = {5119-5139},
          doi = {10.1093/mnras/stad3952},
archivePrefix = {arXiv},
       eprint = {2307.04805},
 primaryClass = {astro-ph.GA},
       adsurl = {https://ui.adsabs.harvard.edu/abs/2024MNRAS.528.5119A}
}

@ARTICLE{ArcaSedda-DRAGON2b,
       author = {{Arca Sedda}, Manuel and {Kamlah}, Albrecht W.~H. and {Spurzem}, Rainer and {Rizzuto}, Francesco Paolo and {Naab}, Thorsten and {Giersz}, Mirek and {Berczik}, Peter},
        title = "{The DRAGON-II simulations - II. Formation mechanisms, mass, and spin of intermediate-mass black holes in star clusters with up to 1 million stars}",
      journal = {\mnras},
         year = 2023,
        month = nov,
       volume = {526},
       number = {1},
        pages = {429-442},
          doi = {10.1093/mnras/stad2292},
archivePrefix = {arXiv},
       eprint = {2307.04806},
 primaryClass = {astro-ph.GA},
       adsurl = {https://ui.adsabs.harvard.edu/abs/2023MNRAS.526..429A}
}

@ARTICLE{Banerjee2020,
       author = {{Banerjee}, S. and {Belczynski}, K. and {Fryer}, C.~L. and {Berczik}, P. and {Hurley}, J.~R. and {Spurzem}, R. and {Wang}, L.},
        title = "{BSE versus StarTrack: Implementations of new wind, remnant-formation, and natal-kick schemes in NBODY7 and their astrophysical consequences}",
      journal = {\aap},
         year = 2020,
        month = jul,
       volume = {639},
          eid = {A41},
        pages = {A41},
          doi = {10.1051/0004-6361/201935332},
archivePrefix = {arXiv},
       eprint = {1902.07718},
 primaryClass = {astro-ph.SR},
       adsurl = {https://ui.adsabs.harvard.edu/abs/2020A&A...639A..41B}
}

@ARTICLE{Barber2025,
       author = {{Barber}, Jordan and {Antonini}, Fabio},
        title = "{Formation and evolution of binary black holes in N-body simulations of star clusters with up to two million stars}",
      journal = {\mnras},
         year = 2025,
        month = apr,
       volume = {538},
       number = {2},
        pages = {639-658},
          doi = {10.1093/mnras/staf279},
archivePrefix = {arXiv},
       eprint = {2410.03832},
 primaryClass = {astro-ph.GA},
       adsurl = {https://ui.adsabs.harvard.edu/abs/2025MNRAS.538..639B}
}

@ARTICLE{Baumgardt2011,
       author = {{Baumgardt}, H. and {Klessen}, R.~S.},
        title = "{The role of stellar collisions for the formation of massive stars}",
      journal = {\mnras},
         year = 2011,
        month = may,
       volume = {413},
       number = {3},
        pages = {1810-1818},
          doi = {10.1111/j.1365-2966.2011.18258.x},
archivePrefix = {arXiv},
       eprint = {1009.1189},
 primaryClass = {astro-ph.GA},
       adsurl = {https://ui.adsabs.harvard.edu/abs/2011MNRAS.413.1810B}
}

@ARTICLE{Baumgardt2017,
       author = {{Baumgardt}, H.},
        title = "{N -body modelling of globular clusters: masses, mass-to-light ratios and intermediate-mass black holes}",
      journal = {\mnras},
         year = 2017,
        month = jan,
       volume = {464},
       number = {2},
        pages = {2174-2202},
          doi = {10.1093/mnras/stw2488},
archivePrefix = {arXiv},
       eprint = {1609.08794},
 primaryClass = {astro-ph.GA},
       adsurl = {https://ui.adsabs.harvard.edu/abs/2017MNRAS.464.2174B}
}

@ARTICLE{Bekki2003,
       author = {{Bekki}, K. and {Freeman}, K.~C.},
        title = "{Formation of {\ensuremath{\omega}} Centauri from an ancient nucleated dwarf galaxy in the young Galactic disc}",
      journal = {\mnras},
         year = 2003,
        month = dec,
       volume = {346},
       number = {2},
        pages = {L11-L15},
          doi = {10.1046/j.1365-2966.2003.07275.x},
archivePrefix = {arXiv},
       eprint = {astro-ph/0310348},
 primaryClass = {astro-ph},
       adsurl = {https://ui.adsabs.harvard.edu/abs/2003MNRAS.346L..11B}
}

@ARTICLE{Belczynski2010,
       author = {{Belczynski}, Krzysztof and {Bulik}, Tomasz and {Fryer}, Chris L. and {Ruiter}, Ashley and {Valsecchi}, Francesca and {Vink}, Jorick S. and {Hurley}, Jarrod R.},
        title = "{On the Maximum Mass of Stellar Black Holes}",
      journal = {\apj},
         year = 2010,
        month = may,
       volume = {714},
       number = {2},
        pages = {1217-1226},
          doi = {10.1088/0004-637X/714/2/1217},
archivePrefix = {arXiv},
       eprint = {0904.2784},
 primaryClass = {astro-ph.SR},
       adsurl = {https://ui.adsabs.harvard.edu/abs/2010ApJ...714.1217B}
}

@ARTICLE{Belczynski2020b,
       author = {{Belczynski}, K. and {Banerjee}, S.},
        title = "{Formation of low-spinning 100 M$_{{\ensuremath{\odot}}}$ black holes}",
      journal = {\aap},
         year = 2020,
        month = aug,
       volume = {640},
          eid = {L20},
        pages = {L20},
          doi = {10.1051/0004-6361/202038427},
archivePrefix = {arXiv},
       eprint = {2002.08050},
 primaryClass = {astro-ph.HE},
       adsurl = {https://ui.adsabs.harvard.edu/abs/2020A&A...640L..20B}
}

@ARTICLE{Belkus2007,
       author = {{Belkus}, H. and {Van Bever}, J. and {Vanbeveren}, D.},
        title = "{The Evolution of Very Massive Stars}",
      journal = {\apj},
         year = 2007,
        month = apr,
       volume = {659},
       number = {2},
        pages = {1576-1581},
          doi = {10.1086/512181},
archivePrefix = {arXiv},
       eprint = {astro-ph/0701334},
 primaryClass = {astro-ph},
       adsurl = {https://ui.adsabs.harvard.edu/abs/2007ApJ...659.1576B}
}

@ARTICLE{Belokurov2022,
       author = {{Belokurov}, Vasily and {Kravtsov}, Andrey},
        title = "{From dawn till disc: Milky Way's turbulent youth revealed by the APOGEE+Gaia data}",
      journal = {\mnras},
         year = 2022,
        month = jul,
       volume = {514},
       number = {1},
        pages = {689-714},
          doi = {10.1093/mnras/stac1267},
archivePrefix = {arXiv},
       eprint = {2203.04980},
 primaryClass = {astro-ph.GA},
       adsurl = {https://ui.adsabs.harvard.edu/abs/2022MNRAS.514..689B}
}

@ARTICLE{Bernard2025,
       author = {{Bernard}, Yann and {Moraux}, Estelle and {Price}, Daniel J. and {Motte}, Fr{\'e}d{\'e}rique and {Louvet}, Fabien and {Joncour}, Isabelle},
        title = "{DAWN. I. Simulating the formation and early evolution of stellar clusters with Phantom N-Body}",
      journal = {arXiv e-prints},
         year = 2025,
        month = aug,
          eid = {arXiv:2508.05296},
        pages = {arXiv:2508.05296},
          doi = {10.48550/arXiv.2508.05296},
archivePrefix = {arXiv},
       eprint = {2508.05296},
 primaryClass = {astro-ph.GA},
       adsurl = {https://ui.adsabs.harvard.edu/abs/2025arXiv250805296B}
}

@BOOK{Binney2008,
       author = {{Binney}, James and {Tremaine}, Scott},
        title = "{Galactic Dynamics: Second Edition}",
         year = "2008",
         publisher = "Princeton University Press, Princeton, NJ USA",
       adsurl = {https://ui.adsabs.harvard.edu/abs/2008gady.book.....B}
}

@ARTICLE{Bradac2025,
       author = {{Brada{\v{c}}}, Maru{\v{s}}a and {Jude{\v{z}}}, Jon and {Willott}, Chris and {Rihtar{\v{s}}i{\v{c}}}, Gregor and {Martis}, Nicholas S. and {Harshan}, Anishya and {Felicioni}, Giordano and {Asada}, Yoshihisa and {Desprez}, Guillaume and {Clowe}, Douglas and {Gonzalez}, Anthony H. and {Jones}, Christine and {Lemaux}, Brian C. and {Markov}, Vladan and {Mowla}, Lamiya and {Noirot}, Ga{\"e}l and {Peter}, Annika H.~G. and {Robertson}, Andrew and {Sarrouh}, Ghassan T.~E. and {Sawicki}, Marcin and {Schrabback}, Tim and {Tripodi}, Roberta},
        title = "{Star Formation under a Cosmic Microscope: Highly magnified z = 11 galaxy behind the Bullet Cluster}",
      journal = {arXiv e-prints},
         year = 2025,
        month = sep,
          eid = {arXiv:2509.20446},
        pages = {arXiv:2509.20446},
          doi = {10.48550/arXiv.2509.20446},
archivePrefix = {arXiv},
       eprint = {2509.20446},
 primaryClass = {astro-ph.GA},
       adsurl = {https://ui.adsabs.harvard.edu/abs/2025arXiv250920446B}
}

@ARTICLE{Bressan2012,
       author = {{Bressan}, Alessandro and {Marigo}, Paola and {Girardi}, L{\'e}o. and {Salasnich}, Bernardo and {Dal Cero}, Claudia and {Rubele}, Stefano and {Nanni}, Ambra},
        title = "{PARSEC: stellar tracks and isochrones with the PAdova and TRieste Stellar Evolution Code}",
      journal = {\mnras},
         year = 2012,
        month = nov,
       volume = {427},
       number = {1},
        pages = {127-145},
          doi = {10.1111/j.1365-2966.2012.21948.x},
archivePrefix = {arXiv},
       eprint = {1208.4498},
 primaryClass = {astro-ph.SR},
       adsurl = {https://ui.adsabs.harvard.edu/abs/2012MNRAS.427..127B}
}

@ARTICLE{Bromm2001,
       author = {{Bromm}, V. and {Ferrara}, A. and {Coppi}, P.~S. and {Larson}, R.~B.},
        title = "{The fragmentation of pre-enriched primordial objects}",
      journal = {\mnras},
         year = 2001,
        month = dec,
       volume = {328},
       number = {3},
        pages = {969-976},
          doi = {10.1046/j.1365-8711.2001.04915.x},
archivePrefix = {arXiv},
       eprint = {astro-ph/0104271},
 primaryClass = {astro-ph},
       adsurl = {https://ui.adsabs.harvard.edu/abs/2001MNRAS.328..969B}
}

@ARTICLE{Brown2021,
       author = {{Brown}, Gillen and {Gnedin}, Oleg Y.},
        title = "{Radii of young star clusters in nearby galaxies}",
      journal = {\mnras},
         year = 2021,
        month = dec,
       volume = {508},
       number = {4},
        pages = {5935-5953},
          doi = {10.1093/mnras/stab2907},
archivePrefix = {arXiv},
       eprint = {2106.12420},
 primaryClass = {astro-ph.GA},
       adsurl = {https://ui.adsabs.harvard.edu/abs/2021MNRAS.508.5935B}
}

@ARTICLE{Bunker2023,
       author = {{Bunker}, Andrew J. and {Saxena}, Aayush and {Cameron}, Alex J. and {Willott}, Chris J. and {Curtis-Lake}, Emma and {Jakobsen}, Peter and {Carniani}, Stefano and {Smit}, Renske and {Maiolino}, Roberto and {Witstok}, Joris and {Curti}, Mirko and {D'Eugenio}, Francesco and {Jones}, Gareth C. and {Ferruit}, Pierre and {Arribas}, Santiago and {Charlot}, Stephane and {Chevallard}, Jacopo and {Giardino}, Giovanna and {de Graaff}, Anna and {Looser}, Tobias J. and {L{\"u}tzgendorf}, Nora and {Maseda}, Michael V. and {Rawle}, Tim and {Rix}, Hans-Walter and {Del Pino}, Bruno Rodr{\'\i}guez and {Alberts}, Stacey and {Egami}, Eiichi and {Eisenstein}, Daniel J. and {Endsley}, Ryan and {Hainline}, Kevin and {Hausen}, Ryan and {Johnson}, Benjamin D. and {Rieke}, George and {Rieke}, Marcia and {Robertson}, Brant E. and {Shivaei}, Irene and {Stark}, Daniel P. and {Sun}, Fengwu and {Tacchella}, Sandro and {Tang}, Mengtao and {Williams}, Christina C. and {Willmer}, Christopher N.~A. and {Baker}, William M. and {Baum}, Stefi and {Bhatawdekar}, Rachana and {Bowler}, Rebecca and {Boyett}, Kristan and {Chen}, Zuyi and {Circosta}, Chiara and {Helton}, Jakob M. and {Ji}, Zhiyuan and {Kumari}, Nimisha and {Lyu}, Jianwei and {Nelson}, Erica and {Parlanti}, Eleonora and {Perna}, Michele and {Sandles}, Lester and {Scholtz}, Jan and {Suess}, Katherine A. and {Topping}, Michael W. and {{\"U}bler}, Hannah and {Wallace}, Imaan E.~B. and {Whitler}, Lily},
        title = "{JADES NIRSpec Spectroscopy of GN-z11: Lyman-{\ensuremath{\alpha}} emission and possible enhanced nitrogen abundance in a z = 10.60 luminous galaxy}",
      journal = {\aap},
         year = 2023,
        month = sep,
       volume = {677},
          eid = {A88},
        pages = {A88},
          doi = {10.1051/0004-6361/202346159},
archivePrefix = {arXiv},
       eprint = {2302.07256},
 primaryClass = {astro-ph.GA},
       adsurl = {https://ui.adsabs.harvard.edu/abs/2023A&A...677A..88B}
}

@ARTICLE{Cameron2023,
       author = {{Cameron}, Alex J. and {Katz}, Harley and {Rey}, Martin P. and {Saxena}, Aayush},
        title = "{Nitrogen enhancements 440 Myr after the big bang: supersolar N/O, a tidal disruption event, or a dense stellar cluster in GN-z11?}",
      journal = {\mnras},
         year = 2023,
        month = aug,
       volume = {523},
       number = {3},
        pages = {3516-3525},
          doi = {10.1093/mnras/stad1579},
archivePrefix = {arXiv},
       eprint = {2302.10142},
 primaryClass = {astro-ph.GA},
       adsurl = {https://ui.adsabs.harvard.edu/abs/2023MNRAS.523.3516C}
}

@ARTICLE{Cao2018,
       author = {{Cao}, Liang and {Lu}, Youjun and {Zhao}, Yuetong},
        title = "{Host galaxy properties of mergers of stellar binary black holes and their implications for advanced LIGO gravitational wave sources}",
      journal = {\mnras},
         year = 2018,
        month = mar,
       volume = {474},
       number = {4},
        pages = {4997-5007},
          doi = {10.1093/mnras/stx3087},
archivePrefix = {arXiv},
       eprint = {1711.09190},
 primaryClass = {astro-ph.GA},
       adsurl = {https://ui.adsabs.harvard.edu/abs/2018MNRAS.474.4997C}
}

@ARTICLE{Cappellari2013,
       author = {{Cappellari}, Michele and {McDermid}, Richard M. and {Alatalo}, Katherine and {Blitz}, Leo and {Bois}, Maxime and {Bournaud}, Fr{\'e}d{\'e}ric and {Bureau}, M. and {Crocker}, Alison F. and {Davies}, Roger L. and {Davis}, Timothy A. and {de Zeeuw}, P.~T. and {Duc}, Pierre-Alain and {Emsellem}, Eric and {Khochfar}, Sadegh and {Krajnovi{\'c}}, Davor and {Kuntschner}, Harald and {Morganti}, Raffaella and {Naab}, Thorsten and {Oosterloo}, Tom and {Sarzi}, Marc and {Scott}, Nicholas and {Serra}, Paolo and {Weijmans}, Anne-Marie and {Young}, Lisa M.},
        title = "{The ATLAS$^{3D}$ project - XX. Mass-size and mass-{\ensuremath{\sigma}} distributions of early-type galaxies: bulge fraction drives kinematics, mass-to-light ratio, molecular gas fraction and stellar initial mass function}",
      journal = {\mnras},
         year = 2013,
        month = jul,
       volume = {432},
       number = {3},
        pages = {1862-1893},
          doi = {10.1093/mnras/stt644},
archivePrefix = {arXiv},
       eprint = {1208.3523},
 primaryClass = {astro-ph.CO},
       adsurl = {https://ui.adsabs.harvard.edu/abs/2013MNRAS.432.1862C}
}

@ARTICLE{DiCarlo2020,
       author = {{Di Carlo}, Ugo N. and {Mapelli}, Michela and {Giacobbo}, Nicola and {Spera}, Mario and {Bouffanais}, Yann and {Rastello}, Sara and {Santoliquido}, Filippo and {Pasquato}, Mario and {Ballone}, Alessandro and {Trani}, Alessandro A. and {Torniamenti}, Stefano and {Haardt}, Francesco},
        title = "{Binary black holes in young star clusters: the impact of metallicity}",
      journal = {\mnras},
         year = 2020,
        month = oct,
       volume = {498},
       number = {1},
        pages = {495-506},
          doi = {10.1093/mnras/staa2286},
archivePrefix = {arXiv},
       eprint = {2004.09525},
 primaryClass = {astro-ph.HE},
       adsurl = {https://ui.adsabs.harvard.edu/abs/2020MNRAS.498..495D}
}

@ARTICLE{DiCarlo2020b,
       author = {{Di Carlo}, Ugo N. and {Mapelli}, Michela and {Bouffanais}, Yann and {Giacobbo}, Nicola and {Santoliquido}, Filippo and {Bressan}, Alessandro and {Spera}, Mario and {Haardt}, Francesco},
        title = "{Binary black holes in the pair instability mass gap}",
      journal = {\mnras},
         year = 2020,
        month = sep,
       volume = {497},
       number = {1},
        pages = {1043-1049},
          doi = {10.1093/mnras/staa1997},
archivePrefix = {arXiv},
       eprint = {1911.01434},
 primaryClass = {astro-ph.HE},
       adsurl = {https://ui.adsabs.harvard.edu/abs/2020MNRAS.497.1043D}
}

@ARTICLE{Castor1975,
       author = {{Castor}, J.~I. and {Abbott}, D.~C. and {Klein}, R.~I.},
        title = "{Radiation-driven winds in Of stars.}",
      journal = {\apj},
         year = 1975,
        month = jan,
       volume = {195},
        pages = {157-174},
          doi = {10.1086/153315},
       adsurl = {https://ui.adsabs.harvard.edu/abs/1975ApJ...195..157C}
}

@ARTICLE{Charbonnel2023,
       author = {{Charbonnel}, C. and {Schaerer}, D. and {Prantzos}, N. and {Ram{\'\i}rez-Galeano}, L. and {Fragos}, T. and {Kuruvanthodi}, A. and {Marques-Chaves}, R. and {Gieles}, M.},
        title = "{N-enhancement in GN-z11: First evidence for supermassive stars nucleosynthesis in proto-globular clusters-like conditions at high redshift?}",
      journal = {\aap},
         year = 2023,
        month = may,
       volume = {673},
          eid = {L7},
        pages = {L7},
          doi = {10.1051/0004-6361/202346410},
archivePrefix = {arXiv},
       eprint = {2303.07955},
 primaryClass = {astro-ph.GA},
       adsurl = {https://ui.adsabs.harvard.edu/abs/2023A&A...673L...7C}
}

@ARTICLE{Chatterjee2017,
       author = {{Chatterjee}, Sourav and {Rodriguez}, Carl L. and {Rasio}, Frederic A.},
        title = "{Binary Black Holes in Dense Star Clusters: Exploring the Theoretical Uncertainties}",
      journal = {\apj},
         year = 2017,
        month = jan,
       volume = {834},
       number = {1},
          eid = {68},
        pages = {68},
          doi = {10.3847/1538-4357/834/1/68},
archivePrefix = {arXiv},
       eprint = {1603.00884},
 primaryClass = {astro-ph.GA},
       adsurl = {https://ui.adsabs.harvard.edu/abs/2017ApJ...834...68C}
}

@ARTICLE{Chattopadhyay2022,
       author = {{Chattopadhyay}, Debatri and {Hurley}, Jarrod and {Stevenson}, Simon and {Raidani}, Arihant},
        title = "{Dynamical double black holes and their host cluster properties}",
      journal = {\mnras},
         year = 2022,
        month = jul,
       volume = {513},
       number = {3},
        pages = {4527-4555},
          doi = {10.1093/mnras/stac1163},
archivePrefix = {arXiv},
       eprint = {2202.08924},
 primaryClass = {astro-ph.GA},
       adsurl = {https://ui.adsabs.harvard.edu/abs/2022MNRAS.513.4527C}
}

@ARTICLE{Chen2015,
       author = {{Chen}, Yang and {Bressan}, Alessandro and {Girardi}, L{\'e}o and {Marigo}, Paola and {Kong}, Xu and {Lanza}, Antonio},
        title = "{PARSEC evolutionary tracks of massive stars up to 350 M$_{{\ensuremath{\odot}}}$ at metallicities 0.0001 {\ensuremath{\leq}} Z {\ensuremath{\leq}} 0.04}",
      journal = {\mnras},
         year = 2015,
        month = sep,
       volume = {452},
       number = {1},
        pages = {1068-1080},
          doi = {10.1093/mnras/stv1281},
archivePrefix = {arXiv},
       eprint = {1506.01681},
 primaryClass = {astro-ph.SR},
       adsurl = {https://ui.adsabs.harvard.edu/abs/2015MNRAS.452.1068C}
}

@ARTICLE{Chilingarian2018,
       author = {{Chilingarian}, Igor V. and {Katkov}, Ivan Yu. and {Zolotukhin}, Ivan Yu. and {Grishin}, Kirill A. and {Beletsky}, Yuri and {Boutsia}, Konstantina and {Osip}, David J.},
        title = "{A Population of Bona Fide Intermediate-mass Black Holes Identified as Low-luminosity Active Galactic Nuclei}",
      journal = {\apj},
         year = 2018,
        month = aug,
       volume = {863},
       number = {1},
          eid = {1},
        pages = {1},
          doi = {10.3847/1538-4357/aad184},
archivePrefix = {arXiv},
       eprint = {1805.01467},
 primaryClass = {astro-ph.GA},
       adsurl = {https://ui.adsabs.harvard.edu/abs/2018ApJ...863....1C}
}

@article{Chin1997,
title = "Symplectic integrators from composite operator factorizations",
journal = "Physics Letters A",
volume = "226",
number = "6",
pages = "344 - 348",
year = "1997",
issn = "0375-9601",
doi = "https://doi.org/10.1016/S0375-9601(97)00003-0",
author = {{Chin}, Siu A.},
}

@ARTICLE{Chin2005,
       author = {{Chin}, Siu A. and {Chen}, C.~R.},
        title = "{Forward Symplectic Integrators for Solving Gravitational Few-Body Problems}",
      journal = {Celestial Mechanics and Dynamical Astronomy},
         year = "2005",
        month = "Mar",
       volume = {91},
       number = {3-4},
        pages = {301-322},
          doi = {10.1007/s10569-004-4622-z}
}

@ARTICLE{Chin2007,
       author = {{Chin}, Siu A.},
        title = "{Forward and non-forward symplectic integrators in solving classical dynamics problems}",
      journal = {arXiv e-prints},
         year = "2007",
        month = "Apr",
          eid = {arXiv:0704.3273},
        pages = {arXiv:0704.3273},
archivePrefix = {arXiv},
       eprint = {0704.3273},
 primaryClass = {physics.comp-ph},
}

@ARTICLE{Chon2022,
       author = {{Chon}, Sunmyon and {Ono}, Haruka and {Omukai}, Kazuyuki and {Schneider}, Raffaella},
        title = "{Impact of the cosmic background radiation on the initial mass function of metal-poor stars}",
      journal = {\mnras},
         year = 2022,
        month = aug,
       volume = {514},
       number = {3},
        pages = {4639-4654},
          doi = {10.1093/mnras/stac1549},
archivePrefix = {arXiv},
       eprint = {2205.15328},
 primaryClass = {astro-ph.GA},
       adsurl = {https://ui.adsabs.harvard.edu/abs/2022MNRAS.514.4639C}
}

@ARTICLE{Chruslinska2019,
       author = {{Chru{\'s}li{\'n}ska}, Martyna and {Nelemans}, Gijs},
        title = "{Metallicity of stars formed throughout the cosmic history based on the observational properties of star-forming galaxies}",
      journal = {\mnras},
         year = 2019,
        month = oct,
       volume = {488},
       number = {4},
        pages = {5300-5326},
          doi = {10.1093/mnras/stz2057},
archivePrefix = {arXiv},
       eprint = {1907.11243},
 primaryClass = {astro-ph.GA},
       adsurl = {https://ui.adsabs.harvard.edu/abs/2019MNRAS.488.5300C}
}

@ARTICLE{Chruslinska2024b,
       author = {{Chru{\'s}li{\'n}ska}, M. and {Pakmor}, R. and {Matthee}, J. and {Matsuno}, T.},
        title = "{Trading oxygen for iron. I. The [O/Fe]-specific star formation rate relation of galaxies}",
      journal = {\aap},
         year = 2024,
        month = jun,
       volume = {686},
          eid = {A186},
        pages = {A186},
          doi = {10.1051/0004-6361/202347602},
archivePrefix = {arXiv},
       eprint = {2308.00023},
 primaryClass = {astro-ph.GA},
       adsurl = {https://ui.adsabs.harvard.edu/abs/2024A&A...686A.186C}
}

@ARTICLE{Chruslinska2025,
       author = {{Chru{\'s}li{\'n}ska}, Martyna and {Curti}, Mirko and {Pakmor}, Ruediger and {De Cia}, Annalisa and {Matthee}, Jorryt and {Bhagwat}, Aniket and {Monty}, Stephanie},
        title = "{Trading oxygen for iron II. Oxygen- versus iron-dependent cosmic star formation history}",
      journal = {arXiv e-prints},
         year = 2025,
        month = nov,
          eid = {arXiv:2511.15782},
        pages = {arXiv:2511.15782},
          doi = {10.48550/arXiv.2511.15782},
archivePrefix = {arXiv},
       eprint = {2511.15782},
 primaryClass = {astro-ph.GA},
       adsurl = {https://ui.adsabs.harvard.edu/abs/2025arXiv251115782C}
}

@ARTICLE{Giacobbo2018,
       author = {{Giacobbo}, Nicola and {Mapelli}, Michela},
        title = "{The progenitors of compact-object binaries: impact of metallicity, common envelope and natal kicks}",
      journal = {\mnras},
         year = 2018,
        month = oct,
       volume = {480},
       number = {2},
        pages = {2011-2030},
          doi = {10.1093/mnras/sty1999},
archivePrefix = {arXiv},
       eprint = {1806.00001},
 primaryClass = {astro-ph.HE},
       adsurl = {https://ui.adsabs.harvard.edu/abs/2018MNRAS.480.2011G}
}

@ARTICLE{Giacobbo2018b,
       author = {{Giacobbo}, Nicola and {Mapelli}, Michela and {Spera}, Mario},
        title = "{Merging black hole binaries: the effects of progenitor's metallicity, mass-loss rate and Eddington factor}",
      journal = {\mnras},
         year = 2018,
        month = mar,
       volume = {474},
       number = {3},
        pages = {2959-2974},
          doi = {10.1093/mnras/stx2933},
archivePrefix = {arXiv},
       eprint = {1711.03556},
 primaryClass = {astro-ph.SR},
       adsurl = {https://ui.adsabs.harvard.edu/abs/2018MNRAS.474.2959G}
}

@ARTICLE{Claeyssens2025,
       author = {{Claeyssens}, Ad{\'e}la{\"\i}de and {Adamo}, Angela and {Messa}, Matteo and {Dessauges-Zavadsky}, Miroslava and {Richard}, Johan and {Kramarenko}, Ivan and {Matthee}, Jorryt and {Naidu}, Rohan P.},
        title = "{Tracing star formation across cosmic time at tens of parsec-scales in the lensing cluster field Abell 2744}",
      journal = {\mnras},
         year = 2025,
        month = mar,
       volume = {537},
       number = {3},
        pages = {2535-2558},
          doi = {10.1093/mnras/staf058},
archivePrefix = {arXiv},
       eprint = {2410.10974},
 primaryClass = {astro-ph.GA},
       adsurl = {https://ui.adsabs.harvard.edu/abs/2025MNRAS.537.2535C}
}

@ARTICLE{Claeyssens2026,
author = {{Claeyssens}, Ad{\'e}la{\"\i}de and {Adamo}, Angela and {Kokorev}, Vasily and {Furtak}, Lukas and {Richard}, Johan and {Beauchesne}, Benjamin and {Dessauges-Zavadsky}, Miroslava and {Atek}, Hakim and {Chisholm}, John and {Endsley}, Ryan and {Fujimoto}, Seiji and {Korber}, Damien and {Pan}, Richard and {Saldana-Lopez}, Alberto and {Schaerer}, Daniel},
        title = "{A first GLIMPSE into star clusters populations across cosmic time}",
      journal = {arXiv e-prints},
         year = 2026,
        month = jan,
          eid = {arXiv:2601.16281},
        pages = {arXiv:2601.16281},
archivePrefix = {arXiv},
       eprint = {2601.16281},
 primaryClass = {astro-ph.GA},
       adsurl = {https://ui.adsabs.harvard.edu/abs/2026arXiv260116281A}
}

@article{Cleveland1988,
author = {Cleveland, W.S. and Devlin, S.J.},
title= {Locally weighted regression: An approach to regression analysis by local fitting},
year = {1988},
doi = {10.2307/2289282},
journal = {Journal of the American Statistical Association},
volume = {83},
number = {403},
pages = {596--610},
}

@ARTICLE{Colbert1999,
       author = {{Colbert}, Edward J.~M. and {Mushotzky}, Richard F.},
        title = "{The Nature of Accreting Black Holes in Nearby Galaxy Nuclei}",
      journal = {\apj},
         year = 1999,
        month = jul,
       volume = {519},
       number = {1},
        pages = {89-107},
          doi = {10.1086/307356},
archivePrefix = {arXiv},
       eprint = {astro-ph/9901023},
 primaryClass = {astro-ph},
       adsurl = {https://ui.adsabs.harvard.edu/abs/1999ApJ...519...89C}
}

@ARTICLE{Costa2021,
       author = {{Costa}, Guglielmo and {Bressan}, Alessandro and {Mapelli}, Michela and {Marigo}, Paola and {Iorio}, Giuliano and {Spera}, Mario},
        title = "{Formation of GW190521 from stellar evolution: the impact of the hydrogen-rich envelope, dredge-up, and $^{12}$C({\ensuremath{\alpha}}, {\ensuremath{\gamma}})$^{16}$O rate on the pair-instability black hole mass gap}",
      journal = {\mnras},
         year = 2021,
        month = mar,
       volume = {501},
       number = {3},
        pages = {4514-4533},
          doi = {10.1093/mnras/staa3916},
archivePrefix = {arXiv},
       eprint = {2010.02242},
 primaryClass = {astro-ph.SR},
       adsurl = {https://ui.adsabs.harvard.edu/abs/2021MNRAS.501.4514C}
}

@ARTICLE{Costa2022,
       author = {{Costa}, Guglielmo and {Ballone}, Alessandro and {Mapelli}, Michela and {Bressan}, Alessandro},
        title = "{Formation of black holes in the pair-instability mass gap: Evolution of a post-collision star}",
      journal = {\mnras},
         year = 2022,
        month = oct,
       volume = {516},
       number = {1},
        pages = {1072-1080},
          doi = {10.1093/mnras/stac2222},
archivePrefix = {arXiv},
       eprint = {2204.03492},
 primaryClass = {astro-ph.SR},
       adsurl = {https://ui.adsabs.harvard.edu/abs/2022MNRAS.516.1072C}
}

@ARTICLE{Costa2025,
       author = {{Costa}, G. and {Shepherd}, K.~G. and {Bressan}, A. and {Addari}, F. and {Chen}, Y. and {Fu}, X. and {Volpato}, G. and {Nguyen}, C.~T. and {Girardi}, L. and {Marigo}, P. and {Mazzi}, A. and {Pastorelli}, G. and {Trabucchi}, M. and {Bossini}, D. and {Zaggia}, S.},
        title = "{Evolutionary tracks, ejecta, and ionizing photons from intermediate-mass to very massive stars with PARSEC}",
      journal = {\aap},
         year = 2025,
        month = feb,
       volume = {694},
          eid = {A193},
        pages = {A193},
          doi = {10.1051/0004-6361/202452573},
archivePrefix = {arXiv},
       eprint = {2501.12917},
 primaryClass = {astro-ph.SR},
       adsurl = {https://ui.adsabs.harvard.edu/abs/2025A&A...694A.193C}
}

@ARTICLE{Crowther2010,
       author = {{Crowther}, Paul A. and {Schnurr}, Olivier and {Hirschi}, Raphael and {Yusof}, Norhasliza and {Parker}, Richard J. and {Goodwin}, Simon P. and {Kassim}, Hasan Abu},
        title = "{The R136 star cluster hosts several stars whose individual masses greatly exceed the accepted 150M$_{solar}$ stellar mass limit}",
      journal = {\mnras},
         year = 2010,
        month = oct,
       volume = {408},
       number = {2},
        pages = {731-751},
          doi = {10.1111/j.1365-2966.2010.17167.x},
archivePrefix = {arXiv},
       eprint = {1007.3284},
 primaryClass = {astro-ph.SR},
       adsurl = {https://ui.adsabs.harvard.edu/abs/2010MNRAS.408..731C}
}

@ARTICLE{Dayal2019,
       author = {{Dayal}, Pratika and {Rossi}, Elena M. and {Shiralilou}, Banafsheh and {Piana}, Olmo and {Choudhury}, Tirthankar Roy and {Volonteri}, Marta},
        title = "{The hierarchical assembly of galaxies and black holes in the first billion years: predictions for the era of gravitational wave astronomy}",
      journal = {\mnras},
         year = 2019,
        month = jun,
       volume = {486},
       number = {2},
        pages = {2336-2350},
          doi = {10.1093/mnras/stz897},
archivePrefix = {arXiv},
       eprint = {1810.11033},
 primaryClass = {astro-ph.GA},
       adsurl = {https://ui.adsabs.harvard.edu/abs/2019MNRAS.486.2336D}
}

@ARTICLE{Dayal2025,
       author = {{Dayal}, Pratika and {Maiolino}, Roberto},
        title = "{The properties of primordially-seeded black holes and their hosts in the first billion years: implications for JWST}",
      journal = {arXiv e-prints},
         year = 2025,
        month = jun,
          eid = {arXiv:2506.08116},
        pages = {arXiv:2506.08116},
          doi = {10.48550/arXiv.2506.08116},
archivePrefix = {arXiv},
       eprint = {2506.08116},
 primaryClass = {astro-ph.GA},
       adsurl = {https://ui.adsabs.harvard.edu/abs/2025arXiv250608116D}
}

@ARTICLE{Dehnen2017a,
       author = {{Dehnen}, Walter and {Hernandez}, David M.},
        title = "{Symplectic fourth-order maps for the collisional N -body problem}",
      journal = {\mnras},
         year = "2017",
        month = "Feb",
       volume = {465},
       number = {1},
        pages = {1201-1217},
          doi = {10.1093/mnras/stw2758},
archivePrefix = {arXiv},
       eprint = {1609.09375},
 primaryClass = {math.NA},
       adsurl = {https://ui.adsabs.harvard.edu/abs/2017MNRAS.465.1201D}
}

@ARTICLE{DellaCroce2024,
       author = {{Della Croce}, A. and {Pascale}, R. and {Giunchi}, E. and {Nipoti}, C. and {Cignoni}, M. and {Dalessandro}, E.},
        title = "{The most stringent upper limit set on the mass of a central black hole in 47 Tucanae using dynamical models}",
      journal = {\aap},
         year = 2024,
        month = feb,
       volume = {682},
          eid = {A22},
        pages = {A22},
          doi = {10.1051/0004-6361/202347569},
archivePrefix = {arXiv},
       eprint = {2310.15221},
 primaryClass = {astro-ph.GA},
       adsurl = {https://ui.adsabs.harvard.edu/abs/2024A&A...682A..22D}
}

@ARTICLE{Denissenkov2014,
       author = {{Denissenkov}, P.~A. and {Hartwick}, F.~D.~A.},
        title = "{Supermassive stars as a source of abundance anomalies of proton-capture elements in globular clusters}",
      journal = {\mnras},
         year = 2014,
        month = jan,
       volume = {437},
       number = {1},
        pages = {L21-L25},
          doi = {10.1093/mnrasl/slt133},
archivePrefix = {arXiv},
       eprint = {1305.5975},
 primaryClass = {astro-ph.SR},
       adsurl = {https://ui.adsabs.harvard.edu/abs/2014MNRAS.437L..21D}
}

@ARTICLE{Dorozsmai2024,
       author = {{Dorozsmai}, Andris and {Toonen}, Silvia},
        title = "{Importance of stable mass transfer and stellar winds for the formation of gravitational wave sources}",
      journal = {\mnras},
         year = 2024,
        month = jun,
       volume = {530},
       number = {4},
        pages = {3706-3739},
          doi = {10.1093/mnras/stae152},
archivePrefix = {arXiv},
       eprint = {2207.08837},
 primaryClass = {astro-ph.SR},
       adsurl = {https://ui.adsabs.harvard.edu/abs/2024MNRAS.530.3706D}
}

@ARTICLE{vanDokkum2025a,
       author = {{van Dokkum}, Pieter and {Brammer}, Gabriel and {Baggen}, Josephine F.~W. and {Keim}, Michael A. and {Natarajan}, Priyamvada and {Pasha}, Imad},
        title = "{The {\ensuremath{\infty}} Galaxy: A Candidate Direct-collapse Supermassive Black Hole between Two Massive, Ringed Nuclei}",
      journal = {\apjl},
         year = 2025,
        month = jul,
       volume = {988},
       number = {1},
          eid = {L6},
        pages = {L6},
          doi = {10.3847/2041-8213/addcfe},
archivePrefix = {arXiv},
       eprint = {2506.15618},
 primaryClass = {astro-ph.GA},
       adsurl = {https://ui.adsabs.harvard.edu/abs/2025ApJ...988L...6V}
}

@ARTICLE{vanDokkum2025b,
       author = {{van Dokkum}, Pieter and {Brammer}, Gabriel and {Jennings}, Connor and {Pasha}, Imad and {Baggen}, Josephine F.~W.},
        title = "{Further Evidence for a Direct-collapse Origin of the Supermassive Black Hole at the Center of the {\ensuremath{\infty}} Galaxy}",
      journal = {\apjl},
         year = 2025,
        month = sep,
       volume = {990},
       number = {2},
          eid = {L48},
        pages = {L48},
          doi = {10.3847/2041-8213/adfb50},
archivePrefix = {arXiv},
       eprint = {2506.15619},
 primaryClass = {astro-ph.GA},
       adsurl = {https://ui.adsabs.harvard.edu/abs/2025ApJ...990L..48V}
}

@ARTICLE{vanDonkelaar2026,
       author = {{van Donkelaar}, Floor and {Mayer}, Lucio and {Capelo}, Pedro R. and {Sijacki}, Debora and {Adamo}, Angela},
        title = "{Cosmic wallflowers: the circumgalactic origins of isolated ultra-compact star clusters at $z>7$}",
      journal = {arXiv e-prints},
         year = 2026,
        month = jan,
          eid = {arXiv:2601.05333},
        pages = {arXiv:2601.05333},
archivePrefix = {arXiv},
       eprint = {2601.05333},
 primaryClass = {astro-ph.GA},
       adsurl = {https://ui.adsabs.harvard.edu/abs/2026arXiv260105333V}
}

@ARTICLE{Dong2012,
       author = {{Dong}, Ruobing and {Greene}, Jenny E. and {Ho}, Luis C.},
        title = "{X-Ray Properties of Intermediate-mass Black Holes in Active Galaxies. III. Spectral Energy Distribution and Possible Evidence for Intrinsically X-Ray-weak Active Galactic Nuclei}",
      journal = {\apj},
         year = 2012,
        month = dec,
       volume = {761},
       number = {1},
          eid = {73},
        pages = {73},
          doi = {10.1088/0004-637X/761/1/73},
archivePrefix = {arXiv},
       eprint = {1210.6653},
 primaryClass = {astro-ph.GA},
       adsurl = {https://ui.adsabs.harvard.edu/abs/2012ApJ...761...73D}
}

@ARTICLE{Ebihara2026,
       author = {{Ebihara}, Sho and {Fujii}, Michiko S. and {Saitoh}, Takayuki R. and {Hirai}, Yutaka and {Isobe}, Yuki and {Nagele}, Chris},
        title = "{Nitrogen enhancement of GN-z11 by metal pollution from supermassive stars}",
      journal = {arXiv e-prints},
         year = 2026,
        month = jan,
          eid = {arXiv:2601.04344},
        pages = {arXiv:2601.04344},
          doi = {10.48550/arXiv.2601.04344},
archivePrefix = {arXiv},
       eprint = {2601.04344},
 primaryClass = {astro-ph.GA},
       adsurl = {https://ui.adsabs.harvard.edu/abs/2026arXiv260104344E}
}

@ARTICLE{Elmegreen1996,
       author = {{Elmegreen}, Bruce G. and {Efremov}, Yuri N.},
        title = "{An Extension of Hierarchical Star Formation to Galactic Scales}",
      journal = {\apj},
         year = 1996,
        month = aug,
       volume = {466},
        pages = {802},
          doi = {10.1086/177554},
       adsurl = {https://ui.adsabs.harvard.edu/abs/1996ApJ...466..802E}
}

@ARTICLE{Farrell2009,
       author = {{Farrell}, Sean A. and {Webb}, Natalie A. and {Barret}, Didier and {Godet}, Olivier and {Rodrigues}, Joana M.},
        title = "{An intermediate-mass black hole of over 500 solar masses in the galaxy ESO243-49}",
      journal = {\nat},
         year = 2009,
        month = jul,
       volume = {460},
       number = {7251},
        pages = {73-75},
          doi = {10.1038/nature08083},
archivePrefix = {arXiv},
       eprint = {1001.0567},
 primaryClass = {astro-ph.HE},
       adsurl = {https://ui.adsabs.harvard.edu/abs/2009Natur.460...73F}
}

@ARTICLE{Fragione2020,
       author = {{Fragione}, Giacomo and {Silk}, Joseph},
        title = "{Repeated mergers and ejection of black holes within nuclear star clusters}",
      journal = {\mnras},
         year = 2020,
        month = nov,
       volume = {498},
       number = {4},
        pages = {4591-4604},
          doi = {10.1093/mnras/staa2629},
archivePrefix = {arXiv},
       eprint = {2006.01867},
 primaryClass = {astro-ph.GA},
       adsurl = {https://ui.adsabs.harvard.edu/abs/2020MNRAS.498.4591F}
}

@ARTICLE{Fragione2022b,
       author = {{Fragione}, Giacomo and {Loeb}, Abraham and {Kocsis}, Bence and {Rasio}, Frederic A.},
        title = "{Merger Rates of Intermediate-mass Black Hole Binaries in Nuclear Star Clusters}",
      journal = {\apj},
         year = 2022,
        month = jul,
       volume = {933},
       number = {2},
          eid = {170},
        pages = {170},
          doi = {10.3847/1538-4357/ac75d0},
archivePrefix = {arXiv},
       eprint = {2204.03745},
 primaryClass = {astro-ph.HE},
       adsurl = {https://ui.adsabs.harvard.edu/abs/2022ApJ...933..170F}
}

@ARTICLE{Fregeau2004,
       author = {{Fregeau}, J.~M. and {Cheung}, P. and {Portegies Zwart}, S.~F. and {Rasio}, F.~A.},
        title = "{Stellar collisions during binary-binary and binary-single star interactions}",
      journal = {\mnras},
         year = 2004,
        month = jul,
       volume = {352},
       number = {1},
        pages = {1-19},
          doi = {10.1111/j.1365-2966.2004.07914.x},
archivePrefix = {arXiv},
       eprint = {astro-ph/0401004},
 primaryClass = {astro-ph},
       adsurl = {https://ui.adsabs.harvard.edu/abs/2004MNRAS.352....1F}
}

@ARTICLE{Freitag2006a,
       author = {{Freitag}, Marc and {Rasio}, Frederic A. and {Baumgardt}, Holger},
        title = "{Runaway collisions in young star clusters - I. Methods and tests}",
      journal = {\mnras},
         year = 2006,
        month = may,
       volume = {368},
       number = {1},
        pages = {121-140},
          doi = {10.1111/j.1365-2966.2006.10095.x},
archivePrefix = {arXiv},
       eprint = {astro-ph/0503129},
 primaryClass = {astro-ph},
       adsurl = {https://ui.adsabs.harvard.edu/abs/2006MNRAS.368..121F}
}

@ARTICLE{Fryer2001,
       author = {{Fryer}, Chris L. and {Kalogera}, Vassiliki},
        title = "{Theoretical Black Hole Mass Distributions}",
      journal = {\apj},
         year = 2001,
        month = jun,
       volume = {554},
       number = {1},
        pages = {548-560},
          doi = {10.1086/321359},
archivePrefix = {arXiv},
       eprint = {astro-ph/9911312},
 primaryClass = {astro-ph},
       adsurl = {https://ui.adsabs.harvard.edu/abs/2001ApJ...554..548F}
}

@ARTICLE{Fryer2012,
       author = {{Fryer}, Chris L. and {Belczynski}, Krzysztof and {Wiktorowicz}, Grzegorz and {Dominik}, Michal and {Kalogera}, Vicky and {Holz}, Daniel E.},
        title = "{Compact Remnant Mass Function: Dependence on the Explosion Mechanism and Metallicity}",
      journal = {\apj},
         year = 2012,
        month = apr,
       volume = {749},
       number = {1},
          eid = {91},
        pages = {91},
          doi = {10.1088/0004-637X/749/1/91},
archivePrefix = {arXiv},
       eprint = {1110.1726},
 primaryClass = {astro-ph.SR},
       adsurl = {https://ui.adsabs.harvard.edu/abs/2012ApJ...749...91F}
}

@ARTICLE{Fujii2013,
       author = {{Fujii}, M.~S. and {Portegies Zwart}, S.},
        title = "{The growth of massive stars via stellar collisions in ensemble star clusters}",
      journal = {\mnras},
         year = 2013,
        month = apr,
       volume = {430},
       number = {2},
        pages = {1018-1029},
          doi = {10.1093/mnras/sts673},
archivePrefix = {arXiv},
       eprint = {1210.3732},
 primaryClass = {astro-ph.GA},
       adsurl = {https://ui.adsabs.harvard.edu/abs/2013MNRAS.430.1018F}
}

@ARTICLE{Fujii2024,
       author = {{Fujii}, Michiko S. and {Wang}, Long and {Tanikawa}, Ataru and {Hirai}, Yutaka and {Saitoh}, Takayuki R.},
        title = "{Simulations predict intermediate-mass black hole formation in globular clusters}",
      journal = {Science},
         year = 2024,
        month = jun,
       volume = {384},
       number = {6703},
        pages = {1488-1492},
          doi = {10.1126/science.adi4211},
archivePrefix = {arXiv},
       eprint = {2406.06772},
 primaryClass = {astro-ph.GA},
       adsurl = {https://ui.adsabs.harvard.edu/abs/2024Sci...384.1488F}
}

@ARTICLE{Fujimoto2024,
       author = {{Fujimoto}, S. and {Ouchi}, M. and {Kohno}, K. and {Valentino}, F. and {Gim{\'e}nez-Arteaga}, C. and {Brammer}, G.~B. and {Furtak}, L.~J. and {Kohandel}, M. and {Oguri}, M. and {Pallottini}, A. and {Richard}, J. and {Zitrin}, A. and {Bauer}, F.~E. and {Boylan-Kolchin}, M. and {Dessauges-Zavadsky}, M. and {Egami}, E. and {Finkelstein}, S.~L. and {Ma}, Z. and {Smail}, I. and {Watson}, D. and {Hutchison}, T.~A. and {Rigby}, J.~R. and {Welch}, B.~D. and {Ao}, Y. and {Bradley}, L.~D. and {Caminha}, G.~B. and {Caputi}, K.~I. and {Espada}, D. and {Endsley}, R. and {Fudamoto}, Y. and {Gonz{\'a}lez-L{\'o}pez}, J. and {Hatsukade}, B. and {Koekemoer}, A.~M. and {Kokorev}, V. and {Laporte}, N. and {Lee}, M. and {Magdis}, G.~E. and {Ono}, Y. and {Rizzo}, F. and {Shibuya}, T. and {Shimasaku}, K. and {Sun}, F. and {Toft}, S. and {Umehata}, H. and {Wang}, T. and {Yajima}, H.},
        title = "{Primordial Rotating Disk Composed of $\geq$15 Dense Star-Forming Clumps at Cosmic Dawn}",
      journal = {arXiv e-prints},
         year = 2024,
        month = feb,
          eid = {arXiv:2402.18543},
        pages = {arXiv:2402.18543},
          doi = {10.48550/arXiv.2402.18543},
archivePrefix = {arXiv},
       eprint = {2402.18543},
 primaryClass = {astro-ph.GA},
       adsurl = {https://ui.adsabs.harvard.edu/abs/2024arXiv240218543F}
}

@ARTICLE{Gaburov2008,
       author = {{Gaburov}, E. and {Gualandris}, A. and {Portegies Zwart}, S.},
        title = "{On the onset of runaway stellar collisions in dense star clusters - I. Dynamics of the first collision}",
      journal = {\mnras},
         year = 2008,
        month = feb,
       volume = {384},
       number = {1},
        pages = {376-385},
          doi = {10.1111/j.1365-2966.2007.12731.x},
archivePrefix = {arXiv},
       eprint = {0707.0406},
 primaryClass = {astro-ph},
       adsurl = {https://ui.adsabs.harvard.edu/abs/2008MNRAS.384..376G}
}

@ARTICLE{Garmany1985,
       author = {{Garmany}, C.~D. and {Conti}, P.~S.},
        title = "{Stellar winds from hot stars in the Magellanic clouds.}",
      journal = {\apj},
         year = 1985,
        month = jun,
       volume = {293},
        pages = {407-413},
          doi = {10.1086/163247},
       adsurl = {https://ui.adsabs.harvard.edu/abs/1985ApJ...293..407G}
}

@ARTICLE{Gebhardt2002,
       author = {{Gebhardt}, Karl and {Rich}, R.~M. and {Ho}, Luis C.},
        title = "{A 20,000 M$_{solar}$ Black Hole in the Stellar Cluster G1}",
      journal = {\apjl},
         year = 2002,
        month = oct,
       volume = {578},
       number = {1},
        pages = {L41-L45},
          doi = {10.1086/342980},
archivePrefix = {arXiv},
       eprint = {astro-ph/0209313},
 primaryClass = {astro-ph},
       adsurl = {https://ui.adsabs.harvard.edu/abs/2002ApJ...578L..41G}
}

@ARTICLE{Genzel2011,
       author = {{Genzel}, R. and {Newman}, S. and {Jones}, T. and {F{\"o}rster Schreiber}, N.~M. and {Shapiro}, K. and {Genel}, S. and {Lilly}, S.~J. and {Renzini}, A. and {Tacconi}, L.~J. and {Bouch{\'e}}, N. and {Burkert}, A. and {Cresci}, G. and {Buschkamp}, P. and {Carollo}, C.~M. and {Ceverino}, D. and {Davies}, R. and {Dekel}, A. and {Eisenhauer}, F. and {Hicks}, E. and {Kurk}, J. and {Lutz}, D. and {Mancini}, C. and {Naab}, T. and {Peng}, Y. and {Sternberg}, A. and {Vergani}, D. and {Zamorani}, G.},
        title = "{The Sins Survey of z \raisebox{-0.5ex}\textasciitilde 2 Galaxy Kinematics: Properties of the Giant Star-forming Clumps}",
      journal = {\apj},
         year = 2011,
        month = jun,
       volume = {733},
       number = {2},
          eid = {101},
        pages = {101},
          doi = {10.1088/0004-637X/733/2/101},
archivePrefix = {arXiv},
       eprint = {1011.5360},
 primaryClass = {astro-ph.CO},
       adsurl = {https://ui.adsabs.harvard.edu/abs/2011ApJ...733..101G}
}

@ARTICLE{Gieles2018,
       author = {{Gieles}, Mark and {Charbonnel}, Corinne and {Krause}, Martin G.~H. and {H{\'e}nault-Brunet}, Vincent and {Agertz}, Oscar and {Lamers}, Henny J.~G.~L.~M. and {Bastian}, Nathan and {Gualandris}, Alessia and {Zocchi}, Alice and {Petts}, James A.},
        title = "{Concurrent formation of supermassive stars and globular clusters: implications for early self-enrichment}",
      journal = {\mnras},
         year = 2018,
        month = aug,
       volume = {478},
       number = {2},
        pages = {2461-2479},
          doi = {10.1093/mnras/sty1059},
archivePrefix = {arXiv},
       eprint = {1804.04682},
 primaryClass = {astro-ph.GA},
       adsurl = {https://ui.adsabs.harvard.edu/abs/2018MNRAS.478.2461G}
}

@ARTICLE{Gieles2025,
       author = {{Gieles}, Mark and {Padoan}, Paolo and {Charbonnel}, Corinne and {Vink}, Jorick S. and {Ram{\'\i}rez-Galeano}, Laura},
        title = "{Globular cluster formation from inertial inflows: accreting extremely massive stars as the origin of abundance anomalies}",
      journal = {\mnras},
         year = 2025,
        month = nov,
       volume = {544},
       number = {1},
        pages = {483-512},
          doi = {10.1093/mnras/staf1314},
archivePrefix = {arXiv},
       eprint = {2501.12138},
 primaryClass = {astro-ph.GA},
       adsurl = {https://ui.adsabs.harvard.edu/abs/2025MNRAS.544..483G}
}

@ARTICLE{Glebbeek2008,
       author = {{Glebbeek}, E. and {Pols}, O.~R.},
        title = "{Evolution of stellar collision products in open clusters. II. A grid of low-mass collisions}",
      journal = {\aap},
         year = 2008,
        month = sep,
       volume = {488},
       number = {3},
        pages = {1017-1025},
          doi = {10.1051/0004-6361:200809931},
archivePrefix = {arXiv},
       eprint = {0806.0865},
 primaryClass = {astro-ph},
       adsurl = {https://ui.adsabs.harvard.edu/abs/2008A&A...488.1017G}
}

@ARTICLE{Glebbeek2009,
       author = {{Glebbeek}, E. and {Gaburov}, E. and {de Mink}, S.~E. and {Pols}, O.~R. and {Portegies Zwart}, S.~F.},
        title = "{The evolution of runaway stellar collision products}",
      journal = {\aap},
         year = 2009,
        month = apr,
       volume = {497},
       number = {1},
        pages = {255-264},
          doi = {10.1051/0004-6361/200810425},
archivePrefix = {arXiv},
       eprint = {0902.1753},
 primaryClass = {astro-ph.SR},
       adsurl = {https://ui.adsabs.harvard.edu/abs/2009A&A...497..255G}
}

@ARTICLE{Glebbeek2013,
       author = {{Glebbeek}, Evert and {Gaburov}, Evghenii and {Portegies Zwart}, Simon and {Pols}, Onno R.},
        title = "{Structure and evolution of high-mass stellar mergers}",
      journal = {\mnras},
         year = 2013,
        month = oct,
       volume = {434},
       number = {4},
        pages = {3497-3510},
          doi = {10.1093/mnras/stt1268},
archivePrefix = {arXiv},
       eprint = {1307.2445},
 primaryClass = {astro-ph.SR},
       adsurl = {https://ui.adsabs.harvard.edu/abs/2013MNRAS.434.3497G}
}

@ARTICLE{GonzalezPrieto2022,
       author = {{Gonz{\'a}lez Prieto}, Elena and {Kremer}, Kyle and {Fragione}, Giacomo and {Martinez}, Miguel A.~S. and {Weatherford}, Newlin C. and {Zevin}, Michael and {Rasio}, Frederic A.},
        title = "{Intermediate-mass Black Holes on the Run from Young Star Clusters}",
      journal = {\apj},
         year = 2022,
        month = dec,
       volume = {940},
       number = {2},
          eid = {131},
        pages = {131},
          doi = {10.3847/1538-4357/ac9b0f},
archivePrefix = {arXiv},
       eprint = {2208.07881},
 primaryClass = {astro-ph.HE},
       adsurl = {https://ui.adsabs.harvard.edu/abs/2022ApJ...940..131G}
}

@ARTICLE{GonzalezPrieto2024,
       author = {{Gonz{\'a}lez Prieto}, Elena and {Weatherford}, Newlin C. and {Fragione}, Giacomo and {Kremer}, Kyle and {Rasio}, Frederic A.},
        title = "{Intermediate-mass Black Hole Progenitors from Stellar Collisions in Dense Star Clusters}",
      journal = {\apj},
         year = 2024,
        month = jul,
       volume = {969},
       number = {1},
          eid = {29},
        pages = {29},
          doi = {10.3847/1538-4357/ad43d6},
archivePrefix = {arXiv},
       eprint = {2404.11646},
 primaryClass = {astro-ph.GA},
       adsurl = {https://ui.adsabs.harvard.edu/abs/2024ApJ...969...29G}
}

@ARTICLE{Goswami2012,
       author = {{Goswami}, Sanghamitra and {Umbreit}, Stefan and {Bierbaum}, Matt and {Rasio}, Frederic A.},
        title = "{Formation of Massive Black Holes in Dense Star Clusters. II. Initial Mass Function and Primordial Mass Segregation}",
      journal = {\apj},
         year = 2012,
        month = jun,
       volume = {752},
       number = {1},
          eid = {43},
        pages = {43},
          doi = {10.1088/0004-637X/752/1/43},
archivePrefix = {arXiv},
       eprint = {1105.5884},
 primaryClass = {astro-ph.GA},
       adsurl = {https://ui.adsabs.harvard.edu/abs/2012ApJ...752...43G}
}

@ARTICLE{Gratton2019,
       author = {{Gratton}, Raffaele and {Bragaglia}, Angela and {Carretta}, Eugenio and {D'Orazi}, Valentina and {Lucatello}, Sara and {Sollima}, Antonio},
        title = "{What is a globular cluster? An observational perspective}",
      journal = {\aapr},
         year = 2019,
        month = nov,
       volume = {27},
       number = {1},
          eid = {8},
        pages = {8},
          doi = {10.1007/s00159-019-0119-3},
archivePrefix = {arXiv},
       eprint = {1911.02835},
 primaryClass = {astro-ph.SR},
       adsurl = {https://ui.adsabs.harvard.edu/abs/2019A&ARv..27....8G}
}

@ARTICLE{Greene2007,
       author = {{Greene}, Jenny E. and {Ho}, Luis C.},
        title = "{X-Ray Properties of Intermediate-Mass Black Holes in Active Galaxies}",
      journal = {\apj},
         year = 2007,
        month = feb,
       volume = {656},
       number = {1},
        pages = {84-92},
          doi = {10.1086/509064},
archivePrefix = {arXiv},
       eprint = {astro-ph/0608061},
 primaryClass = {astro-ph},
       adsurl = {https://ui.adsabs.harvard.edu/abs/2007ApJ...656...84G}
}

@ARTICLE{Greene2020,
       author = {{Greene}, Jenny E. and {Strader}, Jay and {Ho}, Luis C.},
        title = "{Intermediate-Mass Black Holes}",
      journal = {\araa},
         year = 2020,
        month = aug,
       volume = {58},
        pages = {257-312},
          doi = {10.1146/annurev-astro-032620-021835},
archivePrefix = {arXiv},
       eprint = {1911.09678},
 primaryClass = {astro-ph.GA},
       adsurl = {https://ui.adsabs.harvard.edu/abs/2020ARA&A..58..257G}
}

@ARTICLE{Gurkan2004,
       author = {{G{\"u}rkan}, M. Atakan and {Freitag}, Marc and {Rasio}, Frederic A.},
        title = "{Formation of Massive Black Holes in Dense Star Clusters. I. Mass Segregation and Core Collapse}",
      journal = {\apj},
         year = 2004,
        month = apr,
       volume = {604},
       number = {2},
        pages = {632-652},
          doi = {10.1086/381968},
archivePrefix = {arXiv},
       eprint = {astro-ph/0308449},
 primaryClass = {astro-ph},
       adsurl = {https://ui.adsabs.harvard.edu/abs/2004ApJ...604..632G}
}

@ARTICLE{Haemmerle2018,
       author = {{Haemmerl{\'e}}, Lionel and {Woods}, T.~E. and {Klessen}, Ralf S. and {Heger}, Alexander and {Whalen}, Daniel J.},
        title = "{The evolution of supermassive Population III stars}",
      journal = {\mnras},
         year = 2018,
        month = feb,
       volume = {474},
       number = {2},
        pages = {2757-2773},
          doi = {10.1093/mnras/stx2919},
archivePrefix = {arXiv},
       eprint = {1705.09301},
 primaryClass = {astro-ph.SR},
       adsurl = {https://ui.adsabs.harvard.edu/abs/2018MNRAS.474.2757H}
}

@ARTICLE{Haster2016a,
       author = {{Haster}, Carl-Johan and {Wang}, Zhilu and {Berry}, Christopher P.~L. and {Stevenson}, Simon and {Veitch}, John and {Mandel}, Ilya},
        title = "{Inference on gravitational waves from coalescences of stellar-mass compact objects and intermediate-mass black holes}",
      journal = {\mnras},
         year = 2016,
        month = apr,
       volume = {457},
       number = {4},
        pages = {4499-4506},
          doi = {10.1093/mnras/stw233},
archivePrefix = {arXiv},
       eprint = {1511.01431},
 primaryClass = {astro-ph.HE},
       adsurl = {https://ui.adsabs.harvard.edu/abs/2016MNRAS.457.4499H}
}

@ARTICLE{Henon1971,
       author = {{H{\'e}non}, M.~H.},
        title = "{The Monte Carlo Method (Papers appear in the Proceedings of IAU Colloquium No. 10 Gravitational N-Body Problem (ed. by Myron Lecar), R. Reidel Publ. Co. , Dordrecht-Holland.)}",
      journal = {\apss},
         year = 1971,
        month = nov,
       volume = {14},
       number = {1},
        pages = {151-167},
          doi = {10.1007/BF00649201},
       adsurl = {https://ui.adsabs.harvard.edu/abs/1971Ap&SS..14..151H}
}

@ARTICLE{Hosokawa2012,
       author = {{Hosokawa}, Takashi and {Omukai}, Kazuyuki and {Yorke}, Harold W.},
        title = "{Rapidly Accreting Supergiant Protostars: Embryos of Supermassive Black Holes?}",
      journal = {\apj},
         year = 2012,
        month = sep,
       volume = {756},
       number = {1},
          eid = {93},
        pages = {93},
          doi = {10.1088/0004-637X/756/1/93},
archivePrefix = {arXiv},
       eprint = {1203.2613},
 primaryClass = {astro-ph.CO},
       adsurl = {https://ui.adsabs.harvard.edu/abs/2012ApJ...756...93H}
}

@ARTICLE{Hoyer2025,
       author = {{Hoyer}, Nils and {Bonoli}, Silvia and {Bastian}, Nate and {Herrero-Carri{\'o}n}, Diego and {Neumayer}, Nadine and {Izquierdo-Villalba}, David and {Spinoso}, Daniele and {Yates}, Robert M. and {Polkas}, Markos and {Artale}, M. Celeste},
        title = "{Massive Star Clusters in the Semi-Analytical Galaxy Formation Model L-Galaxies 2020}",
      journal = {arXiv e-prints},
         year = 2025,
        month = apr,
          eid = {arXiv:2504.12079},
        pages = {arXiv:2504.12079},
          doi = {10.48550/arXiv.2504.12079},
archivePrefix = {arXiv},
       eprint = {2504.12079},
 primaryClass = {astro-ph.GA},
       adsurl = {https://ui.adsabs.harvard.edu/abs/2025arXiv250412079H}
}

@ARTICLE{Humphreys1994,
       author = {{Humphreys}, Roberta M. and {Davidson}, Kris},
        title = "{The Luminous Blue Variables: Astrophysical Geysers}",
      journal = {\pasp},
         year = 1994,
        month = oct,
       volume = {106},
        pages = {1025},
          doi = {10.1086/133478},
       adsurl = {https://ui.adsabs.harvard.edu/abs/1994PASP..106.1025H}
}

@ARTICLE{Hurley2000,
       author = {{Hurley}, Jarrod R. and {Pols}, Onno R. and {Tout}, Christopher A.},
        title = "{Comprehensive analytic formulae for stellar evolution as a function of mass and metallicity}",
      journal = {\mnras},
         year = 2000,
        month = jul,
       volume = {315},
       number = {3},
        pages = {543-569},
          doi = {10.1046/j.1365-8711.2000.03426.x},
archivePrefix = {arXiv},
       eprint = {astro-ph/0001295},
 primaryClass = {astro-ph},
       adsurl = {https://ui.adsabs.harvard.edu/abs/2000MNRAS.315..543H}
}

@ARTICLE{Hurley2002,
       author = {{Hurley}, Jarrod R. and {Tout}, Christopher A. and {Pols}, Onno R.},
        title = "{Evolution of binary stars and the effect of tides on binary populations}",
      journal = {\mnras},
         year = 2002,
        month = feb,
       volume = {329},
       number = {4},
        pages = {897-928},
          doi = {10.1046/j.1365-8711.2002.05038.x},
archivePrefix = {arXiv},
       eprint = {astro-ph/0201220},
 primaryClass = {astro-ph},
       adsurl = {https://ui.adsabs.harvard.edu/abs/2002MNRAS.329..897H}
}

@ARTICLE{Häberle2024,
       author = {{H{\"a}berle}, Maximilian and {Neumayer}, Nadine and {Seth}, Anil and {Bellini}, Andrea and {Libralato}, Mattia and {Baumgardt}, Holger and {Whitaker}, Matthew and {Dumont}, Antoine and {Alfaro-Cuello}, Mayte and {Anderson}, Jay and {Clontz}, Callie and {Kacharov}, Nikolay and {Kamann}, Sebastian and {Feldmeier-Krause}, Anja and {Milone}, Antonino and {Nitschai}, Maria Selina and {Pechetti}, Renuka and {van de Ven}, Glenn},
        title = "{Fast-moving stars around an intermediate-mass black hole in {\ensuremath{\omega}} Centauri}",
      journal = {\nat},
         year = 2024,
        month = jul,
       volume = {631},
       number = {8020},
        pages = {285-288},
          doi = {10.1038/s41586-024-07511-z},
archivePrefix = {arXiv},
       eprint = {2405.06015},
 primaryClass = {astro-ph.GA},
       adsurl = {https://ui.adsabs.harvard.edu/abs/2024Natur.631..285H}
}

@ARTICLE{Ibata2009,
       author = {{Ibata}, R. and {Bellazzini}, M. and {Chapman}, S.~C. and {Dalessandro}, E. and {Ferraro}, F. and {Irwin}, M. and {Lanzoni}, B. and {Lewis}, G.~F. and {Mackey}, A.~D. and {Miocchi}, P. and {Varghese}, A.},
        title = "{Density and Kinematic Cusps in M54 at the Heart of the Sagittarius Dwarf Galaxy: Evidence for A {}10$^{4}$ M $_{sun}$ Black Hole?}",
      journal = {\apjl},
         year = 2009,
        month = jul,
       volume = {699},
       number = {2},
        pages = {L169-L173},
          doi = {10.1088/0004-637X/699/2/L169},
archivePrefix = {arXiv},
       eprint = {0906.4894},
 primaryClass = {astro-ph.GA},
       adsurl = {https://ui.adsabs.harvard.edu/abs/2009ApJ...699L.169I}
}

@ARTICLE{Inayoshi2020,
       author = {{Inayoshi}, Kohei and {Visbal}, Eli and {Haiman}, Zolt{\'a}n},
        title = "{The Assembly of the First Massive Black Holes}",
      journal = {\araa},
         year = 2020,
        month = aug,
       volume = {58},
        pages = {27-97},
          doi = {10.1146/annurev-astro-120419-014455},
archivePrefix = {arXiv},
       eprint = {1911.05791},
 primaryClass = {astro-ph.GA},
       adsurl = {https://ui.adsabs.harvard.edu/abs/2020ARA&A..58...27I}
}

@ARTICLE{Iorio2023,
       author = {{Iorio}, Giuliano and {Mapelli}, Michela and {Costa}, Guglielmo and {Spera}, Mario and {Escobar}, Gast{\'o}n J. and {Sgalletta}, Cecilia and {Trani}, Alessandro A. and {Korb}, Erika and {Santoliquido}, Filippo and {Dall'Amico}, Marco and {Gaspari}, Nicola and {Bressan}, Alessandro},
        title = "{Compact object mergers: exploring uncertainties from stellar and binary evolution with SEVN}",
      journal = {\mnras},
         year = 2023,
        month = sep,
       volume = {524},
       number = {1},
        pages = {426-470},
          doi = {10.1093/mnras/stad1630},
archivePrefix = {arXiv},
       eprint = {2211.11774},
 primaryClass = {astro-ph.HE},
       adsurl = {https://ui.adsabs.harvard.edu/abs/2023MNRAS.524..426I}
}

@ARTICLE{Isobe2023,
       author = {{Isobe}, Yuki and {Ouchi}, Masami and {Tominaga}, Nozomu and {Watanabe}, Kuria and {Nakajima}, Kimihiko and {Umeda}, Hiroya and {Yajima}, Hidenobu and {Harikane}, Yuichi and {Fukushima}, Hajime and {Xu}, Yi and {Ono}, Yoshiaki and {Zhang}, Yechi},
        title = "{JWST Identification of Extremely Low C/N Galaxies with [N/O] {\ensuremath{\gtrsim}} 0.5 at z 6-10 Evidencing the Early CNO-cycle Enrichment and a Connection with Globular Cluster Formation}",
      journal = {\apj},
         year = 2023,
        month = dec,
       volume = {959},
       number = {2},
          eid = {100},
        pages = {100},
          doi = {10.3847/1538-4357/ad09be},
archivePrefix = {arXiv},
       eprint = {2307.00710},
 primaryClass = {astro-ph.GA},
       adsurl = {https://ui.adsabs.harvard.edu/abs/2023ApJ...959..100I}
}

@ARTICLE{Isobe2025,
       author = {{Isobe}, Yuki and {Maiolino}, Roberto and {D'Eugenio}, Francesco and {Curti}, Mirko and {Ji}, Xihan and {Juod{\v{z}}balis}, Ignas and {Scholtz}, Jan and {Feltre}, Anne and {Charlot}, St{\'e}phane and {{\"U}bler}, Hannah and {Bunker}, Andrew J. and {Carniani}, Stefano and {Curtis-Lake}, Emma and {Ji}, Zhiyuan and {Kumari}, Nimisha and {Rinaldi}, Pierluigi and {Robertson}, Brant and {Willott}, Chris and {Witstok}, Joris},
        title = "{JADES: Average Nitrogen Enhancement in High-Redshift Broad-Line Active Galactic Nuclei}",
      journal = {arXiv e-prints},
         year = 2025,
        month = feb,
          eid = {arXiv:2502.12091},
        pages = {arXiv:2502.12091},
          doi = {10.48550/arXiv.2502.12091},
archivePrefix = {arXiv},
       eprint = {2502.12091},
 primaryClass = {astro-ph.GA},
       adsurl = {https://ui.adsabs.harvard.edu/abs/2025arXiv250212091I}
}

@ARTICLE{Ji2025,
       author = {{Ji}, Xihan and {Belokurov}, Vasily and {Maiolino}, Roberto and {Monty}, Stephanie and {Isobe}, Yuki and {Kravtsov}, Andrey and {McClymont}, William and {{\"U}bler}, Hannah},
        title = "{Connecting JWST discovered N/O-enhanced galaxies to globular clusters: Evidence from chemical imprints}",
      journal = {arXiv e-prints},
         year = 2025,
        month = may,
          eid = {arXiv:2505.12505},
        pages = {arXiv:2505.12505},
          doi = {10.48550/arXiv.2505.12505},
archivePrefix = {arXiv},
       eprint = {2505.12505},
 primaryClass = {astro-ph.GA},
       adsurl = {https://ui.adsabs.harvard.edu/abs/2025arXiv250512505J}
}

@ARTICLE{Jalali2012,
       author = {{Jalali}, B. and {Baumgardt}, H. and {Kissler-Patig}, M. and {Gebhardt}, K. and {Noyola}, E. and {L{\"u}tzgendorf}, N. and {de Zeeuw}, P.~T.},
        title = "{A Dynamical N-body model for the central region of {\ensuremath{\omega}} Centauri}",
      journal = {\aap},
         year = 2012,
        month = feb,
       volume = {538},
          eid = {A19},
        pages = {A19},
          doi = {10.1051/0004-6361/201116923},
archivePrefix = {arXiv},
       eprint = {1111.5011},
 primaryClass = {astro-ph.GA},
       adsurl = {https://ui.adsabs.harvard.edu/abs/2012A&A...538A..19J}
}

@ARTICLE{Ji2024,
       author = {{Ji}, Xihan and {{\"U}bler}, Hannah and {Maiolino}, Roberto and {D'Eugenio}, Francesco and {Arribas}, Santiago and {Bunker}, Andrew J. and {Charlot}, St{\'e}phane and {Perna}, Michele and {Rodr{\'\i}guez Del Pino}, Bruno and {B{\"o}ker}, Torsten and {Cresci}, Giovanni and {Curti}, Mirko and {Kumari}, Nimisha and {Lamperti}, Isabella},
        title = "{GA-NIFS: an extremely nitrogen-loud and chemically stratified galaxy at z   5.55}",
      journal = {\mnras},
         year = 2024,
        month = nov,
       volume = {535},
       number = {1},
        pages = {881-908},
          doi = {10.1093/mnras/stae2375},
archivePrefix = {arXiv},
       eprint = {2404.04148},
 primaryClass = {astro-ph.GA},
       adsurl = {https://ui.adsabs.harvard.edu/abs/2024MNRAS.535..881J}
}

@ARTICLE{Juodzbalis2024,
       author = {{Juod{\v{z}}balis}, Ignas and {Maiolino}, Roberto and {Baker}, William M. and {Tacchella}, Sandro and {Scholtz}, Jan and {D'Eugenio}, Francesco and {Witstok}, Joris and {Schneider}, Raffaella and {Trinca}, Alessandro and {Valiante}, Rosa and {DeCoursey}, Christa and {Curti}, Mirko and {Carniani}, Stefano and {Chevallard}, Jacopo and {de Graaff}, Anna and {Arribas}, Santiago and {Bennett}, Jake S. and {Bourne}, Martin A. and {Bunker}, Andrew J. and {Charlot}, St{\'e}phane and {Jiang}, Brian and {Koudmani}, Sophie and {Perna}, Michele and {Robertson}, Brant and {Sijacki}, Debora and {{\"U}bler}, Hannah and {Williams}, Christina C. and {Willott}, Chris},
        title = "{A dormant overmassive black hole in the early Universe}",
      journal = {\nat},
         year = 2024,
        month = dec,
       volume = {636},
       number = {8043},
        pages = {594-597},
          doi = {10.1038/s41586-024-08210-5},
archivePrefix = {arXiv},
       eprint = {2403.03872},
 primaryClass = {astro-ph.GA},
       adsurl = {https://ui.adsabs.harvard.edu/abs/2024Natur.636..594J}
}

@ARTICLE{Juodzbalis2025,
       author = {{Juod{\v{z}}balis}, Ignas and {Marconcini}, Cosimo and {D'Eugenio}, Francesco and {Maiolino}, Roberto and {Marconi}, Alessandro and {{\"U}bler}, Hannah and {Scholtz}, Jan and {Ji}, Xihan and {Arribas}, Santiago and {Bennett}, Jake S. and {Bromm}, Volker and {Bunker}, Andrew J. and {Carniani}, Stefano and {Charlot}, St{\'e}phane and {Cresci}, Giovanni and {Dayal}, Pratika and {Egami}, Eiichi and {Fabian}, Andrew and {Inayoshi}, Kohei and {Isobe}, Yuki and {Ivey}, Lucy and {Jones}, Gareth C. and {Koudmani}, Sophie and {Laporte}, Nicolas and {Liu}, Boyuan and {Lyu}, Jianwei and {Mazzolari}, Giovanni and {Monty}, Stephanie and {Parlanti}, Eleonora and {P{\'e}rez-Gonz{\'a}lez}, Pablo G. and {Perna}, Michele and {Robertson}, Brant and {Schneider}, Raffaella and {Sijacki}, Debora and {Tacchella}, Sandro and {Trinca}, Alessandro and {Valiante}, Rosa and {Volonteri}, Marta and {Witstok}, Joris and {Zhang}, Saiyang},
        title = "{A direct black hole mass measurement in a Little Red Dot at the Epoch of Reionization}",
      journal = {arXiv e-prints},
         year = 2025,
        month = aug,
          eid = {arXiv:2508.21748},
        pages = {arXiv:2508.21748},
          doi = {10.48550/arXiv.2508.21748},
archivePrefix = {arXiv},
       eprint = {2508.21748},
 primaryClass = {astro-ph.GA},
       adsurl = {https://ui.adsabs.harvard.edu/abs/2025arXiv250821748J}
}

@ARTICLE{Kaaret2001,
       author = {{Kaaret}, P. and {Prestwich}, A.~H. and {Zezas}, A. and {Murray}, S.~S. and {Kim}, D. -W. and {Kilgard}, R.~E. and {Schlegel}, E.~M. and {Ward}, M.~J.},
        title = "{Chandra High-Resolution Camera observations of the luminous X-ray source in the starburst galaxy M82}",
      journal = {\mnras},
         year = 2001,
        month = feb,
       volume = {321},
       number = {2},
        pages = {L29-L32},
          doi = {10.1046/j.1365-8711.2001.04064.x},
archivePrefix = {arXiv},
       eprint = {astro-ph/0009211},
 primaryClass = {astro-ph},
       adsurl = {https://ui.adsabs.harvard.edu/abs/2001MNRAS.321L..29K}
}

@ARTICLE{Kamann2016,
       author = {{Kamann}, S. and {Husser}, T. -O. and {Brinchmann}, J. and {Emsellem}, E. and {Weilbacher}, P.~M. and {Wisotzki}, L. and {Wendt}, M. and {Krajnovi{\'c}}, D. and {Roth}, M.~M. and {Bacon}, R. and {Dreizler}, S.},
        title = "{MUSE crowded field 3D spectroscopy of over 12 000 stars in the globular cluster NGC 6397. II. Probing the internal dynamics and the presence of a central black hole}",
      journal = {\aap},
         year = 2016,
        month = apr,
       volume = {588},
          eid = {A149},
        pages = {A149},
          doi = {10.1051/0004-6361/201527065},
archivePrefix = {arXiv},
       eprint = {1602.01643},
 primaryClass = {astro-ph.SR},
       adsurl = {https://ui.adsabs.harvard.edu/abs/2016A&A...588A.149K}
}

@ARTICLE{Kamlah2022,
       author = {{Kamlah}, A.~W.~H. and {Spurzem}, R. and {Berczik}, P. and {Arca Sedda}, M. and {Flammini Dotti}, F. and {Neumayer}, N. and {Pang}, X. and {Shu}, Q. and {Tanikawa}, A. and {Giersz}, M.},
        title = "{The impact of stellar evolution on rotating star clusters: the gravothermal-gravogyro catastrophe and the formation of a bar of black holes}",
      journal = {\mnras},
         year = 2022,
        month = nov,
       volume = {516},
       number = {3},
        pages = {3266-3283},
          doi = {10.1093/mnras/stac2281},
archivePrefix = {arXiv},
       eprint = {2205.04470},
 primaryClass = {astro-ph.GA},
       adsurl = {https://ui.adsabs.harvard.edu/abs/2022MNRAS.516.3266K}
}

@ARTICLE{Kim2020,
       author = {{Kim}, Minjin and {L{\'o}pez}, Kristhell M. and {Jonker}, Peter G. and {Ho}, Luis C. and {Im}, Myungshin},
        title = "{Stellar properties of the host galaxy of an ultraluminous X-ray source in NGC 5252}",
      journal = {\mnras},
         year = 2020,
        month = mar,
       volume = {493},
       number = {1},
        pages = {L76-L80},
          doi = {10.1093/mnrasl/slaa011},
archivePrefix = {arXiv},
       eprint = {2001.07927},
 primaryClass = {astro-ph.GA},
       adsurl = {https://ui.adsabs.harvard.edu/abs/2020MNRAS.493L..76K}
}

@BOOK{Kippenhahn2013,
       author = {{Kippenhahn}, Rudolf and {Weigert}, Alfred and {Weiss}, Achim},
        title = "{Stellar Structure and Evolution}",
         year = 2013,
    publisher = "{Springer Berlin Heidelberg}",
          doi = {10.1007/978-3-642-30304-3},
       adsurl = {https://ui.adsabs.harvard.edu/abs/2013sse..book.....K}
}

@ARTICLE{Kochanek1992,
       author = {{Kochanek}, Christopher S.},
        title = "{The Dynamical Evolution of Tidal Capture Binaries}",
      journal = {\apj},
         year = 1992,
        month = feb,
       volume = {385},
        pages = {604},
          doi = {10.1086/170966},
       adsurl = {https://ui.adsabs.harvard.edu/abs/1992ApJ...385..604K}
}

@ARTICLE{Kremer2019,
       author = {{Kremer}, Kyle and {Lu}, Wenbin and {Rodriguez}, Carl L. and {Lachat}, Mitchell and {Rasio}, Frederic A.},
        title = "{Tidal Disruptions of Stars by Black Hole Remnants in Dense Star Clusters}",
      journal = {\apj},
         year = 2019,
        month = aug,
       volume = {881},
       number = {1},
          eid = {75},
        pages = {75},
          doi = {10.3847/1538-4357/ab2e0c},
archivePrefix = {arXiv},
       eprint = {1904.06353},
 primaryClass = {astro-ph.HE},
       adsurl = {https://ui.adsabs.harvard.edu/abs/2019ApJ...881...75K}
}

@ARTICLE{Kremer2020b,
       author = {{Kremer}, Kyle and {Spera}, Mario and {Becker}, Devin and {Chatterjee}, Sourav and {Di Carlo}, Ugo N. and {Fragione}, Giacomo and {Rodriguez}, Carl L. and {Ye}, Claire S. and {Rasio}, Frederic A.},
        title = "{Populating the Upper Black Hole Mass Gap through Stellar Collisions in Young Star Clusters}",
      journal = {\apj},
         year = 2020,
        month = nov,
       volume = {903},
       number = {1},
          eid = {45},
        pages = {45},
          doi = {10.3847/1538-4357/abb945},
archivePrefix = {arXiv},
       eprint = {2006.10771},
 primaryClass = {astro-ph.HE},
       adsurl = {https://ui.adsabs.harvard.edu/abs/2020ApJ...903...45K}
}

@ARTICLE{Kroupa2001,
       author = {{Kroupa}, Pavel},
        title = "{On the variation of the initial mass function}",
      journal = {\mnras},
         year = 2001,
        month = apr,
       volume = {322},
       number = {2},
        pages = {231-246},
          doi = {10.1046/j.1365-8711.2001.04022.x},
archivePrefix = {arXiv},
       eprint = {astro-ph/0009005},
 primaryClass = {astro-ph},
       adsurl = {https://ui.adsabs.harvard.edu/abs/2001MNRAS.322..231K}
}

@ARTICLE{Krumholz2019,
       author = {{Krumholz}, Mark R. and {McKee}, Christopher F. and {Bland-Hawthorn}, Joss},
        title = "{Star Clusters Across Cosmic Time}",
      journal = {\araa},
         year = 2019,
        month = aug,
       volume = {57},
        pages = {227-303},
          doi = {10.1146/annurev-astro-091918-104430},
archivePrefix = {arXiv},
       eprint = {1812.01615},
 primaryClass = {astro-ph.GA},
       adsurl = {https://ui.adsabs.harvard.edu/abs/2019ARA&A..57..227K}
}

@ARTICLE{Kuruvanthodi2023,
       author = {{Kuruvanthodi}, A. and {Schaerer}, D. and {Messa}, M. and {Adamo}, A. and {Usher}, C. and {Charbonnel}, C. and {Marques-Chaves}, R.},
        title = "{Search strategies for supermassive stars in young clusters and application to nearby galaxies}",
      journal = {\aap},
         year = 2023,
        month = jun,
       volume = {674},
          eid = {A140},
        pages = {A140},
          doi = {10.1051/0004-6361/202245695},
archivePrefix = {arXiv},
       eprint = {2303.08087},
 primaryClass = {astro-ph.GA},
       adsurl = {https://ui.adsabs.harvard.edu/abs/2023A&A...674A.140K}
}

@ARTICLE{Lahen2020,
       author = {{Lah{\'e}n}, Natalia and {Naab}, Thorsten and {Johansson}, Peter H. and {Elmegreen}, Bruce and {Hu}, Chia-Yu and {Walch}, Stefanie and {Steinwandel}, Ulrich P. and {Moster}, Benjamin P.},
        title = "{The GRIFFIN Project{\textemdash}Formation of Star Clusters with Individual Massive Stars in a Simulated Dwarf Galaxy Starburst}",
      journal = {\apj},
         year = 2020,
        month = mar,
       volume = {891},
       number = {1},
          eid = {2},
        pages = {2},
          doi = {10.3847/1538-4357/ab7190},
archivePrefix = {arXiv},
       eprint = {1911.05093},
 primaryClass = {astro-ph.GA},
       adsurl = {https://ui.adsabs.harvard.edu/abs/2020ApJ...891....2L}
}

@ARTICLE{Lahen2024,
       author = {{Lah{\'e}n}, Natalia and {Naab}, Thorsten and {Sz{\'e}csi}, Dorottya},
        title = "{Star clusters forming in a low-metallicity starburst - rapid self-enrichment by (very) massive stars}",
      journal = {\mnras},
         year = 2024,
        month = may,
       volume = {530},
       number = {1},
        pages = {645-667},
          doi = {10.1093/mnras/stae904},
archivePrefix = {arXiv},
       eprint = {2402.09518},
 primaryClass = {astro-ph.GA},
       adsurl = {https://ui.adsabs.harvard.edu/abs/2024MNRAS.530..645L}
}

@ARTICLE{Lahen2025a,
       author = {{Lah{\'e}n}, Natalia and {Rantala}, Antti and {Naab}, Thorsten and {Partmann}, Christian and {Johansson}, Peter H. and {Hislop}, Jessica May},
        title = "{The formation, evolution, and disruption of star clusters with improved gravitational dynamics in simulated dwarf galaxies}",
      journal = {\mnras},
         year = 2025,
        month = apr,
       volume = {538},
       number = {3},
        pages = {2129-2148},
          doi = {10.1093/mnras/staf350},
archivePrefix = {arXiv},
       eprint = {2410.01891},
 primaryClass = {astro-ph.GA},
       adsurl = {https://ui.adsabs.harvard.edu/abs/2025MNRAS.538.2129L}
}

@ARTICLE{Lahen2025b,
       author = {{Lah{\'e}n}, Natalia and {Naab}, Thorsten and {Rantala}, Antti and {Partmann}, Christian},
        title = "{Mergers all the way down: stellar collisions and kinematics of a dense hierarchically forming massive star cluster in a dwarf starburst}",
      journal = {arXiv e-prints},
         year = 2025,
        month = apr,
          eid = {arXiv:2504.18620},
        pages = {arXiv:2504.18620},
archivePrefix = {arXiv},
       eprint = {2504.18620},
 primaryClass = {astro-ph.GA},
       adsurl = {https://ui.adsabs.harvard.edu/abs/2025arXiv250418620L}
}

@ARTICLE{Lanzoni2013,
       author = {{Lanzoni}, B. and {Mucciarelli}, A. and {Origlia}, L. and {Bellazzini}, M. and {Ferraro}, F.~R. and {Valenti}, E. and {Miocchi}, P. and {Dalessandro}, E. and {Pallanca}, C. and {Massari}, D.},
        title = "{The Velocity Dispersion Profile of NGC 6388 from Resolved-star Spectroscopy: No Evidence of a Central Cusp and New Constraints on the Black Hole Mass}",
      journal = {\apj},
         year = 2013,
        month = jun,
       volume = {769},
       number = {2},
          eid = {107},
        pages = {107},
          doi = {10.1088/0004-637X/769/2/107},
archivePrefix = {arXiv},
       eprint = {1304.2953},
 primaryClass = {astro-ph.SR},
       adsurl = {https://ui.adsabs.harvard.edu/abs/2013ApJ...769..107L}
}

@ARTICLE{Lee1999,
       author = {{Lee}, Y. -W. and {Joo}, J. -M. and {Sohn}, Y. -J. and {Rey}, S. -C. and {Lee}, H. -C. and {Walker}, A.~R.},
        title = "{Multiple stellar populations in the globular cluster {\ensuremath{\omega}} Centauri as tracers of a merger event}",
      journal = {\nat},
         year = 1999,
        month = nov,
       volume = {402},
       number = {6757},
        pages = {55-57},
          doi = {10.1038/46985},
archivePrefix = {arXiv},
       eprint = {astro-ph/9911137},
 primaryClass = {astro-ph},
       adsurl = {https://ui.adsabs.harvard.edu/abs/1999Natur.402...55L}
}

@ARTICLE{Leonard1989,
       author = {{Leonard}, Peter J.~T.},
        title = "{Stellar Collisions in Globular Clusters and the Blue Straggler Problem}",
      journal = {\aj},
         year = 1989,
        month = jul,
       volume = {98},
        pages = {217},
          doi = {10.1086/115138},
       adsurl = {https://ui.adsabs.harvard.edu/abs/1989AJ.....98..217L}
}

@ARTICLE{Levy2024,
       author = {{Levy}, Rebecca C. and {Bolatto}, Alberto D. and {Mayya}, Divakara and {Cuevas-Otahola}, Bolivia and {Tarantino}, Elizabeth and {Boyer}, Martha L. and {Boogaard}, Leindert A. and {B{\"o}ker}, Torsten and {Cronin}, Serena A. and {Dale}, Daniel A. and {Donaghue}, Keaton and {Emig}, Kimberly L. and {Fisher}, Deanne B. and {Glover}, Simon C.~O. and {Herrera-Camus}, Rodrigo and {Jim{\'e}nez-Donaire}, Mar{\'\i}a J. and {Klessen}, Ralf S. and {Lenki{\'c}}, Laura and {Leroy}, Adam K. and {De Looze}, Ilse and {Meier}, David S. and {Mills}, Elisabeth A.~C. and {Ott}, Juergen and {Rela{\~n}o}, M{\'o}nica and {Veilleux}, Sylvain and {Villanueva}, Vicente and {Walter}, Fabian and {van der Werf}, Paul P.},
        title = "{JWST Observations of Starbursts: Massive Star Clusters in the Central Starburst of M82}",
      journal = {\apjl},
         year = 2024,
        month = oct,
       volume = {973},
       number = {2},
          eid = {L55},
        pages = {L55},
          doi = {10.3847/2041-8213/ad7af3},
archivePrefix = {arXiv},
       eprint = {2408.04135},
 primaryClass = {astro-ph.GA},
       adsurl = {https://ui.adsabs.harvard.edu/abs/2024ApJ...973L..55L}
}

@ARTICLE{Liu2024,
       author = {{Liu}, Shuai and {Wang}, Long and {Hu}, Yi-Ming and {Tanikawa}, Ataru and {Trani}, Alessandro A.},
        title = "{Merging hierarchical triple black hole systems with intermediate-mass black holes in population III star clusters}",
      journal = {\mnras},
         year = 2024,
        month = sep,
       volume = {533},
       number = {2},
        pages = {2262-2281},
          doi = {10.1093/mnras/stae1946},
archivePrefix = {arXiv},
       eprint = {2311.05393},
 primaryClass = {astro-ph.GA},
       adsurl = {https://ui.adsabs.harvard.edu/abs/2024MNRAS.533.2262L}
}

@ARTICLE{Lombardi2002,
       author = {{Lombardi}, Jr., James C. and {Warren}, Jessica S. and {Rasio}, Frederic A. and {Sills}, Alison and {Warren}, Aaron R.},
        title = "{Stellar Collisions and the Interior Structure of Blue Stragglers}",
      journal = {\apj},
         year = 2002,
        month = apr,
       volume = {568},
       number = {2},
        pages = {939-953},
          doi = {10.1086/339060},
archivePrefix = {arXiv},
       eprint = {astro-ph/0107388},
 primaryClass = {astro-ph},
       adsurl = {https://ui.adsabs.harvard.edu/abs/2002ApJ...568..939L}
}

@ARTICLE{Lutzgendorf2011,
       author = {{L{\"u}tzgendorf}, N. and {Kissler-Patig}, M. and {Noyola}, E. and {Jalali}, B. and {de Zeeuw}, P.~T. and {Gebhardt}, K. and {Baumgardt}, H.},
        title = "{Kinematic signature of an intermediate-mass black hole in the globular cluster NGC 6388}",
      journal = {\aap},
         year = 2011,
        month = sep,
       volume = {533},
          eid = {A36},
        pages = {A36},
          doi = {10.1051/0004-6361/201116618},
archivePrefix = {arXiv},
       eprint = {1107.4243},
 primaryClass = {astro-ph.GA},
       adsurl = {https://ui.adsabs.harvard.edu/abs/2011A&A...533A..36L}
}

@ARTICLE{Lutzgendorf2013b,
       author = {{L{\"u}tzgendorf}, N. and {Baumgardt}, H. and {Kruijssen}, J.~M.~D.},
        title = "{N-body simulations of globular clusters in tidal fields: Effects of intermediate-mass black holes}",
      journal = {\aap},
         year = 2013,
        month = oct,
       volume = {558},
          eid = {A117},
        pages = {A117},
          doi = {10.1051/0004-6361/201321927},
archivePrefix = {arXiv},
       eprint = {1309.0451},
 primaryClass = {astro-ph.GA},
       adsurl = {https://ui.adsabs.harvard.edu/abs/2013A&A...558A.117L}
}

@ARTICLE{Mackey2003,
       author = {{Mackey}, A.~D. and {Gilmore}, G.~F.},
        title = "{Surface brightness profiles and structural parameters for 53 rich stellar clusters in the Large Magellanic Cloud}",
      journal = {\mnras},
         year = 2003,
        month = jan,
       volume = {338},
       number = {1},
        pages = {85-119},
          doi = {10.1046/j.1365-8711.2003.06021.x},
archivePrefix = {arXiv},
       eprint = {astro-ph/0209031},
 primaryClass = {astro-ph},
       adsurl = {https://ui.adsabs.harvard.edu/abs/2003MNRAS.338...85M}
}

@ARTICLE{Madau2001,
       author = {{Madau}, Piero and {Rees}, Martin J.},
        title = "{Massive Black Holes as Population III Remnants}",
      journal = {\apjl},
         year = 2001,
        month = apr,
       volume = {551},
       number = {1},
        pages = {L27-L30},
          doi = {10.1086/319848},
archivePrefix = {arXiv},
       eprint = {astro-ph/0101223},
 primaryClass = {astro-ph},
       adsurl = {https://ui.adsabs.harvard.edu/abs/2001ApJ...551L..27M}
}

@ARTICLE{Maiolino2024a,
       author = {{Maiolino}, Roberto and {Scholtz}, Jan and {Witstok}, Joris and {Carniani}, Stefano and {D'Eugenio}, Francesco and {de Graaff}, Anna and {{\"U}bler}, Hannah and {Tacchella}, Sandro and {Curtis-Lake}, Emma and {Arribas}, Santiago and {Bunker}, Andrew and {Charlot}, St{\'e}phane and {Chevallard}, Jacopo and {Curti}, Mirko and {Looser}, Tobias J. and {Maseda}, Michael V. and {Rawle}, Timothy D. and {Rodr{\'\i}guez del Pino}, Bruno and {Willott}, Chris J. and {Egami}, Eiichi and {Eisenstein}, Daniel J. and {Hainline}, Kevin N. and {Robertson}, Brant and {Williams}, Christina C. and {Willmer}, Christopher N.~A. and {Baker}, William M. and {Boyett}, Kristan and {DeCoursey}, Christa and {Fabian}, Andrew C. and {Helton}, Jakob M. and {Ji}, Zhiyuan and {Jones}, Gareth C. and {Kumari}, Nimisha and {Laporte}, Nicolas and {Nelson}, Erica J. and {Perna}, Michele and {Sandles}, Lester and {Shivaei}, Irene and {Sun}, Fengwu},
        title = "{A small and vigorous black hole in the early Universe}",
      journal = {\nat},
         year = 2024,
        month = mar,
       volume = {627},
       number = {8002},
        pages = {59-63},
          doi = {10.1038/s41586-024-07052-5},
archivePrefix = {arXiv},
       eprint = {2305.12492},
 primaryClass = {astro-ph.GA},
       adsurl = {https://ui.adsabs.harvard.edu/abs/2024Natur.627...59M}
}

@ARTICLE{Maiolino2024b,
       author = {{Maiolino}, Roberto and {Scholtz}, Jan and {Curtis-Lake}, Emma and {Carniani}, Stefano and {Baker}, William and {de Graaff}, Anna and {Tacchella}, Sandro and {{\"U}bler}, Hannah and {D'Eugenio}, Francesco and {Witstok}, Joris and {Curti}, Mirko and {Arribas}, Santiago and {Bunker}, Andrew J. and {Charlot}, St{\'e}phane and {Chevallard}, Jacopo and {Eisenstein}, Daniel J. and {Egami}, Eiichi and {Ji}, Zhiyuan and {Jones}, Gareth C. and {Lyu}, Jianwei and {Rawle}, Tim and {Robertson}, Brant and {Rujopakarn}, Wiphu and {Perna}, Michele and {Sun}, Fengwu and {Venturi}, Giacomo and {Williams}, Christina C. and {Willott}, Chris},
        title = "{JADES: The diverse population of infant black holes at 4 < z < 11: Merging, tiny, poor, but mighty}",
      journal = {\aap},
         year = 2024,
        month = nov,
       volume = {691},
          eid = {A145},
        pages = {A145},
          doi = {10.1051/0004-6361/202347640},
archivePrefix = {arXiv},
       eprint = {2308.01230},
 primaryClass = {astro-ph.GA},
       adsurl = {https://ui.adsabs.harvard.edu/abs/2024A&A...691A.145M}
}

@ARTICLE{Maiolino2025,
       author = {{Maiolino}, Roberto and {Uebler}, Hannah and {D'Eugenio}, Francesco and {Scholtz}, Jan and {Juodzbalis}, Ignas and {Ji}, Xihan and {Perna}, Michele and {Bromm}, Volker and {Dayal}, Pratika and {Koudmani}, Sophie and {Liu}, Boyuan and {Schneider}, Raffaella and {Sijacki}, Debora and {Valiante}, Rosa and {Trinca}, Alessandro and {Zhang}, Saiyang and {Volonteri}, Marta and {Inayoshi}, Kohei and {Carniani}, Stefano and {Nakajima}, Kimihiko and {Isobe}, Yuki and {Witstok}, Joris and {Jones}, Gareth C. and {Tacchella}, Sandro and {Arribas}, Santiago and {Bunker}, Andrew and {Cataldi}, Elisa and {Charlot}, Stephane and {Cresci}, Giovanni and {Curti}, Mirko and {Fabian}, Andrew C. and {Katz}, Harley and {Kumari}, Nimisha and {Laporte}, Nicolas and {Mazzolari}, Giovanni and {Robertson}, Brant and {Sun}, Fengwu and {Rodriguez Del Pino}, Bruno and {Venturi}, Giacomo},
        title = "{A black hole in a near-pristine galaxy 700 million years after the Big Bang}",
      journal = {arXiv e-prints},
         year = 2025,
        month = may,
          eid = {arXiv:2505.22567},
        pages = {arXiv:2505.22567},
          doi = {10.48550/arXiv.2505.22567},
archivePrefix = {arXiv},
       eprint = {2505.22567},
 primaryClass = {astro-ph.GA},
       adsurl = {https://ui.adsabs.harvard.edu/abs/2025arXiv250522567M}
}

@ARTICLE{Mannerkoski2023,
       author = {{Mannerkoski}, Matias and {Rawlings}, Alexander and {Johansson}, Peter H. and {Naab}, Thorsten and {Rantala}, Antti and {Springel}, Volker and {Irodotou}, Dimitrios and {Liao}, Shihong},
        title = "{KETJU - resolving small-scale supermassive black hole dynamics in GADGET-4}",
      journal = {\mnras},
         year = 2023,
        month = sep,
       volume = {524},
       number = {3},
        pages = {4062-4082},
          doi = {10.1093/mnras/stad2139},
archivePrefix = {arXiv},
       eprint = {2306.04963},
 primaryClass = {astro-ph.IM},
       adsurl = {https://ui.adsabs.harvard.edu/abs/2023MNRAS.524.4062M}
}

@ARTICLE{Mapelli2013,
       author = {{Mapelli}, M. and {Zampieri}, L. and {Ripamonti}, E. and {Bressan}, A.},
        title = "{Dynamics of stellar black holes in young star clusters with different metallicities - I. Implications for X-ray binaries}",
      journal = {\mnras},
         year = 2013,
        month = mar,
       volume = {429},
       number = {3},
        pages = {2298-2314},
          doi = {10.1093/mnras/sts500},
archivePrefix = {arXiv},
       eprint = {1211.6441},
 primaryClass = {astro-ph.HE},
       adsurl = {https://ui.adsabs.harvard.edu/abs/2013MNRAS.429.2298M}
}

@ARTICLE{Mapelli2013b,
       author = {{Mapelli}, M. and {Bressan}, A.},
        title = "{Impact of metallicity on the evolution of young star clusters}",
      journal = {\mnras},
         year = 2013,
        month = apr,
       volume = {430},
       number = {4},
        pages = {3120-3127},
          doi = {10.1093/mnras/stt119},
archivePrefix = {arXiv},
       eprint = {1301.4227},
 primaryClass = {astro-ph.GA},
       adsurl = {https://ui.adsabs.harvard.edu/abs/2013MNRAS.430.3120M}
}

@ARTICLE{Mapelli2016,
       author = {{Mapelli}, Michela},
        title = "{Massive black hole binaries from runaway collisions: the impact of metallicity}",
      journal = {\mnras},
         year = 2016,
        month = jul,
       volume = {459},
       number = {4},
        pages = {3432-3446},
          doi = {10.1093/mnras/stw869},
archivePrefix = {arXiv},
       eprint = {1604.03559},
 primaryClass = {astro-ph.GA},
       adsurl = {https://ui.adsabs.harvard.edu/abs/2016MNRAS.459.3432M}
}

@ARTICLE{Mapelli2020,
       author = {{Mapelli}, Michela and {Spera}, Mario and {Montanari}, Enrico and {Limongi}, Marco and {Chieffi}, Alessandro and {Giacobbo}, Nicola and {Bressan}, Alessandro and {Bouffanais}, Yann},
        title = "{Impact of the Rotation and Compactness of Progenitors on the Mass of Black Holes}",
      journal = {\apj},
         year = 2020,
        month = jan,
       volume = {888},
       number = {2},
          eid = {76},
        pages = {76},
          doi = {10.3847/1538-4357/ab584d},
archivePrefix = {arXiv},
       eprint = {1909.01371},
 primaryClass = {astro-ph.HE},
       adsurl = {https://ui.adsabs.harvard.edu/abs/2020ApJ...888...76M}
}

@ARTICLE{Marks2012b,
       author = {{Marks}, Michael and {Kroupa}, Pavel and {Dabringhausen}, J{\"o}rg and {Pawlowski}, Marcel S.},
        title = "{Evidence for top-heavy stellar initial mass functions with increasing density and decreasing metallicity}",
      journal = {\mnras},
         year = 2012,
        month = may,
       volume = {422},
       number = {3},
        pages = {2246-2254},
          doi = {10.1111/j.1365-2966.2012.20767.x},
archivePrefix = {arXiv},
       eprint = {1202.4755},
 primaryClass = {astro-ph.GA},
       adsurl = {https://ui.adsabs.harvard.edu/abs/2012MNRAS.422.2246M}
}

@ARTICLE{Marks2012,
       author = {{Marks}, M. and {Kroupa}, P.},
        title = "{Inverse dynamical population synthesis. Constraining the initial conditions of young stellar clusters by studying their binary populations}",
      journal = {\aap},
         year = 2012,
        month = jul,
       volume = {543},
          eid = {A8},
        pages = {A8},
          doi = {10.1051/0004-6361/201118231},
archivePrefix = {arXiv},
       eprint = {1205.1508},
 primaryClass = {astro-ph.GA},
       adsurl = {https://ui.adsabs.harvard.edu/abs/2012A&A...543A...8M}
}

@ARTICLE{MarquesChaves2024,
       author = {{Marques-Chaves}, R. and {Schaerer}, D. and {Kuruvanthodi}, A. and {Korber}, D. and {Prantzos}, N. and {Charbonnel}, C. and {Weibel}, A. and {Izotov}, Y.~I. and {Messa}, M. and {Brammer}, G. and {Dessauges-Zavadsky}, M. and {Oesch}, P.},
        title = "{Extreme N-emitters at high redshift: Possible signatures of supermassive stars and globular cluster or black hole formation in action}",
      journal = {\aap},
         year = 2024,
        month = jan,
       volume = {681},
          eid = {A30},
        pages = {A30},
          doi = {10.1051/0004-6361/202347411},
archivePrefix = {arXiv},
       eprint = {2307.04234},
 primaryClass = {astro-ph.GA},
       adsurl = {https://ui.adsabs.harvard.edu/abs/2024A&A...681A..30M}
}

@ARTICLE{Matsumoto2001,
       author = {{Matsumoto}, H. and {Tsuru}, T.~G. and {Koyama}, K. and {Awaki}, H. and {Canizares}, C.~R. and {Kawai}, N. and {Matsushita}, S. and {Kawabe}, R.},
        title = "{Discovery of a Luminous, Variable, Off-Center Source in the Nucleus of M82 with the Chandra High-Resolution Camera}",
      journal = {\apjl},
         year = 2001,
        month = jan,
       volume = {547},
       number = {1},
        pages = {L25-L28},
          doi = {10.1086/318878},
archivePrefix = {arXiv},
       eprint = {astro-ph/0009250},
 primaryClass = {astro-ph},
       adsurl = {https://ui.adsabs.harvard.edu/abs/2001ApJ...547L..25M}
}

@ARTICLE{Mazzolo2014,
       author = {{Mazzolo}, G. and {Salemi}, F. and {Drago}, M. and {Necula}, V. and {Pankow}, C. and {Prodi}, G.~A. and {Re}, V. and {Tiwari}, V. and {Vedovato}, G. and {Yakushin}, I. and {Klimenko}, S.},
        title = "{Prospects for intermediate mass black hole binary searches with advanced gravitational-wave detectors}",
      journal = {\prd},
         year = 2014,
        month = sep,
       volume = {90},
       number = {6},
          eid = {063002},
        pages = {063002},
          doi = {10.1103/PhysRevD.90.063002},
archivePrefix = {arXiv},
       eprint = {1404.7757},
 primaryClass = {gr-qc},
       adsurl = {https://ui.adsabs.harvard.edu/abs/2014PhRvD..90f3002M}
}

@ARTICLE{McLaughlin2005,
       author = {{McLaughlin}, Dean E. and {van der Marel}, Roeland P.},
        title = "{Resolved Massive Star Clusters in the Milky Way and Its Satellites: Brightness Profiles and a Catalog of Fundamental Parameters}",
      journal = {\apjs},
         year = 2005,
        month = dec,
       volume = {161},
       number = {2},
        pages = {304-360},
          doi = {10.1086/497429},
archivePrefix = {arXiv},
       eprint = {astro-ph/0605132},
 primaryClass = {astro-ph},
       adsurl = {https://ui.adsabs.harvard.edu/abs/2005ApJS..161..304M}
}

@ARTICLE{Mehta2022,
       author = {{Mehta}, Ajit Kumar and {Buonanno}, Alessandra and {Gair}, Jonathan and {Miller}, M. Coleman and {Farag}, Ebraheem and {deBoer}, R.~J. and {Wiescher}, M. and {Timmes}, F.~X.},
        title = "{Observing Intermediate-mass Black Holes and the Upper Stellar-mass gap with LIGO and Virgo}",
      journal = {\apj},
         year = 2022,
        month = jan,
       volume = {924},
       number = {1},
          eid = {39},
        pages = {39},
          doi = {10.3847/1538-4357/ac3130},
archivePrefix = {arXiv},
       eprint = {2105.06366},
 primaryClass = {gr-qc},
       adsurl = {https://ui.adsabs.harvard.edu/abs/2022ApJ...924...39M}
}

@ARTICLE{Mezcua2013,
       author = {{Mezcua}, M. and {Roberts}, T.~P. and {Sutton}, A.~D. and {Lobanov}, A.~P.},
        title = "{Radio observations of extreme ULXs: revealing the most powerful ULX radio nebula ever or the jet of an intermediate-mass black hole?}",
      journal = {\mnras},
         year = 2013,
        month = dec,
       volume = {436},
       number = {4},
        pages = {3128-3134},
          doi = {10.1093/mnras/stt1794},
archivePrefix = {arXiv},
       eprint = {1309.5721},
 primaryClass = {astro-ph.HE},
       adsurl = {https://ui.adsabs.harvard.edu/abs/2013MNRAS.436.3128M}
}

@ARTICLE{Mezcua2015,
       author = {{Mezcua}, M. and {Roberts}, T.~P. and {Lobanov}, A.~P. and {Sutton}, A.~D.},
        title = "{The powerful jet of an off-nuclear intermediate-mass black hole in the spiral galaxy NGC 2276}",
      journal = {\mnras},
         year = 2015,
        month = apr,
       volume = {448},
       number = {2},
        pages = {1893-1899},
          doi = {10.1093/mnras/stv143},
archivePrefix = {arXiv},
       eprint = {1501.04897},
 primaryClass = {astro-ph.GA},
       adsurl = {https://ui.adsabs.harvard.edu/abs/2015MNRAS.448.1893M}
}

@ARTICLE{Mezcua2017,
       author = {{Mezcua}, Mar},
        title = "{Observational evidence for intermediate-mass black holes}",
      journal = {International Journal of Modern Physics D},
         year = 2017,
        month = jan,
       volume = {26},
       number = {11},
          eid = {1730021},
        pages = {1730021},
          doi = {10.1142/S021827181730021X},
archivePrefix = {arXiv},
       eprint = {1705.09667},
 primaryClass = {astro-ph.GA},
       adsurl = {https://ui.adsabs.harvard.edu/abs/2017IJMPD..2630021M}
}

@ARTICLE{Mezcua2018,
       author = {{Mezcua}, M. and {Civano}, F. and {Marchesi}, S. and {Suh}, H. and {Fabbiano}, G. and {Volonteri}, M.},
        title = "{Intermediate-mass black holes in dwarf galaxies out to redshift {\ensuremath{\sim}}2.4 in the Chandra COSMOS-Legacy Survey}",
      journal = {\mnras},
         year = 2018,
        month = aug,
       volume = {478},
       number = {2},
        pages = {2576-2591},
          doi = {10.1093/mnras/sty1163},
archivePrefix = {arXiv},
       eprint = {1802.01567},
 primaryClass = {astro-ph.GA},
       adsurl = {https://ui.adsabs.harvard.edu/abs/2018MNRAS.478.2576M}
}

@ARTICLE{Miller2002,
       author = {{Miller}, M. Coleman and {Hamilton}, Douglas P.},
        title = "{Production of intermediate-mass black holes in globular clusters}",
      journal = {\mnras},
         year = 2002,
        month = feb,
       volume = {330},
       number = {1},
        pages = {232-240},
          doi = {10.1046/j.1365-8711.2002.05112.x},
archivePrefix = {arXiv},
       eprint = {astro-ph/0106188},
 primaryClass = {astro-ph},
       adsurl = {https://ui.adsabs.harvard.edu/abs/2002MNRAS.330..232C}
}

@ARTICLE{Moe2017,
       author = {{Moe}, Maxwell and {Di Stefano}, Rosanne},
        title = "{Mind Your Ps and Qs: The Interrelation between Period (P) and Mass-ratio (Q) Distributions of Binary Stars}",
      journal = {\apjs},
         year = 2017,
        month = jun,
       volume = {230},
       number = {2},
          eid = {15},
        pages = {15},
          doi = {10.3847/1538-4365/aa6fb6},
archivePrefix = {arXiv},
       eprint = {1606.05347},
 primaryClass = {astro-ph.SR},
       adsurl = {https://ui.adsabs.harvard.edu/abs/2017ApJS..230...15M}
}

@ARTICLE{Mowla2024,
       author = {{Mowla}, Lamiya and {Iyer}, Kartheik and {Asada}, Yoshihisa and {Desprez}, Guillaume and {Tan}, Vivian Yun Yan and {Martis}, Nicholas and {Sarrouh}, Ghassan and {Strait}, Victoria and {Abraham}, Roberto and {Brada{\v{c}}}, Maru{\v{s}}a and {Brammer}, Gabriel and {Muzzin}, Adam and {Pacifici}, Camilla and {Ravindranath}, Swara and {Sawicki}, Marcin and {Willott}, Chris and {Estrada-Carpenter}, Vince and {Jahan}, Nusrath and {Noirot}, Ga{\"e}l and {Matharu}, Jasleen and {Rihtar{\v{s}}i{\v{c}}}, Gregor and {Zabl}, Johannes},
        title = "{Formation of a low-mass galaxy from star clusters in a 600-million-year-old Universe}",
      journal = {\nat},
         year = 2024,
        month = dec,
       volume = {636},
       number = {8042},
        pages = {332-336},
          doi = {10.1038/s41586-024-08293-0},
archivePrefix = {arXiv},
       eprint = {2402.08696},
 primaryClass = {astro-ph.GA},
       adsurl = {https://ui.adsabs.harvard.edu/abs/2024Natur.636..332M}
}

@ARTICLE{Muijres2012,
       author = {{Muijres}, L. and {Vink}, J.~S. and {de Koter}, A. and {Hirschi}, R. and {Langer}, N. and {Yoon}, S.-C.},
        title = "{Mass-loss predictions for evolved very metal-poor massive stars}",
      journal = {\aap},
         year = 2012,
        month = oct,
       volume = {546},
          eid = {A42},
        pages = {A42},
          doi = {10.1051/0004-6361/201118666},
archivePrefix = {arXiv},
       eprint = {1209.5934},
 primaryClass = {astro-ph.SR},
       adsurl = {https://ui.adsabs.harvard.edu/abs/2012A&A...546A..42M}
}

@ARTICLE{Naab2009,
       author = {{Naab}, Thorsten and {Johansson}, Peter H. and {Ostriker}, Jeremiah P.},
        title = "{Minor Mergers and the Size Evolution of Elliptical Galaxies}",
      journal = {\apjl},
         year = 2009,
        month = jul,
       volume = {699},
       number = {2},
        pages = {L178-L182},
          doi = {10.1088/0004-637X/699/2/L178},
archivePrefix = {arXiv},
       eprint = {0903.1636},
 primaryClass = {astro-ph.CO},
       adsurl = {https://ui.adsabs.harvard.edu/abs/2009ApJ...699L.178N}
}

@ARTICLE{Naidu2025,
       author = {{Naidu}, Rohan P. and {Oesch}, Pascal A. and {Brammer}, Gabriel and {Weibel}, Andrea and {Li}, Yijia and {Matthee}, Jorryt and {Chisholm}, John and {Pollock}, Clara L. and {Heintz}, Kasper E. and {Johnson}, Benjamin D. and {Shen}, Xuejian and {Hviding}, Raphael E. and {Leja}, Joel and {Tacchella}, Sandro and {Ganguly}, Arpita and {Witten}, Callum and {Atek}, Hakim and {Belli}, Sirio and {Bose}, Sownak and {Bouwens}, Rychard and {Dayal}, Pratika and {Decarli}, Roberto and {de Graaff}, Anna and {Fudamoto}, Yoshinobu and {Giovinazzo}, Emma and {Greene}, Jenny E. and {Illingworth}, Garth and {Inoue}, Akio K. and {Kane}, Sarah G. and {Labbe}, Ivo and {Leonova}, Ecaterina and {Marques-Chaves}, Rui and {Meyer}, Romain A. and {Nelson}, Erica J. and {Roberts-Borsani}, Guido and {Schaerer}, Daniel and {Simcoe}, Robert A. and {Stefanon}, Mauro and {Sugahara}, Yuma and {Toft}, Sune and {van der Wel}, Arjen and {van Dokkum}, Pieter and {Walter}, Fabian and {Watson}, Darach and {Weaver}, John R. and {Whitaker}, Katherine E.},
        title = "{A Cosmic Miracle: A Remarkably Luminous Galaxy at $z_{\rm{spec}}=14.44$ Confirmed with JWST}",
      journal = {arXiv e-prints},
         year = 2025,
        month = may,
          eid = {arXiv:2505.11263},
        pages = {arXiv:2505.11263},
          doi = {10.48550/arXiv.2505.11263},
archivePrefix = {arXiv},
       eprint = {2505.11263},
 primaryClass = {astro-ph.GA},
       adsurl = {https://ui.adsabs.harvard.edu/abs/2025arXiv250511263N}
}

@ARTICLE{Nandal2025,
       author = {{Nandal}, Devesh and {Chon}, Sunmyon},
        title = "{Growth of Metal-Enriched Supermassive Stars by Accretion and Collisions}",
      journal = {arXiv e-prints},
         year = 2025,
        month = nov,
          eid = {arXiv:2511.08516},
        pages = {arXiv:2511.08516},
          doi = {10.48550/arXiv.2511.08516},
archivePrefix = {arXiv},
       eprint = {2511.08516},
 primaryClass = {astro-ph.SR},
       adsurl = {https://ui.adsabs.harvard.edu/abs/2025arXiv251108516N}
}

@ARTICLE{Nguyen2022,
       author = {{Nguyen}, C.~T. and {Costa}, G. and {Girardi}, L. and {Volpato}, G. and {Bressan}, A. and {Chen}, Y. and {Marigo}, P. and {Fu}, X. and {Goudfrooij}, P.},
        title = "{PARSEC V2.0: Stellar tracks and isochrones of low- and intermediate-mass stars with rotation}",
      journal = {\aap},
         year = 2022,
        month = sep,
       volume = {665},
          eid = {A126},
        pages = {A126},
          doi = {10.1051/0004-6361/202244166},
archivePrefix = {arXiv},
       eprint = {2207.08642},
 primaryClass = {astro-ph.SR},
       adsurl = {https://ui.adsabs.harvard.edu/abs/2022A&A...665A.126N}
}

@ARTICLE{Niuwenhuijzen1990,
       author = {{Nieuwenhuijzen}, H. and {de Jager}, C.},
        title = "{Parametrization of stellar rates of mass loss as functions of the fundamental stellar parameters M, L, and R.}",
      journal = {\aap},
         year = 1990,
        month = may,
       volume = {231},
        pages = {134-136},
       adsurl = {https://ui.adsabs.harvard.edu/abs/1990A&A...231..134N}
}

@ARTICLE{Noyola2010,
       author = {{Noyola}, Eva and {Gebhardt}, Karl and {Kissler-Patig}, Markus and {L{\"u}tzgendorf}, Nora and {Jalali}, Behrang and {de Zeeuw}, P. Tim and {Baumgardt}, Holger},
        title = "{Very Large Telescope Kinematics for Omega Centauri: Further Support for a Central Black Hole}",
      journal = {\apjl},
         year = 2010,
        month = aug,
       volume = {719},
       number = {1},
        pages = {L60-L64},
          doi = {10.1088/2041-8205/719/1/L60},
archivePrefix = {arXiv},
       eprint = {1007.4559},
 primaryClass = {astro-ph.GA},
       adsurl = {https://ui.adsabs.harvard.edu/abs/2010ApJ...719L..60N}
}

@INPROCEEDINGS{Offner2023,
       author = {{Offner}, S.~S.~R. and {Moe}, M. and {Kratter}, K.~M. and {Sadavoy}, S.~I. and {Jensen}, E.~L.~N. and {Tobin}, J.~J.},
        title = "{The Origin and Evolution of Multiple Star Systems}",
    booktitle = {Protostars and Planets VII},
         year = 2023,
       editor = {{Inutsuka}, S. and {Aikawa}, Y. and {Muto}, T. and {Tomida}, K. and {Tamura}, M.},
       series = {Astronomical Society of the Pacific Conference Series},
       volume = {534},
        month = jul,
        pages = {275},
          doi = {10.48550/arXiv.2203.10066},
archivePrefix = {arXiv},
       eprint = {2203.10066},
 primaryClass = {astro-ph.SR},
       adsurl = {https://ui.adsabs.harvard.edu/abs/2023ASPC..534..275O}
}

@ARTICLE{Omelyan2006,
       author = {{Omelyan}, I.~P.},
        title = "{Extrapolated gradientlike algorithms for molecular dynamics and celestial mechanics simulations}",
      journal = {\pre},
         year = 2006,
        month = sep,
       volume = {74},
       number = {3},
          eid = {036703},
        pages = {036703},
          doi = {10.1103/PhysRevE.74.036703},
       adsurl = {https://ui.adsabs.harvard.edu/abs/2006PhRvE..74c6703O}
}

@ARTICLE{Paiella2025,
       author = {{Paiella}, Lavinia and {Arca Sedda}, Manuel and {Mestichelli}, Benedetta and {Ugolini}, Cristiano},
        title = "{Seeds to success: growing heavy black holes in dense star clusters}",
      journal = {arXiv e-prints},
         year = 2025,
        month = oct,
          eid = {arXiv:2511.00200},
        pages = {arXiv:2511.00200},
          doi = {10.48550/arXiv.2511.00200},
archivePrefix = {arXiv},
       eprint = {2511.00200},
 primaryClass = {astro-ph.GA},
       adsurl = {https://ui.adsabs.harvard.edu/abs/2025arXiv251100200P}
}

@ARTICLE{Pakmor2022,
       author = {{Pakmor}, R{\"u}diger and {Simpson}, Christine M. and {van de Voort}, Freeke and {Hernquist}, Lars and {van Son}, Lieke and {Chru{\'s}li{\'n}ska}, Martyna and {Bieri}, Rebekka and {de Mink}, Selma E. and {Springel}, Volker},
        title = "{Formation and fate of low-metallicity stars in TNG50}",
      journal = {\mnras},
         year = 2022,
        month = may,
       volume = {512},
       number = {3},
        pages = {3602-3615},
          doi = {10.1093/mnras/stac717},
archivePrefix = {arXiv},
       eprint = {2203.07383},
 primaryClass = {astro-ph.GA},
       adsurl = {https://ui.adsabs.harvard.edu/abs/2022MNRAS.512.3602P}
}

@ARTICLE{Partmann2025,
       author = {{Partmann}, Christian and {Naab}, Thorsten and {Lah{\'e}n}, Natalia and {Rantala}, Antti and {Hirschmann}, Michaela and {Hislop}, Jessica M. and {Petersson}, Jonathan and {Johansson}, Peter H.},
        title = "{The importance of nuclear star clusters for massive black hole growth and nuclear star formation in simulated low-mass galaxies}",
      journal = {\mnras},
         year = 2025,
        month = feb,
       volume = {537},
       number = {2},
        pages = {956-977},
          doi = {10.1093/mnras/staf002},
archivePrefix = {arXiv},
       eprint = {2409.18096},
 primaryClass = {astro-ph.GA},
       adsurl = {https://ui.adsabs.harvard.edu/abs/2025MNRAS.537..956P}
}

@ARTICLE{Pasham2014,
       author = {{Pasham}, Dheeraj R. and {Strohmayer}, Tod E. and {Mushotzky}, Richard F.},
        title = "{A 400-solar-mass black hole in the galaxy M82}",
      journal = {\nat},
         year = 2014,
        month = sep,
       volume = {513},
       number = {7516},
        pages = {74-76},
          doi = {10.1038/nature13710},
archivePrefix = {arXiv},
       eprint = {1501.03180},
 primaryClass = {astro-ph.HE},
       adsurl = {https://ui.adsabs.harvard.edu/abs/2014Natur.513...74P}
}

@ARTICLE{Patruno2006,
       author = {{Patruno}, A. and {Portegies Zwart}, S. and {Dewi}, J. and {Hopman}, C.},
        title = "{The ultraluminous X-ray source in M82: an intermediate-mass black hole with a giant companion}",
      journal = {\mnras},
         year = 2006,
        month = jul,
       volume = {370},
       number = {1},
        pages = {L6-L9},
          doi = {10.1111/j.1745-3933.2006.00176.x},
archivePrefix = {arXiv},
       eprint = {astro-ph/0602230},
 primaryClass = {astro-ph},
       adsurl = {https://ui.adsabs.harvard.edu/abs/2006MNRAS.370L...6P}
}

@ARTICLE{Pauldrach2012,
       author = {{Pauldrach}, A.~W.~A. and {Vanbeveren}, D. and {Hoffmann}, T.~L.},
        title = "{Radiation-driven winds of hot luminous stars XVI. Expanding atmospheres of massive and very massive stars and the evolution of dense stellar clusters}",
      journal = {\aap},
         year = 2012,
        month = feb,
       volume = {538},
          eid = {A75},
        pages = {A75},
          doi = {10.1051/0004-6361/201117621},
archivePrefix = {arXiv},
       eprint = {1107.0654},
 primaryClass = {astro-ph.SR},
       adsurl = {https://ui.adsabs.harvard.edu/abs/2012A&A...538A..75P}
}

@ARTICLE{Pechetti2022,
       author = {{Pechetti}, Renuka and {Seth}, Anil and {Kamann}, Sebastian and {Caldwell}, Nelson and {Strader}, Jay and {den Brok}, Mark and {Luetzgendorf}, Nora and {Neumayer}, Nadine and {Voggel}, Karina},
        title = "{Detection of a 100,000 M $_{{\ensuremath{\odot}}}$ black hole in M31's Most Massive Globular Cluster: A Tidally Stripped Nucleus}",
      journal = {\apj},
         year = 2022,
        month = jan,
       volume = {924},
       number = {2},
          eid = {48},
        pages = {48},
          doi = {10.3847/1538-4357/ac339f},
archivePrefix = {arXiv},
       eprint = {2111.08720},
 primaryClass = {astro-ph.GA},
       adsurl = {https://ui.adsabs.harvard.edu/abs/2022ApJ...924...48P}
}

@ARTICLE{Petersson2025,
       author = {{Petersson}, Jonathan and {Hirschmann}, Michaela and {Tress}, Robin G. and {Farcy}, Marion and {Glover}, Simon C.~O. and {Klessen}, Ralf S. and {Naab}, Thorsten and {Partmann}, Christian and {Whitworth}, David J.},
        title = "{The Noctua Suite of Simulations -- The Difficulty of Growing Massive Black Holes in Low-Mass Dwarf Galaxies}",
      journal = {arXiv e-prints},
         year = 2025,
        month = apr,
          eid = {arXiv:2504.08035},
        pages = {arXiv:2504.08035},
          doi = {10.48550/arXiv.2504.08035},
archivePrefix = {arXiv},
       eprint = {2504.08035},
 primaryClass = {astro-ph.GA},
       adsurl = {https://ui.adsabs.harvard.edu/abs/2025arXiv250408035P}
}

@ARTICLE{Plummer1911,
       author = {{Plummer}, H.~C.},
        title = "{On the problem of distribution in globular star clusters}",
      journal = {\mnras},
         year = "1911",
        month = "Mar",
       volume = {71},
        pages = {460-470},
          doi = {10.1093/mnras/71.5.460},
       adsurl = {https://ui.adsabs.harvard.edu/abs/1911MNRAS..71..460P}
}

@ARTICLE{Puls2008,
       author = {{Puls}, Joachim and {Vink}, Jorick S. and {Najarro}, Francisco},
        title = "{Mass loss from hot massive stars}",
      journal = {\aapr},
         year = 2008,
        month = dec,
       volume = {16},
       number = {3-4},
        pages = {209-325},
          doi = {10.1007/s00159-008-0015-8},
archivePrefix = {arXiv},
       eprint = {0811.0487},
 primaryClass = {astro-ph},
       adsurl = {https://ui.adsabs.harvard.edu/abs/2008A&ARv..16..209P}
}

@ARTICLE{PortegiesZwart1999,
       author = {{Portegies Zwart}, S.~F. and {Makino}, J. and {McMillan}, S.~L.~W. and {Hut}, P.},
        title = "{Star cluster ecology. III. Runaway collisions in young compact star clusters}",
      journal = {\aap},
         year = 1999,
        month = aug,
       volume = {348},
        pages = {117-126},
          doi = {10.48550/arXiv.astro-ph/9812006},
archivePrefix = {arXiv},
       eprint = {astro-ph/9812006},
 primaryClass = {astro-ph},
       adsurl = {https://ui.adsabs.harvard.edu/abs/1999A&A...348..117P}
}

@ARTICLE{PortegiesZwart2004,
       author = {{Portegies Zwart}, Simon F. and {Baumgardt}, Holger and {Hut}, Piet and {Makino}, Junichiro and {McMillan}, Stephen L.~W.},
        title = "{Formation of massive black holes through runaway collisions in dense young star clusters}",
      journal = {\nat},
         year = 2004,
        month = apr,
       volume = {428},
       number = {6984},
        pages = {724-726},
          doi = {10.1038/nature02448},
archivePrefix = {arXiv},
       eprint = {astro-ph/0402622},
 primaryClass = {astro-ph},
       adsurl = {https://ui.adsabs.harvard.edu/abs/2004Natur.428..724P}
}

@ARTICLE{PortegiesZwart2010,
       author = {{Portegies Zwart}, Simon F. and {McMillan}, Stephen L.~W. and {Gieles}, Mark},
        title = "{Young Massive Star Clusters}",
      journal = {\araa},
         year = 2010,
        month = sep,
       volume = {48},
        pages = {431-493},
          doi = {10.1146/annurev-astro-081309-130834},
archivePrefix = {arXiv},
       eprint = {1002.1961},
 primaryClass = {astro-ph.GA},
       adsurl = {https://ui.adsabs.harvard.edu/abs/2010ARA&A..48..431P}
}

@ARTICLE{Press1977,
       author = {{Press}, W.~H. and {Teukolsky}, S.~A.},
        title = "{On formation of close binaries by two-body tidal capture.}",
      journal = {\apj},
         year = 1977,
        month = apr,
       volume = {213},
        pages = {183-192},
          doi = {10.1086/155143},
       adsurl = {https://ui.adsabs.harvard.edu/abs/1977ApJ...213..183P}
}

@ARTICLE{RamirezGaleano2025,
       author = {{Ram{\'\i}rez-Galeano}, Laura and {Charbonnel}, Corinne and {Fragos}, Tassos and {Tazakkati}, Zouba{\"\i}r and {Roman-Garza}, Jaime and {Gieles}, Mark},
        title = "{Collision-induced mass loss and mass gain on an extremely massive star: An analytical approach and a static proto-globular cluster test-case}",
      journal = {\aap},
         year = 2025,
        month = jul,
       volume = {699},
          eid = {A223},
        pages = {A223},
          doi = {10.1051/0004-6361/202453462},
archivePrefix = {arXiv},
       eprint = {2506.12132},
 primaryClass = {astro-ph.SR},
       adsurl = {https://ui.adsabs.harvard.edu/abs/2025A&A...699A.223R}
}

@ARTICLE{Rantala2017,
   author = {{Rantala}, A. and {Pihajoki}, P. and {Johansson}, P.~H. and 
	{Naab}, T. and {Lah{\'e}n}, N. and {Sawala}, T.},
    title = "{Post-Newtonian Dynamical Modeling of Supermassive Black Holes in Galactic-scale Simulations}",
  journal = {\apj},
archivePrefix = "arXiv",
   eprint = {1611.07028},
     year = 2017,
    month = may,
   volume = 840,
      eid = {53},
    pages = {53},
      doi = {10.3847/1538-4357/aa6d65},
   adsurl = {https://ui.adsabs.harvard.edu/abs/2017ApJ...840...53R}
}

@ARTICLE{Rantala2020,
       author = {{Rantala}, Antti and {Pihajoki}, Pauli and {Mannerkoski}, Matias and
         {Johansson}, Peter H. and {Naab}, Thorsten},
        title = "{MSTAR - a fast parallelized algorithmically regularized integrator with minimum spanning tree coordinates}",
      journal = {\mnras},
         year = 2020,
        month = mar,
       volume = {492},
       number = {3},
        pages = {4131-4148},
          doi = {10.1093/mnras/staa084},
archivePrefix = {arXiv},
       eprint = {2001.03180},
 primaryClass = {astro-ph.IM},
       adsurl = {https://ui.adsabs.harvard.edu/abs/2020MNRAS.492.4131R}
}

@ARTICLE{Rantala2021,
       author = {{Rantala}, Antti and {Naab}, Thorsten and {Springel}, Volker},
        title = "{frost: a momentum-conserving CUDA implementation of a hierarchical fourth-order forward symplectic integrator}",
      journal = {\mnras},
         year = 2021,
        month = apr,
       volume = {502},
       number = {4},
        pages = {5546-5562},
          doi = {10.1093/mnras/stab057},
archivePrefix = {arXiv},
       eprint = {2011.14984},
 primaryClass = {astro-ph.IM},
       adsurl = {https://ui.adsabs.harvard.edu/abs/2021MNRAS.502.5546R}
}

@ARTICLE{Rantala2023,
       author = {{Rantala}, Antti and {Naab}, Thorsten and {Rizzuto}, Francesco Paolo and {Mannerkoski}, Matias and {Partmann}, Christian and {Lautensch{\"u}tz}, Kristina},
        title = "{BIFROST: simulating compact subsystems in star clusters using a hierarchical fourth-order forward symplectic integrator code}",
      journal = {\mnras},
         year = 2023,
        month = jul,
       volume = {522},
       number = {4},
        pages = {5180-5203},
          doi = {10.1093/mnras/stad1360},
archivePrefix = {arXiv},
       eprint = {2210.02472},
 primaryClass = {astro-ph.IM},
       adsurl = {https://ui.adsabs.harvard.edu/abs/2023MNRAS.522.5180R}
}

@ARTICLE{Rantala2024b,
       author = {{Rantala}, Antti and {Naab}, Thorsten and {Lah{\'e}n}, Natalia},
        title = "{FROST-CLUSTERS - I. Hierarchical star cluster assembly boosts intermediate-mass black hole formation}",
      journal = {\mnras},
         year = 2024,
        month = jul,
       volume = {531},
       number = {3},
        pages = {3770-3799},
          doi = {10.1093/mnras/stae1413},
archivePrefix = {arXiv},
       eprint = {2403.10602},
 primaryClass = {astro-ph.GA},
       adsurl = {https://ui.adsabs.harvard.edu/abs/2024MNRAS.531.3770R}
}

@ARTICLE{Rantala2025a,
       author = {{Rantala}, Antti and {Naab}, Thorsten},
        title = "{A rapid channel for the collisional formation and gravitational wave driven mergers of supermassive black hole seeds at high redshift}",
      journal = {arXiv e-prints},
         year = 2025,
        month = mar,
          eid = {arXiv:2503.21879},
        pages = {arXiv:2503.21879},
          doi = {10.48550/arXiv.2503.21879},
archivePrefix = {arXiv},
       eprint = {2503.21879},
 primaryClass = {astro-ph.GA},
       adsurl = {https://ui.adsabs.harvard.edu/abs/2025arXiv250321879R}
}

@ARTICLE{Rantala2025b,
       author = {{Rantala}, Antti and {Lah{\'e}n}, Natalia and {Naab}, Thorsten and {Escobar}, Gast{\'o}n J. and {Iorio}, Giuliano},
        title = "{FROST-CLUSTERS {\textendash} II. Massive stars, binaries, and triples boost supermassive black hole seed formation in assembling star clusters}",
      journal = {\mnras},
         year = 2025,
        month = nov,
       volume = {543},
       number = {3},
        pages = {2130-2158},
          doi = {10.1093/mnras/staf1519},
archivePrefix = {arXiv},
       eprint = {2506.04330},
 primaryClass = {astro-ph.GA},
       adsurl = {https://ui.adsabs.harvard.edu/abs/2025MNRAS.543.2130R}
}

@ARTICLE{Rantala2026b,
       author = {{Rantala}, Antti},
        title = "{Supermassive stars with embedded stellar black hole cores: dense assembling star clusters as faint multiple Little Red Dot systems}",
      journal = {arXiv e-prints},
         year = 2026,
        month = apr,
          eid = {arXiv:2604.22924},
        pages = {arXiv:2604.22924},
          doi = {10.48550/arXiv.2604.22924},
archivePrefix = {arXiv},
       eprint = {2604.22924},
 primaryClass = {astro-ph.GA},
       adsurl = {https://ui.adsabs.harvard.edu/abs/2026arXiv260422924R}
}

@ARTICLE{Rastello2025,
       author = {{Rastello}, Sara and {Iorio}, Giuliano and {Gieles}, Mark and {Wang}, Long},
        title = "{Micro-Tidal Disruption Events in Young Star Clusters}",
      journal = {arXiv e-prints},
         year = 2025,
        month = sep,
          eid = {arXiv:2509.07067},
        pages = {arXiv:2509.07067},
          doi = {10.48550/arXiv.2509.07067},
archivePrefix = {arXiv},
       eprint = {2509.07067},
 primaryClass = {astro-ph.HE},
       adsurl = {https://ui.adsabs.harvard.edu/abs/2025arXiv250907067R}
}

@ARTICLE{Reali2024,
       author = {{Reali}, Luca and {Cotesta}, Roberto and {Antonelli}, Andrea and {Kritos}, Konstantinos and {Strokov}, Vladimir and {Berti}, Emanuele},
        title = "{Intermediate-mass black hole binary parameter estimation with next-generation ground-based detector networks}",
      journal = {\prd},
         year = 2024,
        month = nov,
       volume = {110},
       number = {10},
          eid = {103002},
        pages = {103002},
          doi = {10.1103/PhysRevD.110.103002},
archivePrefix = {arXiv},
       eprint = {2406.01687},
 primaryClass = {gr-qc},
       adsurl = {https://ui.adsabs.harvard.edu/abs/2024PhRvD.110j3002R}
}

@ARTICLE{Rees1984,
       author = {{Rees}, Martin J.},
        title = "{Black Hole Models for Active Galactic Nuclei}",
      journal = {\araa},
         year = 1984,
        month = jan,
       volume = {22},
        pages = {471-506},
          doi = {10.1146/annurev.aa.22.090184.002351},
       adsurl = {https://ui.adsabs.harvard.edu/abs/1984ARA&A..22..471R}
}

@ARTICLE{Regan2017,
       author = {{Regan}, John A. and {Visbal}, Eli and {Wise}, John H. and {Haiman}, Zolt{\'a}n and {Johansson}, Peter H. and {Bryan}, Greg L.},
        title = "{Rapid formation of massive black holes in close proximity to embryonic protogalaxies}",
      journal = {Nature Astronomy},
         year = 2017,
        month = mar,
       volume = {1},
          eid = {0075},
        pages = {0075},
          doi = {10.1038/s41550-017-0075},
archivePrefix = {arXiv},
       eprint = {1703.03805},
 primaryClass = {astro-ph.GA},
       adsurl = {https://ui.adsabs.harvard.edu/abs/2017NatAs...1E..75R}
}

@ARTICLE{Reines2022,
       author = {{Reines}, Amy E.},
        title = "{Hunting for massive black holes in dwarf galaxies}",
      journal = {Nature Astronomy},
         year = 2022,
        month = jan,
       volume = {6},
        pages = {26-34},
          doi = {10.1038/s41550-021-01556-0},
archivePrefix = {arXiv},
       eprint = {2201.10569},
 primaryClass = {astro-ph.GA},
       adsurl = {https://ui.adsabs.harvard.edu/abs/2022NatAs...6...26R}
}

@ARTICLE{Reinoso2018,
       author = {{Reinoso}, B. and {Schleicher}, D.~R.~G. and {Fellhauer}, M. and {Klessen}, R.~S. and {Boekholt}, T.~C.~N.},
        title = "{Collisions in primordial star clusters. Formation pathway for intermediate mass black holes}",
      journal = {\aap},
         year = 2018,
        month = jun,
       volume = {614},
          eid = {A14},
        pages = {A14},
          doi = {10.1051/0004-6361/201732224},
archivePrefix = {arXiv},
       eprint = {1801.05891},
 primaryClass = {astro-ph.GA},
       adsurl = {https://ui.adsabs.harvard.edu/abs/2018A&A...614A..14R}
}

@ARTICLE{Reinoso2023,
       author = {{Reinoso}, Basti{\'a}n and {Klessen}, Ralf S. and {Schleicher}, Dominik and {Glover}, Simon C.~O. and {Solar}, P.},
        title = "{Formation of supermassive stars in the first star clusters}",
      journal = {\mnras},
         year = 2023,
        month = may,
       volume = {521},
       number = {3},
        pages = {3553-3569},
          doi = {10.1093/mnras/stad790},
archivePrefix = {arXiv},
       eprint = {2303.07827},
 primaryClass = {astro-ph.GA},
       adsurl = {https://ui.adsabs.harvard.edu/abs/2023MNRAS.521.3553R}
}

@ARTICLE{Rinaldi2024,
       author = {{Rinaldi}, Stefano and {Del Pozzo}, Walter and {Mapelli}, Michela and {Lorenzo-Medina}, Ana and {Dent}, Thomas},
        title = "{Evidence of evolution of the black hole mass function with redshift}",
      journal = {\aap},
         year = 2024,
        month = apr,
       volume = {684},
          eid = {A204},
        pages = {A204},
          doi = {10.1051/0004-6361/202348161},
archivePrefix = {arXiv},
       eprint = {2310.03074},
 primaryClass = {astro-ph.HE},
       adsurl = {https://ui.adsabs.harvard.edu/abs/2024A&A...684A.204R}
}

@ARTICLE{Rizzuto2021,
       author = {{Rizzuto}, Francesco Paolo and {Naab}, Thorsten and {Spurzem}, Rainer and {Giersz}, Mirek and {Ostriker}, J.~P. and {Stone}, N.~C. and {Wang}, Long and {Berczik}, Peter and {Rampp}, M.},
        title = "{Intermediate mass black hole formation in compact young massive star clusters}",
      journal = {\mnras},
         year = 2021,
        month = mar,
       volume = {501},
       number = {4},
        pages = {5257-5273},
          doi = {10.1093/mnras/staa3634},
archivePrefix = {arXiv},
       eprint = {2008.09571},
 primaryClass = {astro-ph.GA},
       adsurl = {https://ui.adsabs.harvard.edu/abs/2021MNRAS.501.5257R}
}

@ARTICLE{Rizzuto2022,
       author = {{Rizzuto}, Francesco Paolo and {Naab}, Thorsten and {Spurzem}, Rainer and {Arca-Sedda}, Manuel and {Giersz}, Mirek and {Ostriker}, Jeremiah Paul and {Banerjee}, Sambaran},
        title = "{Black hole mergers in compact star clusters and massive black hole formation beyond the mass gap}",
      journal = {\mnras},
         year = 2022,
        month = may,
       volume = {512},
       number = {1},
        pages = {884-898},
          doi = {10.1093/mnras/stac231},
archivePrefix = {arXiv},
       eprint = {2108.11457},
 primaryClass = {astro-ph.GA},
       adsurl = {https://ui.adsabs.harvard.edu/abs/2022MNRAS.512..884R}
}

@ARTICLE{Rizzuto2023,
       author = {{Rizzuto}, Francesco Paolo and {Naab}, Thorsten and {Rantala}, Antti and {Johansson}, Peter H. and {Ostriker}, Jeremiah P. and {Stone}, Nicholas C. and {Liao}, Shihong and {Irodotou}, Dimitrios},
        title = "{The growth of intermediate mass black holes through tidal captures and tidal disruption events}",
      journal = {\mnras},
         year = 2023,
        month = may,
       volume = {521},
       number = {2},
        pages = {2930-2948},
          doi = {10.1093/mnras/stad734},
archivePrefix = {arXiv},
       eprint = {2211.13320},
 primaryClass = {astro-ph.GA},
       adsurl = {https://ui.adsabs.harvard.edu/abs/2023MNRAS.521.2930R}
}

@ARTICLE{RomanGarza2026,
       author = {{Roman-Garza}, J. and {Fragos}, T. and {Charbonnel}, C. and {Ram{\'\i}rez-Galeano}, L. and {Kruckow}, M. and {Farag}, E.},
        title = "{Massive stellar cannibals: How stellar mergers drive mass-loss in extremely massive stars}",
      journal = {arXiv e-prints},
         year = 2026,
        month = feb,
          eid = {arXiv:2602.02141},
        pages = {arXiv:2602.02141},
archivePrefix = {arXiv},
       eprint = {2602.02141},
 primaryClass = {astro-ph.SR},
       adsurl = {https://ui.adsabs.harvard.edu/abs/2026arXiv260202141R}
}

@ARTICLE{Ryon2017,
       author = {{Ryon}, J.~E. and {Gallagher}, J.~S. and {Smith}, L.~J. and {Adamo}, A. and {Calzetti}, D. and {Bright}, S.~N. and {Cignoni}, M. and {Cook}, D.~O. and {Dale}, D.~A. and {Elmegreen}, B.~E. and {Fumagalli}, M. and {Gouliermis}, D.~A. and {Grasha}, K. and {Grebel}, E.~K. and {Kim}, H. and {Messa}, M. and {Thilker}, D. and {Ubeda}, L.},
        title = "{Effective Radii of Young, Massive Star Clusters in Two LEGUS Galaxies}",
      journal = {\apj},
         year = 2017,
        month = jun,
       volume = {841},
       number = {2},
          eid = {92},
        pages = {92},
          doi = {10.3847/1538-4357/aa719e},
archivePrefix = {arXiv},
       eprint = {1705.02692},
 primaryClass = {astro-ph.GA},
       adsurl = {https://ui.adsabs.harvard.edu/abs/2017ApJ...841...92R}
}

@ARTICLE{Sabhahit2023,
       author = {{Sabhahit}, Gautham N. and {Vink}, Jorick S. and {Sander}, Andreas A.~C. and {Higgins}, Erin R.},
        title = "{Very massive stars and pair-instability supernovae: mass-loss framework for low metallicity}",
      journal = {\mnras},
         year = 2023,
        month = sep,
       volume = {524},
       number = {1},
        pages = {1529-1546},
          doi = {10.1093/mnras/stad1888},
archivePrefix = {arXiv},
       eprint = {2306.11785},
 primaryClass = {astro-ph.SR},
       adsurl = {https://ui.adsabs.harvard.edu/abs/2023MNRAS.524.1529S}
}

@ARTICLE{Sakurai2017,
       author = {{Sakurai}, Yuya and {Yoshida}, Naoki and {Fujii}, Michiko S. and {Hirano}, Shingo},
        title = "{Formation of intermediate-mass black holes through runaway collisions in the first star clusters}",
      journal = {\mnras},
         year = 2017,
        month = dec,
       volume = {472},
       number = {2},
        pages = {1677-1684},
          doi = {10.1093/mnras/stx2044},
archivePrefix = {arXiv},
       eprint = {1704.06130},
 primaryClass = {astro-ph.GA},
       adsurl = {https://ui.adsabs.harvard.edu/abs/2017MNRAS.472.1677S}
}

@ARTICLE{Samsing2018,
       author = {{Samsing}, Johan and {Leigh}, Nathan W.~C. and {Trani}, Alessandro A.},
        title = "{Implementing tidal and gravitational wave energy losses in few-body codes: A fast and easy drag force model}",
      journal = {\mnras},
         year = 2018,
        month = dec,
       volume = {481},
       number = {4},
        pages = {5436-5444},
          doi = {10.1093/mnras/sty2247},
archivePrefix = {arXiv},
       eprint = {1803.08215},
 primaryClass = {astro-ph.HE},
       adsurl = {https://ui.adsabs.harvard.edu/abs/2018MNRAS.481.5436S}
}

@ARTICLE{Soltan1982,
       author = {{Soltan}, A.},
        title = "{Masses of quasars.}",
      journal = {\mnras},
         year = 1982,
        month = jul,
       volume = {200},
        pages = {115-122},
          doi = {10.1093/mnras/200.1.115},
       adsurl = {https://ui.adsabs.harvard.edu/abs/1982MNRAS.200..115S}
}

%%%%%%%%%%%%%%%%%%%%%%%%%%%%%%%%%%%%%%%%%%%%%%%%%%
%%%%%%%%%%%%%%%%% APPENDICES %%%%%%%%%%%%%%%%%%%%%

\appendix

\section{The cumulative mass loss in multiple stellar collisions}\label{appendix: cumuloss}

\begin{figure}
\includegraphics[width=1.0\columnwidth]{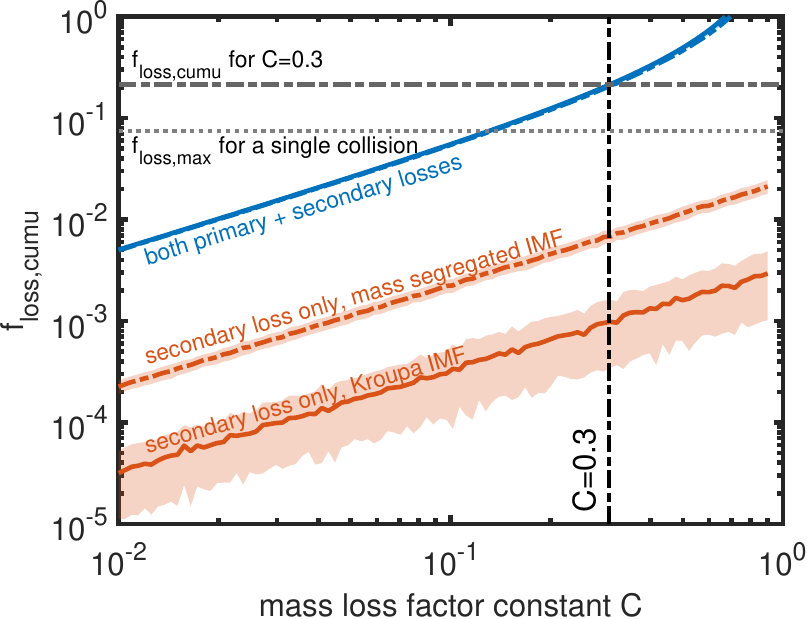}
\caption{The cumulative mass loss fraction $f_\mathrm{loss,cumu}$ as a function of the mass loss factor constant $C$ of Eq. \eqref{eq: phidef}. When including mass loss from both the primary and secondary colliding stars, the cumulative fractional mass loss in a collision cascade can be higher compared to the maximum mass loss fraction in a single collision.}
\label{fig: floss-appendix}
\end{figure}

As an example of cumulative mass loss in a large number of stellar collisions, we consider an extremely massive star with $m_\mathrm{1,init}=\msol{1000}$ doubling its mass to reach $m_\mathrm{1,final}=\msol{2000}$ via collisional mass growth. In the collisional mass loss model adopted for this study, each collision with a secondary star with a mass of $m_\mathrm{2}$ leads to a mass loss of $m_\mathrm{loss} = f_\mathrm{loss}(m_\mathrm{1}+m_\mathrm{2})$. Following Eq. \eqref{eq: phidef} $f_\mathbf{loss} = Cq/(1+q)^2$ with a fiducial free parameter value of $C=0.3$. We evaluate the total mass loss $m_\mathrm{loss,tot} = \sum_\mathrm{i} f_\mathrm{loss,i}(m_\mathrm{1,i}+m_\mathrm{2,i})$ during a series of collisions doubling the mass of $m_\mathrm{1,init}$ and finally evaluate $f_\mathrm{loss,cumu} = m_\mathrm{loss,tot}/m_\mathrm{1,final}$. We examine two different mass functions for the secondary stars: the \cite{Kroupa2001} IMF ($\msol{0.08} \leq m_\star \leq \msol{150}$) and one produced by mass segregation in which massive stars beyond $\gtrsim \msol{5}$ are over-represented (see e.g. figure 11 of \citealt{Rantala2024b}). For the recipe including mass loss from both of the colliding stars (primary and secondary), the cumulative mass loss factor $f_\mathrm{loss,cumu}$ is almost independent of the mass function of the secondary stars $m_\mathrm{2}$. This is because most of the ejected mass originates from the primary star. We show the $f_\mathrm{loss,cumu}$ as a function of the constant $C$ of the model in Fig. \ref{fig: floss-appendix}. As expected, increasing the value of $C$ increases $f_\mathrm{loss,cumu}$. For the fiducial value of $C=0.3$, the cumulative mass loss fraction $f_\mathrm{loss,cumu}\sim0.214$, approximately $\sim2.85$ times the maximum mass loss fraction $f_\mathrm{loss,max}=0.075$ in a single equal-mass collision. Compared to a collision with $q=\msol{150}/ \msol{1000}$, $f_\mathrm{loss,cumu}$ is $6.29$ times larger. Overall, the cumulative mass loss fractions increase with increasing $m_\mathrm{1,final}/m_\mathrm{1,init}$.

For \nbody{} simulation models that only include mass loss from the secondary star the situation is different. In Fig. \ref{fig: floss-appendix} we show that in the case of secondary star losses only the cumulative mass loss fraction depends on the mass function of the colliding stars in the absence of mass loss from the primary. The on average more massive stars from the mass segregated IMF lead to larger mass loss compared to the standard Kroupa IMF. Overall, models with primary star mass loss lead to more efficient collisional mass loss by $1$--$2$ orders of magnitude depending on the mass function of the colliding stars. For $C=0.3$, $f_\mathrm{loss,cumu}\sim10^{-3}$ for Kroupa IMF and $\sim7\times10^{-3}$ for the mass segregated mass function.

%%%%%%%%%%%%%%%%%%%%%%%%%%%%%%%%%%%%%%%%%%%%%%%%%%
\section{Massive star wind loss rate uncertainties}\label{appedix: wind}

\begin{figure}
\includegraphics[width=1.0\columnwidth]{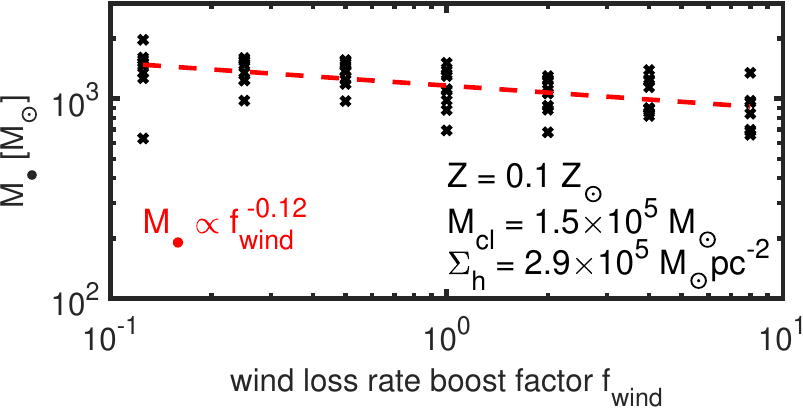}
\caption{The effect of weakened or boosted winds on the maximum final IMBH masses in the isolated models. In the simulations the wild mass loss rate of \citet{Vink2018} multiplied by the boost factor $f_\mathrm{wind}$. For star clusters with intermediate mass ($M_\mathrm{cl}=\msol{1.5\times10^5}$) and density ($\Sigma_\mathrm{h}=\Sigmasol{2.9\times10^5}$) the final IMBH masses depend on the boost factor relatively weakly as $M_\bullet \propto f_\mathrm{wind}^\mathrm{-0.12}$.}
\label{fig: appendix-windrate-uncertainty}
\end{figure}

In order to estimate how sensitive the IMBH masses in our simulations are to the stellar wind rate uncertainties, we perform $60$ additional isolated star cluster simulations in which we artificially boost or weaken the winds of $>\msol{600}$ stars. The additional models are all based on the isolated setup I3D5Z5 with a cluster mass of $M_\mathrm{cl}=\msol{1.5\times10^5}$, half mass surface density $\Sigma_\mathrm{h}=\Sigmasol{2.9\times10^5}$ and metallicity of $Z=\zsol{0.10}$. In the models we multiply the \cite{Vink2018} wind mass loss rate by a boost factor $f_\mathrm{wind}$ in the range of $0.125 \leq f_\mathrm{wind} \leq 8.0$. The final IMBH masses in the simulations are shown in Fig. \ref{fig: appendix-windrate-uncertainty}. The dependence of the mean IMBH masses for a given $f_\mathrm{wind}$ follows the empirical relation of
\begin{equation}
    \log_\mathrm{10}\left( M_\mathrm{\bullet,mean} \right) = (-0.12\pm0.01) \times \log_\mathrm{10}( f_\mathrm{wind} ) + (3.07\pm0.01)
\end{equation}
while for the maximum IMBH masses we have
\begin{equation}
    \log_\mathrm{10}\left( M_\mathrm{\bullet,max} \right) = (-0.08\pm0.02) \times \log_\mathrm{10}( f_\mathrm{wind} ) + (3.18\pm0.01).
\end{equation}
Both the mean and maximum IMBH masses depend on the wind boost factor relatively weakly, at least for the models I3D5Z5.

%%%%%%%%%%%%%%%
\section{The effect of the maximum wind rate limit}\label{appendix: wind-limit}

\begin{figure}
\includegraphics[width=1.0\columnwidth]{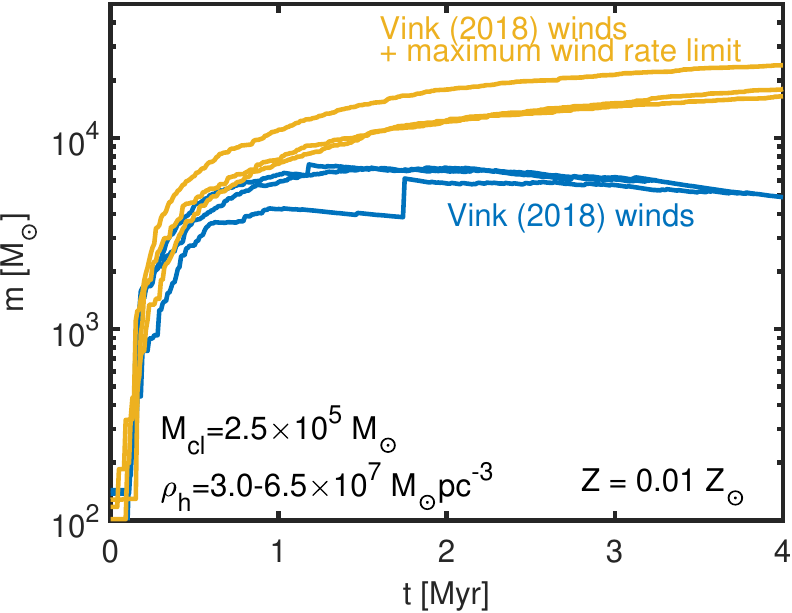}
\caption{The effect of assuming a maximum wind loss rate limit on the collisional stellar growth histories in extremely dense star clusters. With the fiducial \citet{Vink2018} winds the collisional runaway mass growth is quenched at $\sim\msol{7000}$ even at $Z=\zsol{0.01}$, including the maximum wind loss rate of $\max(\dot{m}_\mathrm{wind})=\ratesol{1.5\times10^{-4}}$ leads to monotonic growth into the SMS regime beyond $\msol{25000}$.}
\label{fig: appendix-windrate-limit}
\end{figure}

We perform three additional isolated simulations of the extremely dense central sub-clusters of the models HD9Z1 to study the effect of a maximum wind mass loss rate limit on the maximum masses of the collisionally grown stars. While our models based on \cite{Vink2018} do not include such an upper limit for the wind loss rates, recent \nbody{} studies such as \cite{Vergara2025,Vergara2025b} feature such a limit in their models reaching collisional stellar masses of $\msol{50000}$. In Fig. \ref{fig: postprocess} we showed using simulation post-processing that such a maximum wind rate limit allows for monotonic collisional stellar growth well into the SMS regime ($\gtrsim20000$) while without the limit the maximum stellar masses plateau at $\sim \msol{7000}$ for $Z=\zsol{0.01}$. We proceed to test this scenario in a direct simulation instead of post-processing only. Following \cite{Vergara2025}, we set $\max(\dot{m}_\mathrm{wind})=\ratesol{1.5\times10^{-4}}$ in the additional test simulations. The mass growth histories of three most massive stars in the models, with and without the maximum wind rate limit, are displayed in Fig. \ref{fig: appendix-windrate-limit}. In the wind limited models the collisional mass growth proceeds well into the SMS regime, just as in the post-processed models, the most massive SMS reaching $\msol{25000}$ during the first $t=4$ Myr of its life. We do not reach the $\msol{50000}$ star of \cite{Vergara2025} even in out wind rate limited models, which we attribute to the smaller EMS and SMS radii in late stellar evolutionary stages, and the lower host star cluster mass we have.

%%%%%%%%%%%%%%%

% Don't change these lines
\bsp	% typesetting comment
\label{lastpage}
\end{document}